\documentclass[12pt,a4paper]{article}
\usepackage{amsmath,amssymb,bm,epsfig,color,graphicx,cite,braket,mathrsfs,slashed,multirow,subcaption}
\usepackage[compat=1.1.0]{tikz-feynhand}

\usepackage{hyperref,xcolor}
\hypersetup{
    colorlinks=true,
    linktocpage=true,
    linkcolor=blue,
    filecolor=blue,      
    urlcolor=blue,
    citecolor=blue,
    }

\numberwithin{equation}{section}

\renewcommand{\thefootnote}{\fnsymbol{footnote}}

\begin{document}

\title{
\begin{flushright}
\begin{minipage}{0.2\linewidth}
\small
CTPU-PTC-22-23 \\*[40pt]
\end{minipage}
\end{flushright}
{\Large \bf
A review of neutrino decoupling \\
 from the early universe to the current universe
\\*[20pt]}}

\author{
Kensuke Akita$^{1}$\footnote{
E-mail: \href{mailto:kensuke8a1@ibs.re.kr}{kensuke8a1@ibs.re.kr}}\ \ and\ \ 
Masahide Yamaguchi$^{2}$\footnote{
E-mail: \href{mailto:gucci@phys.titech.ac.jp}{gucci@phys.titech.ac.jp}}\\*[20pt]
$^1${\it \small
Center for Theoretical Physics of the Universe, Institute for Basic Science,
Daejeon 34126, Korea} \\
$^2${\it \small
Department of Physics, Tokyo Institute of Technology,
Tokyo 152-8551, Japan} \\*[50pt]
}

\date{
\centerline{\small \bf Abstract}
\begin{minipage}{0.9\linewidth}
\medskip \medskip \small 
We review the distortions of spectra
of relic neutrinos due to the interactions with electrons,
positrons, and neutrinos in the early universe.
We solve integro-differential kinetic equations for the neutrino density matrix,
including vacuum three-flavor neutrino oscillations, oscillations in electron and positron background, a collision term
and finite temperature corrections to electron mass and electromagnetic plasma up to the next-to-leading order $\mathcal{O}(e^3)$.
After that, we estimate the effects of the spectral distortions in  neutrino decoupling
 on the number density and energy density of the Cosmic Neutrino Background (C$\nu$B) in the current universe,
and discuss the implications of these effects on the capture rates in direct detection of the C$\nu$B on tritium, with emphasis on the PTOLEMY-type experiment.  
In addition, we find a precise value of the
effective number of neutrinos, $N_{\rm eff}=3.044$. 
However, QED corrections to weak interaction rates at order $\mathcal{O}(e^2 G_F^2)$ and forward scattering of neutrinos via their self-interactions have not been precisely taken into account in the whole literature so far.
Recent studies suggest that these neglections might induce uncertainties of $\pm(10^{-3} - 10^{-4})$ in $N_{\rm eff}$.
\end{minipage}
}

\maketitle
\thispagestyle{empty}
\clearpage
\tableofcontents
\clearpage

\renewcommand{\thefootnote}{\arabic{footnote}}
\setcounter{footnote}{0}

\section{Introduction}

The successful hot big bang model after inflation predicts that
neutrinos produced in the early universe still exist in the current
universe. 
After the temperature of the universe dropped  below $T \sim 2\ {\rm MeV}$,
weak interactions became ineffective and neutrinos would have decoupled from thermal plasma. 
Analogous to photons that make up the Cosmic Microwave Background (CMB),
these decoupled neutrinos are called the Cosmic Neutrino Background (C$\nu$B).
The existence of these relic neutrinos is confirmed indirectly by the
observations of primordial abundances of light elements from the Big Bang
Nucleosynthesis (BBN), the anisotropies of the CMB
 and the distribution of Large Scale Structure (LSS) of
the universe.
In particular, observations from the Planck satellite impose the severe constraint on the effective number of relativistic species 
$N_{\rm eff}$, which describes the total neutrino energy in the Standard Model (SM), and the sum of the neutrino masses at $95\%$ CL as \cite{Planck:2018vyg}
\begin{align}
&N_{\rm eff} \equiv \frac{8}{7}\left(\frac{11}{4} \right)^{4/3}\left[\frac{\rho_r}{\rho_{\gamma}}-1 \right] = 2.99^{+0.34}_{-0.33} \ \ \ \  {\rm and}\ \ \ \  \sum m_{\nu}< 0.12\ {\rm eV},
\end{align}
where $\rho_\gamma$ and $\rho_r$ are the energy densities of photons and radiation, which is composed of photons and neutrinos in the SM, respectively.

Future observations of the C$\nu$B will be developed both indirectly and directly.
In fact, CMB-S4 observations are expected to determine $N_{\rm eff}$ with a very good precision of $\sim 0.03$ at 68 $\%$ C.L. \cite{Abazajian:2019eic}.
Thus, an estimation of $N_{\rm eff}$ in the SM with $10^{-3}$ precision will be important towards the future CMB-S4 observation.  
In addition, although it is still very difficult to observe the C$\nu$B in a direct way at present, 
it is inconceivable that the C$\nu$B will never be directly observed. 
Among the various discussions on the direct observations, 
the most promising method of direct detection of the C$\nu$B is neutrino capture on $\beta$-decaying nuclei \cite{Weinberg:1962zza, Cocco:2007za}, 
$\nu+n\rightarrow p+e^-$, where there is no threshold energy for relic cosmic neutrinos.
In both cases, the theoretical prediction of the relic neutrino spectrum is a crucial ingredient 
since the radiation energy density in $N_{\rm eff}$ and the direct detection rates depend on the spectrum, and their deviations from the SM suggest physics beyond the SM.

Soon after the decoupling of neutrinos, $e^{\pm}$-pairs 
start to annihilate and heat photons when the temperature
of the universe is $T\sim m_e = 0.511\ {\rm MeV}$. 
If neutrinos decoupled instantaneously and all electrons and positrons annihilated into
photons, the ratio for the temperatures of cosmic photons and neutrinos would be $T_{\gamma}/T_{\nu}=(11/4)^{1/3}
\simeq 1.40102$, due to entropy conservation of the universe.
However, the temperatures of neutrino decoupling and $e^{\pm}$-annihilations
are so close that $e^{\pm}$-pairs slightly annihilate into neutrinos, which leads to non-thermal distortions in neutrino spectra and a less increase in the photon temperature.
These modifications are also parametrized by an increase of $N_{\rm eff}$ from 3. 

The non-thermal distortions of relic neutrino spectra and the precise value of $N_{\rm eff}$ 
have long been studied by solving kinetic equations for neutrinos, which are the Boltzmann equations and the continuity equation.
First, several studies solved the
Boltzmann equations for neutrino distribution functions 
\cite{Dodelson:1992km, Dolgov:1992qg, Fields:1992zb, Hannestad:1995rs, Dolgov:1997mb,Dolgov:1998sf,
Esposito:2000hi, Hannestad:2001iy}. Then the kinetic equations were solved with
including finite temperature radiative corrections at leading order $\mathcal{O}(e^2)$
\cite{Fornengo:1997wa, Mangano:2001iu, Birrell:2014uka, Grohs:2015tfy,
Grohs:2017iit, Froustey:2019owm}, and then including three-flavor neutrino oscillations
the Boltzmann equations for a neutrino density matrix formalism were solved
\cite{Mangano:2005cc, deSalas:2016ztq, Gariazzo:2019gyi}. 
A fast and precise method to calculate effective neutrino temperature for all neutrino species and $N_{\rm eff}$ was also proposed\cite{Escudero:2018mvt, EscuderoAbenza:2020cmq}.  
Recently, the authors in ref. \cite{Bennett:2019ewm} pointed out that the finite temperature corrections to electromagnetic plasma at the next-to-leading order $\mathcal{O}(e^3)$ are expected to decrease $N_{\rm eff}$ by $10^{-3}$. 
After that, the present authors found a precise value of $N_{\rm eff}=3.0439\simeq 3.044$ \cite{Akita:2020szl} by solving the Boltzmann equations for the neutrino density matrix including the corrections to electron mass and electromagnetic plasma up to $\mathcal{O}(e^3)$ but neglecting off-diagonal parts derived from self-interactions of neutrinos. 
Later, the authors in refs. \cite{Froustey:2020mcq, Bennett:2020zkv} estimate $N_{\rm eff}=3.0440$ and $3.0440\pm 0.0002$, respectively, including off-diagonal parts of the collision term derived by neutrino self-interactions.
However, QED corrections to weak interaction rates at the order $\mathcal{O}(e^2 G_F^2)$ and forward scattering of neutrinos via their self-interactions have not been precisely taken into account in the above references so far.
Recent studies \cite{EscuderoAbenza:2020cmq, Hansen:2020vgm} suggest that these omissions might still induce uncertainties of $\pm(10^{-3} - 10^{-4})$ in $N_{\rm eff}$.

If we observe the C$\nu$B in a direct way in addition to its indirect observations, we might see neutrino decoupling directly.
In the current universe, since the average momentum of the C$\nu$B is $\langle p_\nu \rangle\sim 0.53~{\rm meV} \ll \sqrt{\Delta
m^2_{21}}, \sqrt{|\Delta m^2_{31}|}$, two massive neutrinos at least are
non-relativistic. Under such a situation, it is quite nontrivial to quantize neutrinos in the flavor basis.
To reveal the contribution of $e^{\pm}$-annihilation in neutrino decoupling to the spectrum of the C$\nu$B, 
we calculated the spectra, number densities and energy densities for relic neutrinos in the mass-diagonal basis in the current homogeneous and isotropic universe \cite{Akita:2020szl, Akita:2020jbo}.
 
In this article, we present a review of the distorted spectra of relic cosmic neutrinos 
from neutrino decoupling to the current universe based on refs.~\cite{Akita:2020szl, Akita:2020jbo}.
First, in section \ref{sec:2}, we describe the
kinetic equations for cosmic neutrinos.
In section \ref{sec:3}, we present our results of relic neutrino spectra and $N_{\rm eff}$.
Here we also discuss the uncertainties in $N_{\rm eff}$.
In section \ref{sec:4}, we calculate the number density and energy density of the C$\nu$B in the present universe. 
In section \ref{sec:5}, the impact of the distortions of the spectra in neutrino decoupling on neutrino capture experiments is also discussed. One of such experiments, which is called the PTOLEMY-type experiment \cite{Betti:2018bjv, PTOLEMY:2019hkd}, uses 100 g of tritium \cite{Akita:2020jbo, Lazauskas:2007da, Blennow:2008fh, Long:2014zva, Roulet:2018fyh} as a target through the reaction, $\nu_i+\mathrm{^3H}\rightarrow e^- + \mathrm{^3He}$. 
Tritium is an appropriate candidate for the target due to its availability, high neutrino capture cross section, low Q-value and long half lifetime of $t_{1/2}=12.32$ years. 
Here we also include the effects of gravitational clustering of the C$\nu$B by our Galaxy and nearby galaxies based on the results in ref.~\cite{Mertsch:2019qjv}. Finally, conclusions and discussion are given in section\ \ref{sec:con}.

\clearpage
\section{Kinetic equations for neutrinos in their decoupling} 
\label{sec:2}
To follow relic neutrino spectra from neutrino decoupling to the current homogeneous and isotropic universe, we first discuss the field operators and the density matrix for relativistic and non-relativistic neutrinos. Then we introduce the kinetic equations for neutrinos, which are the Boltzmann equations for the evolution of the neutrino density matrix known as the quantum kinetic equations. The continuity equations for the evolution of the total energy density are also introduced.

\subsection{Field operators and density matrix}
\label{sec:2.1}

We consider field operators of neutrinos and their
density matrices in a homogeneous and isotropic system. 
With neutrino masses, 
we cannot define annihilation and creation operators for neutrinos 
in flavor basis due to their off-diagonal masses in the conventional way, where we interpret these operators as operators that annihilate and create a state with eigenvalues of energy and momentum.
On the other hand, in the mass-diagonal basis, we can define such annihilation and creation operators, including neutrino masses.
We also compare relic cosmic neutrino spectra obtained in the two bases and confirm their match.

In the ultra-relativistic limit, the
field operators for left-handed flavor neutrinos in terms of 4-component spinors, which are composed of only active states for Majorana neutrinos and both active and sterile states for Dirac neutrinos, are
expanded in terms of plane wave solutions as 
\begin{align}
\bm{\nu}_\alpha(x) = \int \frac{d^3{\bm
 p}}{(2\pi)^3 \sqrt{2p_0}}\left(a_\alpha({\bm{p}},t)u_{\bm{p}}e^{i\bm{p\cdot x}} 
+b_\alpha^{\dag}({\bm{p}},t)v_{\bm{p}} e^{-i\bm{p\cdot x}} \right),
\label{operator}
\end{align} 
where $a_\alpha(\bm{p},t) = e^{iHt}a_\alpha(\bm{p}) e^{-iHt}$ and
$b_\alpha(\bm{p},t) = e^{-iHt}b_\alpha(\bm{p})e^{iHt}$ are annihilation
operators for negative-helicity neutrinos and positive-helicity anti-neutrinos,
respectively, and $H$ is the Hamiltonian. $\alpha$ and $\bm{p}$ are a flavor index and a three
dimensional momentum with $p_0 \simeq |\bm{p}|$,
respectively. $u_{\bm p}\ (v_{\bm p})$ denotes the Dirac spinor for a
massless negative-helicity particle (positive-helicity anti-particle), which is
normalized to be
$u_{\bm{p}}^{\dagger}u_{\bm{p}} = v_{\bm{p}}^{\dagger}v_{\bm{p}} = 2
p_0$.
The annihilation and creation operators satisfy the anti-commutation relations,
\begin{align}
\{ a_\alpha(\bm{p}), a_\beta^{\dag}(\bm{p}') \} = \{b_\alpha(\bm{p}), b_\beta^{\dag}(\bm{p}') \} = \delta_{\alpha\beta}(2\pi)^3\delta^{(3)}(\bm{p}-\bm{p}').
\end{align}
For freely evolving massless neutrinos without any interactions, $a_\alpha^0(\bm{p},t) = a_\alpha(\bm{p}) e^{-ip_0t}$ and
$b_\alpha^0(\bm{p},t) = b_\alpha(\bm{p})e^{-ip_0t}$ and the Dirac spinors satisfy free Dirac equations, 
$
/\hspace{-2.2mm}pu_{\bm p}^0=0,\ /\hspace{-2.2mm}pv_{\bm p}^0=0
$. On the other hand, for free massive neutrinos in the flavor basis, $a_\alpha^0(\bm{p},t)$ and $b_\alpha^0(\bm{p},t)$ cannot be expanded in terms of a plane wave with an eigenvalue of their energy due to off-diagonal neutrino masses. Then we cannot interpret $a_\alpha(\bm{p},t)$ and $b_\alpha(\bm{p},t)$ as annihilation operators except in the ultra-relativistic case.

The density matrices for neutrinos and anti-neutrinos in the flavor basis are defined
through the following expectation values of these operators
concerning the initial states,
\begin{align}
\langle a^{\dag}_\beta(\bm{p},t)a_\alpha(\bm{p}',t) \rangle &= (2\pi)^3\delta^{(3)}(\bm{p}-\bm{p'})\left(\rho_p\right)_{\alpha\beta}, \nonumber \\
\langle b^{\dag}_\alpha(\bm{p},t)b_\beta(\bm{p'},t) \rangle &= (2\pi)^3\delta^{(3)}(\bm{p}-\bm{p'})\left(\bar{\rho}_p\right)_{\alpha\beta},
\label{DM}
\end{align}
where $p=|\bm{p}|$. Due to the reversed order of flavor indices in
$\bar{\rho}_p(t)$, both density matrices transform in the same way under
a unitary transformation of flavor space. Here the diagonal parts are
the usual distribution functions of flavor neutrinos and the
off-diagonal parts represent non-zero in the presence of flavor mixing.

On the other hand, in the mass-diagonal basis, the field operators for the negative helicity neutrinos
\footnote{If we follow the evolution of neutrinos until today,
it is also
easier to follow the evolution of negative-helicity neutrinos
in the mass-diagonal basis
since the helicity states of neutrinos are conserved while
non-relativistic neutrinos are freely streaming.
On the other hand, the chiral states for non-relativistic neutrinos are not conserved.} 
can be expanded as, including neutrino masses, 
\begin{align}
\bm{\nu}_i(x) = \int \frac{d^3{\bm p}}{(2\pi)^3 \sqrt{2E_i}}\left(a_i({\bm{p}},t)u^{(i)}_{\bm{p}}e^{i\bm{p\cdot x}}+b_i^{\dag}({\bm{p}},t)v^{(i)}_{\bm{p}}e^{-i\bm{p\cdot x}} \right),
\end{align}
where $i(=1,2,3)$ denotes a mass eigenstate, $a_i({\bm{p}},t) =
e^{iHt}a_i({\bm{p}})e^{-iHt} , b_i({\bm{p}},t) = e^{-iHt}b_i({\bm{p}})e^{iHt}$,
$E_i = \sqrt{\bm{p}^2+m_i^2}$ and $m_i$ is the neutrino mass in the
mass basis. $u^{(i)}_{\bm p}\ (v^{(i)}_{\bm p})$
denotes the Dirac spinor for negative-helicity particles\
(positive-helicity anti-particles), which is also normalized to be
$u^{(i)}_{\bm{p}}{}^{\dagger}u^{(i)}_{\bm{p}} = 
v^{(i)}_{\bm{p}}{}^{\dagger}v^{(i)}_{\bm{p}} = 2E_i$.
For freely evolving neutrinos, $a_i^0({\bm{p}},t) =
a_i({\bm{p}})e^{-iE_it} , b_i^0({\bm{p}},t) =b_i({\bm{p}})e^{-iE_it}$ and
the Dirac spinors satisfy
$(/\hspace{-2.2mm}p-m_i)u^{(i),0}_{\bm p}=0$ and $(/\hspace{-2.2mm}p+m_i)v^{(i),0}_{\bm p}=0.$
As in the flavor basis, the commutation relations for $a_i({\bm{p}})$ and $b_i({\bm{p}})$, and the density matrix are defined in the same way except for the exchange of the subscripts, $\alpha \leftrightarrow i$.

The diagonalization of the mass matrix for left-handed neutrinos in the flavor basis is achieved through the transformations,
\begin{align}
\bm{\nu}_{\alpha}(x) = \sum_{i=1,2,3}U_{\alpha i}\bm{\nu}_i(x),
\label{FOrelation}
\end{align}
where $U_{\alpha i}$ represents a component of
the Pontecorvo-Maki-Nakagawa-Sakata (PMNS) matrix $U_{\rm PMNS}$.
Due to eq.~(\ref{FOrelation}), in the ultra-relativistic limit, the relation of the density matrices in the flavor and the mass bases is described as
\begin{align}
\left(\rho_p\right)_{\alpha\beta}=\sum_{i,j=1,2,3}U_{\beta j}^{\ast}U_{\alpha i}\left(\rho_p\right)_{ij}
\end{align}
In addition, after neutrino decoupling, the off-diagonal parts of the density matrix in the mass basis are zero, $(\rho_p)_{ij}\simeq 0\ (i\neq j)$,
since all neutrino interactions are ineffective and the oscillations do not occur after neutrino decoupling.
In this case, the relations of distribution function in the two bases are simply\footnote{Note that eq.~(\ref{relation}) is different from eq.~(13) in ref.~\cite{Mangano:2005cc}}
\begin{align}
f_{\nu_\alpha}(\bm{p},t) = \sum_{i=1,2,3} |U_{\alpha i}|^2f_{\nu_i}(\bm{p},t).
\label{relation}
\end{align}
Note that eq.~(\ref{relation}) is only valid when neutrinos are relativistic and decoupled with thermal plasma. Our numerical calculations also confirm eq.~(\ref{relation}).

\subsection{Boltzmann equations }
In this section, we derive the Boltzmann equations for the neutrino density matrix, known as quantum kinetic equations, including neutrino oscillations in vacuum, forward scattering with $e^\pm, \nu, \bar{\nu}$-background, corresponding to neutrino oscillations in matter, and the collision process at tree level.
The resulting Boltzmann equations for neutrinos are summarized in section \ref{sec:2.2.5}, where we will also discuss the approximations we used in our numerical calculations.

\subsubsection{Boltzmann equations in a homogeneous and isotropic system}

The Boltzmann equations for neutrinos, including flavor conversion effects, are derived from the Heisenberg equations for the neutrino density operator,
\begin{align}
\frac{d}{dt}N_{\alpha\beta}(t)=i[H,N_{\alpha\beta}],
\label{Hequation}
\end{align}
where $[\cdot, \cdot]$ represents the
commutator of matrices with a flavor (or mass) index and $N_{\alpha\beta}$ is the neutrino density operator,
\begin{align}
N_{\alpha\beta}=a_\beta^\dag(\bm{p},t)a_\alpha(\bm{p},t).
\end{align}
$H$ is the full Hamiltonian in a system, which can be separated into
\begin{align}
    H=H_{\rm free}+H_{\rm int},
\end{align}
where $H_{\rm free}$ is the free Hamiltonian and $H_{\rm int}$ is the interaction Hamiltonian. 
We assume interactions are enough small that collisions occur individually. Then any fields can be regarded as free ones except during interactions.
When the interaction Hamiltonian can be treated perturbatively, the density operator evolves at the first order of $H_{\rm int}$,
\begin{align}
    N_{\alpha\beta}(t)
    &\simeq N_{\alpha\beta}^0(t)+i\int^t_{t_0} dt'\left[H_{\rm int}^0(t'), N_{\alpha\beta}^0(t) \right],
    \label{FOS}
\end{align}
where $t_0$ is the initial time and $H_{\rm int}^0$ is the interaction Hamiltonian as a function of freely evolving fields, which are solutions of free Dirac equations, and  $N_{\alpha\beta}^0(t)$ is the free density operator evolved as
\begin{align}
    N_{\alpha\beta}^0(t)=e^{iH_{\rm free}(t-t_0) }N_{\alpha\beta}(t_0)e^{-iH_{\rm free}(t-t_0)}
\end{align}
The first order solution (\ref{FOS}) includes only neutrino oscillation in vacuum and forward (momentum conserving) scattering with a medium in the system. 

To take into account momentum changing collisions, we consider the evolution equation for the density operator at second order of $H_{\rm int}$, substituting eq.~(\ref{FOS}) into eq.~(\ref{Hequation}), 
\begin{align}
    \frac{d}{dt}N_{\alpha\beta}(t)\simeq i\left[H_{\rm free}^0(t),N_{\alpha\beta}^0(t)\right]+i\left[H_{\rm int}^0(t),N_{\alpha\beta}^0(t)\right]-\int^t_{t_0}dt'\left[H_{\rm int}^0(t),\left[H_{\rm int}^0(t'),N_{\alpha\beta}^0(t)\right] \right],
    \label{BE111}
\end{align}
and an analogous equation for anti-neutrinos \cite{Sigl:1993ctk}, $\bar{N}_{\alpha\beta}\equiv b^{\dag}_\alpha(\bm{p},t)b_\beta(\bm{p},t)$, which is not solved in this article since we assume no lepton asymmetry. 
Here $H_{\rm free}^0$ is also the free Hamiltonian as a function of freely evolving fields, where we neglect would-be tiny corrections in the presence of interactions. We also ignore the tiny modification of oscillation and forward scattering, $\left[H_{\rm free},\left[H_{\rm int},N_{\alpha\beta}^0\right]\right]$ compared with $\left[H_{\rm free}^0,N_{\alpha\beta}^0(t)\right]$ and $\left[H_{\rm int}^0,N_{\alpha\beta}^0(t)\right]$. Note that the differential equation~(\ref{BE111}) is not closed for both $N_{\rm \alpha\beta}$ and $N_{\rm \alpha\beta}^0$.

To close and simplify the differential equation~(\ref{BE111}), we impose additional approximations.
We may set $t_0=0$ and $t\rightarrow \infty$ in the integral range since the time step of the change of  $N_{\alpha\beta}$, $t$, may be chosen to be small enough compared to the timescale of the evolution of the universe and large enough compared to the timescale of one collision, $t'$.
In addition, at $t=0$, the free density operator coincides with the full one, $N_{\alpha\beta}^0(0)=N_{\alpha\beta}(0)$.
Then eq.~(\ref{BE111}) can be rewritten as
\begin{align}
    \frac{d}{dt}N_{\alpha\beta}^0(0)=i\left[H_{\rm free}^0(0),N_{\alpha\beta}^0(0)\right]+i\left[H_{\rm int}^0(0),N_{\alpha\beta}^0(0)\right]-\frac{1}{2}\int^\infty_{-\infty}dt'\left[H_{\rm int}^0(0),\left[H_{\rm int}^0(t'),N_{\alpha\beta}^0(0)\right] \right].
\end{align}
Thus, the time evolution of the expectation value of $N_{\alpha\beta}^0(0)$ concerning the initial state, $\rho_p(0)$, is given by
\begin{align}
    (2\pi)^3\delta^{(0)}(0)\frac{d}{dt}\rho_p(0)&=i\left\langle\left[H_{\rm free}^0(0),N_{\alpha\beta}^0(0)\right]\right \rangle+i\left\langle\left[H_{\rm int}^0(0),N_{\alpha\beta}^0(0)\right]\right\rangle \nonumber \\
    &\ \ \ \ -\frac{1}{2}\int^\infty_{-\infty}dt'\left\langle\left[H_{\rm int}^0(0),\left[H_{\rm int}^0(t'),N_{\alpha\beta}^0(0)\right] \right]\right \rangle.
    \label{BE112}
\end{align}
eq.~(\ref{BE112}) will be valid at all times, even at $t\neq 0$, if in two or more collisions, the correlation of the particles in each collision is independent. This assumption is called molecular chaos in the derivation of the Boltzmann equation. In general, n-point correlation functions are produced by both forward and non-forward collisions. Under the assumption of molecular chaos, n-point correlation functions are reduced to combinations of two-point correlation functions as in ordinary scattering theory. Here two-point correlation functions correspond to distribution functions and neutrino density matrix.

The first term in the right hand side (RHS) represents neutrino oscillations in vacuum and the second term represents forward scattering of neutrinos with background in the system, which is called refractive effects and corresponds to neutrino oscillations in matter. These two terms do not change neutrino momenta but induce flavor conversions. The third term represents scattering and annihilation including both momentum conserving and changing processes, usually rewritten as 
\begin{align}
    -\frac{1}{2}\int^\infty_{-\infty}dt'\left\langle\left[H_{\rm int}^0(0),\left[H_{\rm int}^0(t'),N_{\alpha\beta}^0(0)\right] \right]\right \rangle\equiv (2\pi)^3\delta^{(3)}(0)C\left[\rho_p(t) \right],
\end{align}
where $C\left[\rho_p(t) \right]$ is called the collision term.
In the following sections, we calculate the formulae of these three terms.
The resulting Boltzmann equations for the neutrino density matrix are summarized in section \ref{sec:2.2.5}.

\subsubsection{Neutrino oscillation in vacuum}

The calculation of the first term in the RHS of eq.~(\ref{BE112}) is well established in the mass basis.
The free Hamiltonian of neutrinos in the mass basis is given by
\begin{align}
    H_{\rm free}=\int d^3x \sum_{i=1}^3\bar{\nu}_i(-i\bm{\gamma}\cdot\nabla+m_i)\nu_i,
\end{align}
where $\bm{\gamma}=(\gamma^1,\gamma^2,\gamma^3)$ are the gamma matrices.
After substituting the free operators for left-handed neutrinos, the free Hamiltonian becomes
\begin{align}
    H_{\rm free}^0=\int d^3p\sum_{i=1}^3\left[a_i^\dag(\bm{p})E_ia_i(\bm{p})+b_i^\dag(\bm{p})E_ib_i(\bm{p}) \right].
\end{align}
The first term in the RHS of eq.~(\ref{BE112}) in the mass basis is written as
\begin{align}
    i\left\langle \left[H_{\rm free}^0, N_{ij}^0(0)\right]\right\rangle&=
    -i(2\pi)^3\delta^{(3)}(0)\left[{\rm diag}(E_1,\ E_2,\ E_3), \rho_{p} \right], \nonumber \\
&\simeq-i(2\pi)^3\delta^{(3)}(0)\left[ \frac{\bm{\mathrm{M}}_{\rm diag}^2}{2p}, \rho_{p} \right],
\end{align}
where $\bm{\mathrm{M}}^2_{\rm diag}={\rm diag}(m_{\nu_1}^2,\ m_{\nu_2}^2,\ m_{\nu_3}^2)$ and $i,j$ denote mass-eigenstates.
In the flavor basis, as in discussed in section \ref{sec:2.1}, it is quite nontrivial to quantize neutrinos in the flavor basis with non-zero masses. 
When we calculate the first term in the RHS of eq.~(\ref{BE112}) in the flavor basis directly, we replace the free annihilation
operators $a_{\alpha}^0(\bm{p},t)$ and $b_{\alpha}^0(\bm{p},t)$ with $a_{\alpha}^{\rm
osc}(\bm{p},t)=(\exp(-i\Omega_{\bm{p}}t))_{\alpha\beta}a_{\beta}(\bm{p})$
and $b_{\alpha}^{\rm
osc}(\bm{p},t)=(\exp(-i\Omega_{\bm{p}}t))_{\alpha\beta}b_{\beta}(\bm{p})$
as in \cite{Sigl:1993ctk}, where $\Omega_{\bm{p}}=\sqrt{\bm{p}^2+\bm{\mathrm{M}}^2}$.
Then we also obtain the first term of eq.~(\ref{BE112})
, following the similar procedure in the mass basis,
\begin{align}
i\left\langle \left[H_{\rm free}^0, N_{\alpha\beta}^0(0)\right]\right\rangle
&\simeq-i(2\pi)^3\delta^{(3)}(0)\left[ \frac{\bm{\mathrm{M}}^2}{2p}, \rho_{p} \right],
\end{align}
where $\bm{\mathrm{M}}^2=U_{\rm PMNS}\bm{\mathrm{M}}^2_{\rm diag}U_{\rm PMNS}^{\dag}$ is the neutrino mass matrix in the flavor basis.
For anti-neutrinos, the corresponding term is obtained by adding a minus sign for the reverse indices in the anti-neutrino density matrix (\ref{DM}), $i\langle [H_{\rm free}^0, \bar{N}_{\alpha\beta}^0(0)] \rangle \simeq i(2\pi)^3\delta^{(3)}(0)[\bm{\mathrm{M}}^2/2p, \bar{\rho}_p]$.

\subsubsection{Forward scattering with $e^\pm, \nu, \bar{\nu}$-background}

\begin{figure}[htpb]    
    \begin{minipage}{0.5\hsize}
\begin{center}      
\begin{tikzpicture}[baseline=+1.85cm]
\begin{feynhand}
 \vertex [particle] (n0) at (-0.7,0.8) {$\nu_\alpha$};
 \vertex [particle] (n1) at (2,0.8);
  \vertex [particle] (n2) at (4.7,0.8) {$\nu_\alpha$};
 \vertex [particle] (i1) at (2,2);
    \vertex [particle] (f1) at (2,4);
    \propag [fermion, half left, looseness=1.64] (i1) to (f1);
    \propag [fermion, half left, looseness=1.64] (f1) to (i1);
    \propag [fermion] (n0) to (n1);
    \propag [fermion] (n1) to (n2);
    \propag [boson] (i1) to [edge label=$Z^0$] (n1);
    \node at (3.9,3)  {$\beta, \nu_\beta, \bar{\nu}_\beta$};
\end{feynhand}
\end{tikzpicture}
\end{center}
    \end{minipage}
    \begin{minipage}{0.5\hsize}
    \begin{center} 
\begin{tikzpicture}[baseline=+1.85cm]
\begin{feynhand}
 \vertex [particle] (n0) at (-0.7,0.8) {$\nu_\alpha$};
 \vertex [particle] (n1) at (0.8,0.8);
 \vertex [particle] (n2) at (3.2,0.8);
  \vertex [particle] (n3) at (4.7,0.8) {$\nu_\alpha$};
    \propag [boson, half left, looseness=1.2] (n1) to (n2);
    \propag [fermion] (n0) to (n1);
    \propag [fermion] (n1) to (n2);
     \propag [fermion] (n2) to (n3);
    \node at (2,2)  {$Z^0, W^\pm$};
    \node at (2,0.4)  {$\alpha, \nu_\alpha, \bar{\nu}_\alpha$};
\end{feynhand}
\end{tikzpicture}
\end{center}
    \end{minipage}
\caption{One-loop thermal contributions to forward scattering of neutrinos in the flavor basis with $\alpha,\beta=e^\pm, \mu^\pm$ and $\tau^\pm$. \underline{Left}: Tadpole diagram with all flavors in the one-loop.  \underline{Right}: Babble diagram with the same flavor in the one-loop.}
\label{fig:forwardoneloop}
\end{figure}
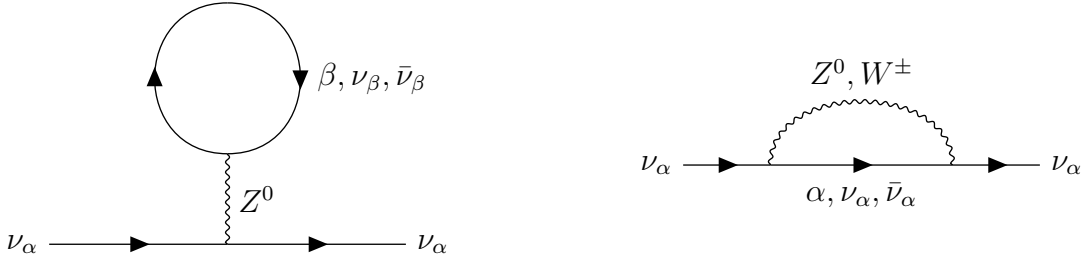

In the following of section~\ref{sec:2}, we consider the flavor basis of neutrinos.
Forward scattering of neutrinos with background in the system called refractive effects modifies neutrino oscillations through the one-loop thermal interaction as given in figure~\ref{fig:forwardoneloop}.
Since the temperature in thermal plasma is $\sim 2\ {\rm MeV}$ in neutrino decoupling, 
particles except for photons, electrons, neutrinos and their anti-particles are already annihilated due to their heavy masses. Then we consider only $e^\pm, \nu, \bar{\nu}$-background.
The interaction Hamiltonian is described as
\begin{align}
    H_{\rm int}&=\frac{g^2}{2}\int d^3xd^4y\frac{d^4p}{(2\pi)^4}e^{-ip(x-y)}\left[D_{\mu\nu}^Z(p)J^{\mu}_{NC}(x)J^{\nu}_{NC}(y)+2D_{\mu\nu}^W(p)J^{\mu\dag}_{CC}(x)J_{CC}^{\nu}(y)\right], \nonumber \\
    &\equiv H_{NC}+H_{CC},
\end{align}
where $D^Z_{\mu\nu}(p)$ and $D^W_{\mu\nu}(p)$ are the full propagator of $Z^0$ boson and $W^\pm$ boson,
\begin{align}
D^{W,Z}_{\mu \nu}(p)&=\left(g_{\mu\nu}-\frac{p_\mu p_\nu}{m_{W,Z}^2} \right)\frac{1}{m_{W,Z}^2-p^2}, \nonumber \\
&\simeq\frac{g_{\mu\nu}}{m_{W,Z}^2}+\frac{g_{\mu\nu}p^2-p_\mu p_\nu}{m_{W,Z}^4}.
\end{align}
Here $g, m_Z, m_W$ are the electroweak coupling constant, the $Z^0$ boson mass and the $W^\pm$ boson mass, respectively.
The neutral current and the charged current are given by
\begin{align}
    J_{NC}^{\mu}&\simeq J_{\nu\nu}^{\mu}+J_{ee}^{L\mu}+J_{ee}^{R\mu}, \nonumber \\
    J_{CC}^\mu &\simeq J_{e\nu_e}^{\mu},
    \label{Currents}
\end{align}
where
\begin{align}
    J_{\nu\nu}^{\mu}&=\frac{1}{4\cos\theta_W}\bar{\bm{\nu}}\gamma^\mu(1-\gamma_5)\bm{\nu}, \ \ \ \
J_{ee}^{L\mu}= \frac{1}{2\cos\theta_W}\left(-\frac{1}{2}+\sin^2\theta_W \right)\bar{\bm{e}}\gamma^\mu(1-\gamma_5)\bm{e},\nonumber \\
J_{ee}^{R\mu} &= \frac{1}{2\cos\theta_W}\sin^2\theta_W\bar{\bm{e}}\gamma^\mu(1+\gamma_5)\bm{e},\ \ \ \ 
J_{e\nu_e}^{\mu}=\frac{1}{2\sqrt{2}}\bar{\bm{\nu}}_e\gamma^\mu(1-\gamma_5)\bm{e},
\end{align}
with
\begin{align}
\bm{\nu}=\begin{pmatrix}
\bm{\nu}_e \\
\bm{\nu}_\mu  \\
\bm{\nu}_\tau \\
\end{pmatrix}.
\end{align}
Here $\theta_W$ is the weak mixing angle, $\bm{e}$ is the field operator for electron and positron and $\bm{\nu}_\alpha$ is the field operator for neutrinos and anti-neutrinos with a flavor $\alpha$.

The interaction Hamiltonian is divided into the two parts corresponding to the neutral current interaction, $H_{NC}\propto J_{NC}^\mu$, and to the charged current interaction, $H_{CC}\propto J_{CC}^\mu$.
For the charged current interactions, the second term in the RHS of eq.~(\ref{BE112}), which represents forward scattering of neutrinos with $e^\pm$-background, is given by \cite{Notzold:1987ik, Sigl:1993ctk}
\begin{align}
     &i\left\langle \left[H_{CC}^0(0), N_{\alpha\beta}^0(0) \right] \right\rangle \nonumber \\
     &= -i(2\pi)^3\delta^{(3)}(0)\left[\sqrt{2}G_F(\bm{\mathrm{N}}_{e^-}-\bm{\mathrm{N}}_{e^+})-\frac{2\sqrt{2}G_Fp}{3m_Z^2}\left(\bm{\mathrm{E}}_{e^-}+\bm{\mathrm{P}_{e^-}}+\bm{\mathrm{E}}_{e^+}+\bm{\mathrm{P}_{e^+}} \right), \rho_p \right],
     \label{CCFS}
\end{align}
where $G_F$ is the Fermi coupling constant and $\bm{\mathrm{N}}_{e^\pm}, \bm{\mathrm{E}}_{e^\pm}$ and $\bm{\mathrm{P}}_{e^\pm}$ are the number density, energy density and pressure for $e^\pm$-background, respectively, which are described in the flavor basis as
\begin{align}
    &\bm{\mathrm{N}}_{e^-}\simeq{\rm diag}(n_{e^-},\ 0,\ 0),\ \ \ \
    \bm{\mathrm{N}}_{e^+}\simeq{\rm diag}(n_{e^+},\ 0,\ 0),\ \ \ \ 
    n_{e^\pm}=2\int\frac{d^3p}{(2\pi)^3}f_{e^\pm}(p), \nonumber \\
    &\bm{\mathrm{E}}_{e^\pm}+\bm{\mathrm{P}}_{e^\pm}\simeq{\rm diag}(\rho_{e^\pm}+P_e{^\pm},\ 0,\ 0),
\ \ \rho_{e^\pm}+P_{e^\pm}=\int\frac{d^3p}{(2\pi)^3}\left(E_e+ \frac{p^2}{3E_e} \right)f_{e^\pm}(p),
\label{CLD}
\end{align}
where $E_e=\sqrt{p^2+m_e^2}$. In the temperature of MeV scale in neutrino decoupling, the densities for muons and tauons are enough suppressed by their heavy masses. We neglect forward scattering of neutrinos with muons and tauons, which corresponds to the second and third diagonal components in eq.~(\ref{CLD}).

For the neutral current interactions, the second term in the RHS of eq.~(\ref{BE112}), which represents forward scattering of neutrinos via neutrino self-interactions, is given by \cite{Notzold:1987ik, Sigl:1993ctk}
\begin{align}
     i\left\langle \left[H_{NC}^0(0), N_{\alpha\beta}^0(0) \right] \right\rangle= -i(2\pi)^3\delta^{(3)}(0)\left[\sqrt{2}G_F(\bm{\mathrm{N}}_\nu-\bm{\mathrm{N}}_{\bar{\nu}})-\frac{8\sqrt{2}G_Fp}{3m_Z^2}\left(\bm{\mathrm{E}}_\nu+\bm{\mathrm{E}}_{\bar{\nu}} \right), \rho_p \right],
      \label{NCFS}
\end{align}
where $\bm{\mathrm{N}}_\nu, \bm{\mathrm{N}}_{\bar{\nu}}, \bm{\mathrm{E}}_\nu$ and $\bm{\mathrm{E}}_{\bar{\nu}}$ are the number and energy densities for the density matrices of $\nu, \bar{\nu}$-background, respectively, which are described in the flavor basis as
\begin{align}
    &\bm{\mathrm{N}}_\nu=\int\frac{d^3p}{(2\pi)^3}\rho_p, \ \ \ \ \ 
   \bm{\mathrm{N}}_{\bar{\nu}}=\int\frac{d^3p}{(2\pi)^3}\bar{\rho}_p, \nonumber \\
   &\bm{\mathrm{E}}_\nu=\int\frac{d^3p}{(2\pi)^3}p\rho_p, \ \ \ \ 
   \bm{\mathrm{E}}_{\bar{\nu}}=\int\frac{d^3p}{(2\pi)^3}p\bar{\rho}_p,
\end{align}
where we neglect neutrino masses since neutrinos are relativistic in neutrino decoupling.

For anti-neutrinos, the corresponding terms, $i\langle [H_{CC}^0(0), \bar{N}^0_{\alpha\beta}(0) \rangle$ and $i\langle [H_{CC}^0(0), \bar{N}^0_{\alpha\beta}(0) \rangle$, are obtained by adding an overall minus sign for the reverse indices in the anti-neutrino density matrix (\ref{DM}) and replacing $\bm{\mathrm{N}}_{e^-}-\bm{\mathrm{N}}_{e^+}\rightarrow -(\bm{\mathrm{N}}_{e^-}-\bm{\mathrm{N}}_{e^+})$ and $\bm{\mathrm{N}}_{\nu}-\bm{\mathrm{N}}_{\bar{\nu}}\rightarrow -(\bm{\mathrm{N}}_{\nu}-\bm{\mathrm{N}}_{\bar{\nu}})$ for an opposite evolution of anti-neutrinos due to the lepton asymmetry in eqs.~(\ref{CCFS}) and (\ref{NCFS}) \cite{Sigl:1993ctk}.

If there is a large lepton asymmetry, the terms proportional to $\bm{\mathrm{N}}_{e^-}-\bm{\mathrm{N}}_{e^+}$ and $\bm{\mathrm{N}}_{\nu}-\bm{\mathrm{N}}_{\bar{\nu}}$ will be important.
Note that even if there is no lepton asymmetry, the off-diagonal parts of $\bm{\mathrm{N}}_{\nu}-\bm{\mathrm{N}}_{\bar{\nu}}$ have non-zero contribution since the density matrices for neutrinos and anti-neutrinos follow the same evolution, $\rho_p=\bar{\rho}_p^{\mathrm{T}} \neq \bar{\rho}_p$, in the case of no lepton asymmetry.

\subsubsection{Collision term}
Finally we discuss the third term in the RHS of eq.~(\ref{BE112}) called the collision term.
The temperature of $\sim 2\ {\rm MeV}$ in neutrino decoupling is much lower than the electroweak scale of $\sim m_Z,m_W$.
After integrating out $Z^0$ and $W^\pm$ bosons in the instantaneous interaction limit, the interaction Hamiltonian in neutrino decoupling can be written as
\begin{align}
H_{\rm int}\simeq\frac{g^2}{2}\int d^3x \biggl[\frac{1}{m_Z^2}J_{NC}^{\mu}(x)J_{NC\mu}(x)+\frac{2}{m_W^2}J_{CC}^{\dag \mu}(x)J_{CC\mu}(x)\biggl].
\label{HI4point}
\end{align}
The interaction Hamiltonian can be divided into the part including both neutrinos and electrons (and their anti-particles), and the one only including neutrinos and anti-neutrinos, $H_{\rm int}\simeq H_{\rm int}^{e\nu}+H_{\rm int}^{\nu}$, while we ignore the part including only electrons and positrons,
\begin{align}
    H_{\rm int}^{e\nu}&=\frac{G_F}{\sqrt{2}}\int dx^3 \left[\bar{\bm{\nu}}\gamma^\mu(1-\gamma_5)Y^L\bm{\nu}\bar{\bm{e}}\gamma_\mu(1-\gamma_5)\bm{e} +\bar{\bm{\nu}}\gamma^\mu(1-\gamma_5)Y^R\bm{\nu}\bar{\bm{e}}\gamma_\mu(1+\gamma_5)\bm{e}\right], \nonumber \\
    H_{\rm int}^\nu&=\frac{G_F}{4\sqrt{2}}\int dx^3 \bar{\bm{\nu}}\gamma^\mu(1-\gamma_5)\bm{\nu}\bar{\bm{\nu}}\gamma_\mu(1-\gamma_5)\bm{\nu},
    \label{enuint}
\end{align}
with
\begin{align}
    Y^L=\begin{pmatrix}
\frac{1}{2}+\sin^2\theta_W & 0 & 0 \\
0 & -\frac{1}{2}+\sin^2\theta_W & 0 \\
0 & 0 & -\frac{1}{2}+\sin^2\theta_W \\
\end{pmatrix}
,\ \ Y^R=\sin^2\theta_W \times\bm{1}.
\label{Yvalue}
\end{align}
Here we have used the following Fierz transformation in the charged currents,
\begin{align}
\bar{\bm{\nu}}_e\gamma^\mu(1-\gamma_5)\bm{e}\bar{\bm{e}}\gamma_\mu(1-\gamma_5)\bm{\nu}_e= \bar{\bm{\nu}}_e\gamma^\mu(1-\gamma_5)\bm{\nu}_e\bar{\bm{e}}\gamma_\mu(1-\gamma_5)\bm{e}.
\label{Fierztransformation}
\end{align}
Due to this contribution of the charged current, only electron-type neutrinos and anti-neutrinos interact with electrons and positrons via different magnitudes of interactions, compared to other flavor neutrinos with $(Y^L)_{11}=(Y^L)_{22(33)}+1$.

The Hamiltonian of eq.~(\ref{enuint}) can be further divided as
\begin{align}
    H_{\rm int}^{e\nu}&=H_{\nu\bar{\nu}\leftrightarrow e^-e^+}+H_{\nu e^{\pm}\leftrightarrow \nu e^{\pm}}+H_{\bar{\nu} e^{\pm}\leftrightarrow \bar{\nu} e^{\pm}}, \nonumber \\
H_{\rm int}^{\nu}&=H_{\nu\nu\leftrightarrow\nu\nu}+H_{\nu\bar{\nu}\leftrightarrow\nu\bar{\nu}},
\end{align}
where $H_{ab\leftrightarrow cd}$ is the term including operators of (anti-)particles, $a,b,c$ and $d$. In the following, we neglect $H_{\bar{\nu} e^{\pm}\leftrightarrow \bar{\nu} e^{\pm}}$ since this Hamiltonian does not contribute the evolution of neutrinos.
 In addition, we only consider contributions proportional to the following terms as a function of freely evolving fields in the collision term in eq.~(\ref{BE112}),
\begin{align}
&[H^0_{\nu\bar{\nu}\leftrightarrow e^-e^+},[H^0_{\nu\bar{\nu}\leftrightarrow e^-e^+}, N_{\alpha\beta}^0]], \ \ \ \ 
[H^0_{\nu e^{\pm}\leftrightarrow \nu e^{\pm}},[H^0_{\nu e^{\pm}\leftrightarrow \nu e^{\pm}}, N_{\alpha\beta}^0]], \nonumber \\
&[H^0_{\nu\nu\leftrightarrow\nu\nu} ,[H^0_{\nu\nu\leftrightarrow\nu\nu}, N_{\alpha\beta}^0]],\ \ \ \ [H^0_{\nu\bar{\nu}\leftrightarrow\nu\bar{\nu}} ,[H^0_{\nu\bar{\nu}\leftrightarrow\nu\bar{\nu}}, N_{\alpha\beta}^0]].
\label{CT111}
\end{align}
The other terms also denote forward scattering, which would give tiny modifications of eqs.~(\ref{CCFS}) and (\ref{NCFS}).
The first term in eq.~(\ref{CT111}) denotes the annihilation of neutrinos and anti-neutrinos into $e^{\pm}$-pairs, which mainly contribute to the distortion of neutrino spectrum in their decoupling. The second term denotes the scattering between neutrinos and electrons (positrons). The third term represents the scattering process including only neutrinos while the fourth term denotes the annihilation and scattering processes of neutrinos and anti-neutrinos.

In a schematic manner, the collision term for two-body reactions $1+2\leftrightarrow 3+4$ at tree level takes the following expressions,
\begin{align}
    (2\pi)^3\delta^{(3)}(0)C[\rho_{p_1}]&=-\frac{1}{2}\int^{\infty}_{-\infty} dt'\langle [H_{\rm int}^0(0), [ H_{\rm int}^0, N_{\alpha\beta}^0]\rangle \nonumber \\
    &=(2\pi)^3\delta^{(3)}(0)\frac{1}{2E_1}\sum\int \frac{d^3p_2}{(2\pi)^32E_2} \frac{d^3p_3}{(2\pi)^32E_3} \frac{d^3p_4}{(2\pi)^32E_4} \nonumber \\
    & \times (2\pi)^4\delta^{(4)}(p_1+p_2-p_3-p_4)F(\rho,f_{e^\pm}, Y^L,Y^R)\left(S|M|^2_{12\rightarrow 34}\right)_{\rm part},
    \label{CI}
\end{align}
where $\rho_i\ (i=1,2,3,4)$ denote the neutrino density matrix, not the energy density, and $E_i\simeq|\bm{p}_i|$ for $\nu$ and $\bar{\nu}$ while $E_i=\sqrt{p_i^2+m_e^2}$ for $e^\pm$. $F(\rho,f_{e^\pm}, Y^L,Y^R)$ is a matrix depending on $\rho$, $f_{e^\pm}$, $Y^L$ and/or $Y^R$.
$\left(S|M|^2_{12 \rightarrow 34}\right)_{\rm part}$ is a part of $S|M|^2_{12\rightarrow 34}$, where $S$ is the symmetric factor and $|M|^2$ is the squared matrix element summed over spins of all particles except for the first one. The formulae of $S|M|^2$ for the relevant reaction in neutrino decoupling are shown in table~\ref{tb:Process}. Nine integrals in the collision term in eq.~(\ref{CI}) can be reduced analytically to two integrals as in appendix \ref{appb}. 

In the following, we rewrite the collision terms $C[\rho_p(t)]$ including eq.~(\ref{CT111}) with neutrino density matrices and the distribution functions of electrons and positrons. The formulae of the collision terms for neutrino density matrix are originally given in refs.~\cite{Sigl:1993ctk, Blaschke:2016xxt}, and for numerical calculations of neutrino spectra, these formulae are developed in refs.~\cite{deSalas:2016ztq, Froustey:2020mcq}.

\begin{table}[htbp]
\begin{center}
  \begin{tabular}{c|c} \hline \hline
    Process  & $2^{-5}G_F^{-2}S|M|^2$  \\ \hline 
    $\nu_e+\bar{\nu}_e\rightarrow \nu_e+\bar{\nu_e}$   & $4(p_1\cdot p_4)(p_2\cdot p_3)$          \\ 
    $\nu_e+\nu_e\rightarrow \nu_e+\nu_e$ & $2(p_1\cdot p_2)(p_3\cdot p_4)$        \\
    $\nu_e+\bar{\nu}_e \rightarrow \nu_{\mu(\tau)}+\bar{\nu}_{\mu(\tau)} $        & $(p_1\cdot p_4)(p_2\cdot p_3)$          \\ 
     $\nu_e+\bar{\nu}_{\mu(\tau)}\rightarrow \nu_e+\bar{\nu}_{\mu(\tau)}$        &   $(p_1\cdot p_4)(p_2\cdot p_3)$        \\ 
    $\nu_e+\nu_{\mu(\tau)}\rightarrow \nu_e+\nu_{\mu(\tau)}$              & $(p_1\cdot p_2)(p_3\cdot p_4)$         \\
    $\nu_e + \bar{\nu}_e \rightarrow e^-+e^+$  & $4[g_L^2(p_1\cdot p_4)(p_2\cdot p_3)+g_R^2(p_1\cdot p_3)(p_2\cdot p_4)+g_Lg_Rm_e^2(p_1\cdot p_2)]$       \\
    $\nu_e+e^-\rightarrow \nu_e+e^-$ & $4[g_L^2(p_1\cdot p_2)(p_3\cdot p_4)+g_R^2(p_1\cdot p_4)(p_2\cdot p_3)-g_Lg_Rm_e^2(p_1\cdot p_3)]$ \\
    $\nu_e+e^+\rightarrow \nu_e+e^+$ & $4[g_R^2(p_1\cdot p_2)(p_3\cdot p_4)+g_L^2(p_1\cdot p_4)(p_2\cdot p_3)-g_Lg_Rm_e^2(p_1\cdot p_3)]$ \\
     \hline \hline
  \end{tabular}
  \caption{Squared matrix elements with the symmetric factor $S|M|^2$ for processes $\nu_e(p_1)+b(p_2)\rightarrow c(p_3)+d(p_4)$. $g_L=\frac{1}{2}+\sin^2\theta^2_W$ and $g_R=\sin^2\theta_W$ correspond $(Y^L)_{11}$ and $(Y^R)_{11}$ in eq.~(\ref{Yvalue}).
  For processes of $\nu_\mu$ and $\nu_\tau$, $\nu_{\mu(\tau)}(p_1)+b(p_2)\rightarrow c(p_3)+d(p_4)$, squared matrix elements are obtained by the substitutions of $g_L\rightarrow g_L-1=-\frac{1}{2}+\sin^2\theta_W$, which corresponds $(Y^L)_{22(33)}$ in eq.~(\ref{Yvalue}) \cite{Dolgov:1997mb}.}
  \label{tb:Process}
  \end{center}
\end{table}


\paragraph{(i) $\bm{\nu(p_1)+\bar{\nu}(p_2)\leftrightarrow e^-(p_3)+e^+(p_4)}$ \\ \\}
The collision term for the annihilation process including $e^\pm$, $\nu(p_1)+\bar{\nu}(p_2)\leftrightarrow e^-(p_3)+e^+(p_4)$, comes from the term proportional to $[H^0_{\nu\bar{\nu}\leftrightarrow e^-e^+},[H^0_{\nu\bar{\nu}\leftrightarrow e^-e^+}, N_{\alpha\beta}^0]$.
We can calculate the corresponding collision terms, which are denoted as $(2\pi)^3\delta^{(3)}(0)C^{\nu\bar{\nu}\leftrightarrow e^-e^+}[\rho_{p_1}(t)]$,
\begin{align}
    &(2\pi)^3\delta^{(3)}(0)C^{\nu\bar{\nu}\leftrightarrow e^-e^+}[\rho_{p_1}(t)] \nonumber \\
    &=-\frac{1}{2}\int^{\infty}_{-\infty} dt'\langle [H^0_{\nu\bar{\nu}\leftrightarrow e^-e^+}(0), [ H^0_{\nu\bar{\nu}\leftrightarrow e^-e^+}(t'), N_{\alpha\beta}^0]\rangle \nonumber \\
&=(2\pi)^3\delta^{(3)}(0)\frac{1}{2}\frac{2^5G_F^2}{2|\bm{p}_1|}\int \frac{d^3p_2}{(2\pi)^32|\bm{p}_2|} \frac{d^3p_3}{(2\pi)^32E_3} \frac{d^3p_4}{(2\pi)^32E_4}(2\pi)^4\delta^{(4)}(p_1+p_2-p_3-p_4) \nonumber \\
&\ \ \ \ \ \ \ \  \times \Bigl[ 4(p_1\cdot p_4)(p_2\cdot p_3)F^{LL}_{\rm ann}\left(\nu^{(1)},\bar{\nu}^{(2)}, e^{-(3)},e^{+(4)}\right) \nonumber \\
&\ \ \ \ \ \ \ \ \ \ \ \  + 4(p_1\cdot p_3)(p_2\cdot p_4)F^{RR}_{\rm ann}\left(\nu^{(1)},\bar{\nu}^{(2)}, e^{-(3)},e^{+(4)}\right) \nonumber \\
&\ \ \ \ \ \ \ \ \ \ \ \ +2(p_1\cdot p_2)m_e^2 \Bigl(F^{LR}_{\rm ann}\left(\nu^{(1)},\bar{\nu}^{(2)}, e^{-(3)},e^{+(4)}\right) + F^{RL}_{\rm ann}\left(\nu^{(1)},\bar{\nu}^{(2)}, e^{-(3)},e^{+(4)}\right) \Bigl) 
\Bigl],
\label{Collision1}
\end{align}
where
\begin{align}
&F^{ab}_{\rm ann}\left(\nu^{(1)},\bar{\nu}^{(2)}, e^{-(3)},e^{+(4)}\right) \nonumber \\
&= f_{e^-}(p_3)f_{e^+}(p_4)
\Bigl( Y^a(1-\bar{\rho}_2))Y^b(1-\rho_1)+(1-\rho_1)Y^b(1-\bar{\rho}_2)Y^a \Bigl) \nonumber \\
&\ \ \ \ -\left(1-f_{e^-}(p_3)\right)\left(1-f_{e^+}(p_4)\right)\Bigl( Y^a\bar{\rho}_1Y^b\rho_1+\rho_1Y^b\bar{\rho}_2Y^a \Bigl).
\label{FAAA}
\end{align}
Here $f_{e^\pm}(p)$ is the distribution function for electrons and positrons, respectively.


\paragraph{(ii) $\bm{\nu(p_1)+e^\pm(p_2)\leftrightarrow \nu(p_3)+e^\pm(p_4)}$ \\ \\}
The collision term for the scatterings including $e^\pm$, $\nu(p_1)+e^\pm(p_2)\leftrightarrow \nu(p_3)+e^\pm(p_4)$, comes from the term proportional to $[H^0_{\nu e^\pm\leftrightarrow \nu e^\pm},[H^0_{\nu e^\pm\leftrightarrow \nu e^\pm}, N_{\alpha\beta}^0]$.
We can similarly calculate the corresponding collision term, which is denoted as $C^{\nu e^-\leftrightarrow \nu e^-}[\rho_{p_1}(t)]$ and $C^{\nu e^+\leftrightarrow \nu e^+}[\rho_{p_1}(t)]$, respectively,
\begin{align}
    &C^{\nu e^-\leftrightarrow \nu e^-}[\rho_{p_1}(t)] \nonumber \\
&= \frac{1}{2}\frac{2^5G_F^2}{2\left|\bm{p}_1\right|}\int \frac{d^3p_2}{(2\pi)^32E_2}\frac{d^3p_3}{(2\pi)^32\left|\bm{p}_3\right|}\frac{d^3p_4}{(2\pi)^32E_4}(2\pi)^4\delta^{(4)}(p_1+p_2-p_3-p_4) \nonumber \\
&\ \ \ \ \ \times  \Bigl[  4(p_1\cdot p_2)(p_3\cdot p_4)F^{LL}_{\rm sc}\left(\nu^{(1)},e^{-(2)}, \nu^{(3)},e^{-(4)}\right) \nonumber \\
&\ \ \ \ \ \ \ \ + 4(p_1\cdot p_4)(p_2\cdot p_3)F^{RR}_{\rm sc}\left(\nu^{(1)},e^{-(2)}, \nu^{(3)},e^{-(4)}\right) \nonumber \\
&\ \ \ \ \ \ \ \ -2(p_1\cdot p_3)m_e^2  \Bigl(F^{LR}_{\rm sc}\left(\nu^{(1)},e^{-(2)}, \nu^{(3)},e^{-(4)}\right) + F^{RL}_{\rm sc}\left(\nu^{(1)},e^{-(2)}, \nu^{(3)},e^{-(4)}\right) \Bigl)
\Bigl],
\label{SCminus}
\end{align}
and
\begin{align}
    &C^{\nu e^+\leftrightarrow \nu e^+}[\rho_{p_1}(t)] \nonumber \\
&= \frac{1}{2}\frac{2^5G_F^2}{2\left|\bm{p}_1\right|}\int \frac{d^3p_2}{(2\pi)^32E_2}\frac{d^3p_3}{(2\pi)^32\left|\bm{p}_3\right|}\frac{d^3p_4}{(2\pi)^32E_4}(2\pi)^4\delta^{(4)}(p_1+p_2-p_3-p_4) \nonumber \\
&\ \ \ \ \ \times  \Bigl[  4(p_1\cdot p_2)(p_3\cdot p_4)F^{RR}_{\rm sc}\left(\nu^{(1)},e^{+(2)}, \nu^{(3)},e^{+(4)}\right) \nonumber \\
&\ \ \ \ \ \ \ \ + 4(p_1\cdot p_4)(p_2\cdot p_3)F^{LL}_{\rm sc}\left(\nu^{(1)},e^{+(2)}, \nu^{(3)},e^{+(4)}\right) \nonumber \\
&\ \ \ \ \ \ \ \ -2(p_1\cdot p_3)m_e^2  \Bigl(F^{LR}_{\rm sc}\left(\nu^{(1)},e^{+(2)}, \nu^{(3)},e^{+(4)}\right) + F^{RL}_{\rm sc}\left(\nu^{(1)},e^{+(2)}, \nu^{(3)},e^{+(4)}\right) \Bigl)
\Bigl],
\label{SCplus}
\end{align}
where
\begin{align}
&F^{ab}_{\rm sc}\left(\nu^{(1)},e^{\pm(2)}, \nu^{(3)},e^{\pm(4)}\right) \nonumber \\
&= f_e^\pm(p_4)(1-f_e^\pm(p_2))\Bigl(Y^a\rho_3Y^b(1-\rho_1) + (1-\rho_1)Y^b\rho_3Y^a \Bigl) \nonumber \\
&-f_e^\pm(p_2)(1-f_e^\pm(p_4))\Bigl( \rho_1Y^b(1-\rho_3)Y^a+Y^a(1-\rho_3)Y^b\rho_1 \Bigl).
\label{FSC}
\end{align}

    
\paragraph{(iii) $\bm{\nu(p_1)+\nu(p_2)\leftrightarrow \nu(p_3)+\nu(p_4)}$ and $\bm{\nu(p_1)+\bar{\nu}(p_2)\leftrightarrow \nu(p_3)+\bar{\nu}(p_4)}$ \\ \\}
The collision terms for the scatterings including only neutrinos and anti-neutrinos, $\nu(p_1)+\nu(p_2)\leftrightarrow \nu(p_3)+\nu(p_4)$ and $\nu(p_1)+\bar{\nu}(p_2)\leftrightarrow \nu(p_3)+\bar{\nu}(p_4)$, come from the term proportional to $[H^0_{\nu\nu\leftrightarrow\nu\nu} ,[H^0_{\nu\nu\leftrightarrow\nu\nu}, N_{\alpha\beta}^0]]$ and $[H^0_{\nu\bar{\nu}\leftrightarrow\nu\bar{\nu}} ,[H^0_{\nu\bar{\nu}\leftrightarrow\nu\bar{\nu}}, N_{\alpha\beta}^0]]$, respectively.
The corresponding collision terms, which are denoted as $C^{\nu \nu\leftrightarrow \nu \nu}[\rho_{p_1}(t)]$ and $C^{\nu \bar{\nu}\leftrightarrow \nu \bar{\nu}}[\rho_{p_1}(t)]$, respectively, are calculated as
\begin{align}
    &C^{\nu \nu\leftrightarrow \nu \nu}[\rho_{p_1}(t)] \nonumber \\
    &=\frac{1}{2}\frac{2^5G_F^2}{2\left|\bm{p}_1\right|}\int \frac{d^3p_2}{(2\pi)^32|\bm{p}_2|}\frac{d^3p_3}{(2\pi)^32\left|\bm{p}_3\right|}\frac{d^3p_4}{(2\pi)^32|\bm{p}_4|}(2\pi)^4\delta^{(4)}(p_1+p_2-p_3-p_4) \nonumber \\
    &\ \ \ \ \times (p_1\cdot p_2)(p_3\cdot p_4)F_{\rm sc}\left(\nu^{(1)},\nu^{(2)}, \nu^{(3)},\nu^{(4)}\right), 
    \label{Ssc} \\
    &C^{\nu \bar{\nu}\leftrightarrow \nu \bar{\nu}}[\rho_{p_1}(t)] \nonumber \\
    &=\frac{1}{2}\frac{2^5G_F^2}{2\left|\bm{p}_1\right|}\int \frac{d^3p_2}{(2\pi)^32|\bm{p}_2|}\frac{d^3p_3}{(2\pi)^32\left|\bm{p}_3\right|}\frac{d^3p_4}{(2\pi)^32|\bm{p}_4|}(2\pi)^4\delta^{(4)}(p_1+p_2-p_3-p_4) \nonumber \\
    &\ \ \ \ \times (p_1\cdot p_4)(p_2\cdot p_3)\left(F_{\rm sc}\left(\nu^{(1)},\bar{\nu}^{(2)}, \nu^{(3)},\bar{\nu}^{(4)}\right)
    + F_{\rm ann}\left(\nu^{(1)},\bar{\nu}^{(2)}, \nu^{(3)},\bar{\nu}^{(4)}\right)
    \right),
    \label{Sann}
\end{align}
where $F_{\rm sc}\left(\nu^{(1)},\nu^{(2)}, \nu^{(3)},\nu^{(4)}\right)$, $F_{\rm sc}\left(\nu^{(1)},\bar{\nu}^{(2)}, \nu^{(3)},\bar{\nu}^{(4)}\right)$ and $F_{\rm ann}\left(\nu^{(1)},\bar{\nu}^{(2)}, \nu^{(3)},\bar{\nu}^{(4)}\right)$
denote contributions from scatterings for $\nu\nu\leftrightarrow \nu\nu$, scatterings and annihilations for  $\nu\bar{\nu}\leftrightarrow \nu\bar{\nu}$, respectively,
\begin{align}
    &F_{\rm sc}\left(\nu^{(1)},\nu^{(2)}, \nu^{(3)},\nu^{(4)}\right) \nonumber \\
    &=\left[\rho_4(1-\rho_2)+{\rm Tr}(...)\right]\rho_3(1-\rho_1)
    +(1-\rho_1)\rho_3\left[(1-\rho_2)\rho_4+{\rm Tr}(...) \right] \nonumber \\
    &\ \ \ \ -\left[(1-\rho_4)\rho_2+{\rm Tr}(...)\right](1-\rho_3)\rho_1
    -\rho_1(1-\rho_3)\left[\rho_2(1-\rho_4)+{\rm Tr}(...) \right], \label{Fself1} \\ 
        &F_{\rm sc}\left(\nu^{(1)},\bar{\nu}^{(2)}, \nu^{(3)},\bar{\nu}^{(4)}\right) \nonumber \\
    &=\left[(1-\bar{\rho}_2)\bar{\rho}_4+{\rm Tr}(...)\right]\rho_3(1-\rho_1)
    +(1-\rho_1)\rho_3\left[\bar{\rho}_4(1-\bar{\rho}_2)+{\rm Tr}(...) \right] \nonumber \\
    &\ \ \ \ -\left[\bar{\rho}_2(1-\bar{\rho}_4)+{\rm Tr}(...)\right](1-\rho_3)\rho_1
    -\rho_1(1-\rho_3)\left[(1-\bar{\rho}_4)\bar{\rho}_2+{\rm Tr}(...) \right], \label{Fself2} \\
    &F_{\rm ann}\left(\nu^{(1)},\bar{\nu}^{(2)}, \nu^{(3)},\bar{\nu}^{(4)}\right) \nonumber \\
    &=[\rho_3\bar{\rho}_4+{\rm Tr}(...)](1-\bar{\rho}_2)(1-\rho_1)+(1-\rho_1)(1-\bar{\rho}_2)[\bar{\rho}_4\rho_3+{\rm Tr}(...)] \nonumber \\
    &\ \ \ \ -[(1-\rho_3)(1-\bar{\rho}_4)+{\rm Tr}(...)]\bar{\rho}_2\rho_1-\rho_1\bar{\rho}_2[(1-\bar{\rho}_4)(1-\rho_3)+{\rm Tr}(...)],
    \label{Fself3}
\end{align}
where $[\alpha+{\rm Tr}(...)]\equiv [\alpha+{\rm Tr}(\alpha)]$.

Finally, we obtain the collision term in eq.~(\ref{BE112}), $C[\rho_p(t)]$, combining eqs.~(\ref{Collision1}), (\ref{SCminus}), (\ref{SCplus}), (\ref{Ssc}) and (\ref{Sann}),
\begin{align}
    C[\rho_p(t)]=C^{\nu\bar{\nu}\leftrightarrow e^-e^+}+C^{\nu e^-\leftrightarrow \nu e^-}+C^{\nu e^+\leftrightarrow \nu e^+}+C^{\nu \nu\leftrightarrow \nu \nu}+C^{\nu \bar{\nu}\leftrightarrow \nu\bar{\nu}}.
    \label{CTS}
\end{align}

The collision terms for anti-neutrinos can be obtained by appropriately replacing the density matrices and momenta, $\rho_i \leftrightarrow \bar{\rho}_i$ and $p_i \leftrightarrow p_j$ \cite{Sigl:1993ctk,Froustey:2022sla}. Changing the collision term for $\nu(p_1)X \rightarrow \nu(p_3)X'$ to $\bar{\nu}(p_1)X \rightarrow \bar{\nu}(p_3)X'$ corresponds replacing $\rho_1\rightarrow \bar{\rho}_1,\ \rho_3 \rightarrow \bar{\rho}_3$ and $p_1 \leftrightarrow p_3$ in this collision term while changing that for $\nu(p_1)\bar{\nu}(p_2)\rightarrow XX'$ to $\bar{\nu}(p_1)\nu(p_2)\rightarrow XX'$ corresponds $\rho_1\rightarrow \bar{\rho}_1,\ \bar{\rho}_2\rightarrow \rho_2$ and $p_1\leftrightarrow p_2$. 
One may consider the transpose in the collision terms is necessary for the reverse indices in the anti-neutrino density matrix (\ref{DM}), but this is not necessary since the collision terms are invariant under the transpose.

\subsection{Continuity equation}
In addition to the Boltzmann equations for the neutrino density
matrix, the energy conservation law must be satisfied,
\begin{align}
\frac{d\rho}{dt}=-3H(\rho+P),
\label{CE2}
\end{align}
where $\rho$ and $P$ are the total energy density and pressure of $\gamma, e^{\pm}, \nu, \bar{\nu}$ around MeV-scale temperature,
respectively. The continuity equation corresponds to the evolution of the photon
temperature $T_{\gamma}$. 

 Though we will discuss finite temperature corrections
from QED to $\rho, P$ and $m_e$ in the next section, in
the ideal gas limit, they are given as follows, which are denoted by
$\rho_{(0)}$ and $P_{(0)}$ respectively,
\begin{align}
\rho_{(0)} &= \frac{\pi^2T_{\gamma}^4}{15}+\frac{2}{\pi^2}\int\frac{dpp^2\sqrt{p^2+m_e^2}}{\exp(\sqrt{p^2+m_e^2}/T_{\gamma})+1}+\sum_{\alpha=e,\mu,\tau}\frac{1}{\pi^2}\int dp~p^3f_{\nu_\alpha}(p), \nonumber \\
P_{(0)} &= \frac{\pi^2T_{\gamma}^4}{45}+\frac{2}{\pi^2}\int\frac{dpp^4}{3\sqrt{p^2+m_e^2}[\exp(\sqrt{p^2+m_e^2}/T_{\gamma})+1]}+\sum_{\alpha=e,\mu,\tau}\frac{1}{3\pi^2}\int dp~p^3f_{\nu_\alpha}(p).
\label{rhoPideal}
\end{align}
The Hubble parameter in eq.~(\ref{CE2}) is calculated using the usual relation,
$3H^2m_{\rm Pl}^2=8\pi\rho$ with $m_{\rm Pl}$ being the Planck mass, where
we ignore the curvature term and the cosmological constant because they
are negligible in the radiation dominated epoch.

\subsection{Finite temperature QED corrections to $m_e$, $\rho$ and $P$ up to $\mathcal{O}(e^3)$}
QED interactions at finite temperature modify the energy density and pressure of electromagnetic plasma from the ideal gas limit. In addition, their interactions change the electron mass (and produce an effective photon mass). 
These corrections affect the kinetic equations for neutrinos discussed in the former sections. The corrections to the electron mass modify the weak interaction rates and the distribution function for $e^\pm$.
Through the direct modifications of $\rho$ and $P$, the expansion rate $H$ is also changed.
Note that QED interactions also modify weak interaction rates in the collision term $C[\rho_p(t)]$ and the Hamiltonian for the forward scattering (\ref{HVM}) at order $\mathcal{O}(e^2G_F^2)$ directly. 
In our numerical calculations, we consider corrections to weak interaction rates only due to the change of $m_e$.
We will discuss other QED corrections to weak interaction rates and their uncertainties in $N_{\rm eff}$ in section \ref{sec:3.3.1}.

The corrections to the grand canonical partition function $Z$ by interactions at finite temperature are well established perturbatively and can be calculated by the similar procedure of the functional integrals of Quantum Field Theory (QFT) at zero temperature after changing $t\rightarrow -i/T$.
$P$ and $\rho$ are described by $Z$ as
\begin{align}
P&=\frac{T}{V}\ln Z, \nonumber \\
\rho&=\frac{T^2}{V}\frac{\partial \ln Z}{\partial T}=-P+T\frac{\partial P}{\partial T},
\end{align}
where $T$ and $V$ are the temperature and volume in the system, respectively. 
Then we can expand $\ln Z$ in powers of the QED coupling constant $e$ as $\ln Z=\sum_{n=1}^{\infty}\ln Z_{(n)}$, where $\ln Z_{(n)}\propto e^n$.
In the isotropic and lepton symmetric universe, the corresponding corrections to $P$ and $\rho$ at $\mathcal{O}(e^2)$, $P_{(2)}, \rho_{(2)},\propto e^2$, are \cite{Kapusta:2006pm}
\begin{align}
P_{(2)}&=-\frac{e^2T_\gamma^2}{12\pi^2}\int_0^{\infty} dp \frac{p^2}{E_p}N_F(p)-\frac{e^2}{8\pi^4}\left(\int^{\infty}_0dp\frac{p^2}{E_p}N_F(p) \right)^2 \nonumber \\
&\ \ \ \ \ \ \ \ +\frac{e^2m_e^2}{16\pi^4}\int_0^{\infty}\int_0^{\infty}dpdp'\frac{pp'}{E_pE_{p'}}\ln\left|\frac{p+p'}{p-p'} \right|N_F(p)N_F(p'), \nonumber \\
\rho_{(2)}&=-P_{(2)}+T_\gamma\frac{\partial P_{(2)}}{\partial T_\gamma},
\label{SecCore}
\end{align}
where $E_p=\sqrt{p^2+m_e^2}$ and $N_F(p)$ is the sum of the distribution functions for $e^\pm$,
\begin{align}
N_F(p)=2\frac{1}{e^{E_p/T_\gamma}+1}.
\end{align}
The next-to-leading order of thermal corrections to $\rho,\ P$ is $\mathcal{O}(e^3)$, not $\mathcal{O}(e^4)$. These non-trivial corrections come from the resummation of ring diagrams in the photon propagator at all orders. The thermal corrections to $P,\ \rho$ at $\mathcal{O}(e^3)$, $P_{(3)}, \rho_{(3)},\propto e^3$, are
\cite{Bennett:2019ewm, Kapusta:2006pm},
\begin{align}
 P_{(3)} = \frac{e^3T_\gamma}{12\pi^4}I^{3/2}(T_\gamma), \nonumber \\
 \rho_{(3)} = \frac{e^3T_\gamma^2}{8\pi^4}I^{1/2}\frac{\partial I}{\partial T_\gamma},
 \label{FTC3}
\end{align} 
where
\begin{align}
I(T_\gamma) = \int^{\infty}_0 dp \left( \frac{p^2+E_p^2}{E_p} \right)N_F(p).
\end{align}
Finally, we read the total energy density and the total pressure of electromagnetic plasma up to $\mathcal{O}(e^3)$ corrections as
\begin{align}
 P &= P_{(0)}+P_{(2)}+P_{(3)}, \nonumber \\
 \rho &= \rho_{(0)} + \rho_{(2)}+\rho_{(3)}. 
 \label{rhoandP}
\end{align}

The thermal corrections to the $e^\pm$ mass at $\mathcal{O}(e^2)$ is given by, through modifications of the $e^\pm$ self energy 
\cite{Heckler:1994tv},
 \begin{align}
 \delta m_{e(2)}^2(p,T_{\gamma}) &= \frac{e^2 T_\gamma^2}{6}+\frac{e^2}{2\pi^2}\int_0^{\infty}dp' \frac{k^2}{E_p'}N_F(p') \nonumber \\
 &\ \ \ \ \ \ \ \  -\frac{e^2m_e^2}{4\pi^2 p}\int^{\infty}_0dp'\frac{p'}{E_{p'}}\log \left| \frac{p+p'}{p-p'} \right|N_F(p').
 \label{delta_me}
 \end{align}
The last logarithmic terms in eqs.~(\ref{SecCore}) and (\ref{delta_me})
give less than $10\%$ corrections to these equations around the decoupling temperature and the average momentum of electrons
\cite{Lopez:1998vk}. 
These terms also give contributions less than $10^{-4}$ to $N_{\rm eff}$ \cite{Bennett:2019ewm, Bennett:2020zkv}. In the following, we neglect the logarithmic corrections.
Note that thermal corrections to $m_e$ at $\mathcal{O}(e^3)$ do not appear because $\mathcal{O}(e^3)$ corrections stem from ring diagrams in the photon propagator. 

\subsection{Summary and approximations}
\label{sec:2.2.5}

In this section we summarize the closed system of the resulting Boltzmann equations for the neutrino density matrix and the continuity equation in neutrino decoupling. We also discuss the approximations we used in our numerical calculations. The following eqs.~(\ref{BE})-(\ref{NO2}) have already been presented in the previous sections.

The closed system of the equations of motion for the neutrino density matrix and the continuity equation, which reads the equation of the evolution for the photon temperature, in the expanding universe are \cite{Sigl:1993ctk, Blaschke:2016xxt}
\begin{align}
\frac{d\rho_p(t)}{dt}&=(\partial_t-Hp\partial_p)\rho_p(t) = -i\left[  \mathcal{H}_p ,\ \rho_p(t)  \right] +C[\rho_p(t)],
\label{BE} \\
\frac{d\rho}{dt}&=-3H(\rho+P),
\label{EC}
\end{align}
and analogous Boltzmann equations for anti-neutrinos \cite{Sigl:1993ctk,Froustey:2022sla}, which is not solved in this article since we assume no lepton asymmetry.
Here $H=\frac{1}{m_{\rm Pl}}\sqrt{\frac{8\pi\rho}{3}}$ is the Hubble parameter, $\mathcal{H}_p$ is the Hamiltonian which governs the neutrino oscillation in vacuum and the forward scattering of neutrinos in the $e^\pm,\ \nu,\ \bar{\nu}$-background, $C[\rho_p(t)]$ is the collision term describing the momentum changing scatterings and annihilations 
, and $[\cdot, \cdot]$ represents the
commutator of matrices with a flavor (or mass) index. 
$\rho$ and $P$ in eq.~(\ref{EC}) are the total energy density and the pressure for $\gamma,e^\pm,\nu,\bar{\nu}$, respectively. Including QED finite temperature corrections up to $\mathcal{O}(e^3)$, $\rho$ and $P$ are given by eq.~(\ref{rhoandP}) (see also eqs.~(\ref{rhoPideal}), (\ref{SecCore}) and (\ref{FTC3}) for the detail of eq.~(\ref{rhoandP})).

The effective Hamiltonian for the neutrino oscillations in vacuum and the forward scattering of neutrinos in the $e^\pm,\ \nu,\ \bar{\nu}$-background is given by
\footnote{For forward scattering with background in an anisotropic universe, see ref.~\cite{Hansen:2020vgm}.}
\begin{align}
\mathcal{H}_p&= \frac{\bm{\mathrm{M}}^2}{2p}+\sqrt{2}G_F(\bm{\mathrm{N}}_{e^-}-\bm{\mathrm{N}}_{e^+})+\sqrt{2}G_F(\bm{\mathrm{N}}_\nu-\bm{\mathrm{N}}_{\bar{\nu}}) \nonumber \\
&\ \ \ \ -\frac{2\sqrt{2}G_Fp}{m_W^2}(\bm{\mathrm{E}}_{e^-}+\bm{\mathrm{P}}_{e^-}+\bm{\mathrm{E}}_{e^+}+\bm{\mathrm{P}}_{e^+})-\frac{8\sqrt{2}G_Fp}{3m_Z^2}(\bm{\mathrm{E}}_\nu+\bm{\mathrm{E}}_{\bar{\nu}}),
\label{HVM}
\end{align}
where $G_F$ is the Fermi coupling constant and $m_{W},\ m_Z$ are the W and Z boson masses, respectively.

The first term
 in the RHS of eq.~(\ref{HVM}) denotes neutrino oscillations in vacuum and $\bm{\mathrm{M}}^2$ is the mass-squared matrix.
 In the flavor basis, we can write $\bm{\mathrm{M}}^2=U_{\rm PMNS}\bm{\mathrm{M}}^2_{\rm diag}U_{\rm PMNS}^{\dag}$,
 where $\bm{\mathrm{M}}^2_{\rm diag}={\rm diag}(m_{\nu_1}^2,\ m_{\nu_2}^2,\ m_{\nu_3}^2)$.
The other terms
 describe the forward scattering of neutrinos in the background of thermal plasma which comes from one-loop thermal contributions to neutrino self energy. $\bm{\mathrm{N}}_{e^\pm},\ \bm{\mathrm{N}}_{\nu, \bar{\nu}},\ \bm{\mathrm{E}}_{e^\pm}, \bm{\mathrm{P}}_{e^\pm},\ \bm{\mathrm{E}}_{\nu,\bar{\nu}}$ are defined in the flavor basis around the temperature of MeV scale as
\begin{align}
&\bm{\mathrm{N}}_{e^-}-\bm{\mathrm{N}}_{e^+}={\rm diag}(n_{e^-}-n_{e^+},\ 0,\ 0),\  \ n_{e^\pm}=2\int\frac{d^3p}{(2\pi)^3}f_{e^\pm}(p), \nonumber \\
&\bm{\mathrm{N}}_{\nu}-\bm{\mathrm{N}}_{\bar{\nu}}=\int\frac{d^3p}{(2\pi)^3}\left(\rho_p(t)-\bar{\rho}_p(t) \right), \nonumber \\
&\bm{\mathrm{E}}_{e^\pm}+\bm{\mathrm{P}}_{e^\pm}={\rm diag}(\rho_{e^\pm}+P_e{^\pm},\ 0,\ 0),
\ \ \rho_{e^\pm}+P_{e^\pm}=\int\frac{d^3p}{(2\pi)^3}\left(E_e+ \frac{p^2}{3E_e} \right)f_{e^\pm}(p), \nonumber \\
&\bm{\mathrm{E}}_{\nu}+\bm{\mathrm{E}}_{\bar{\nu}}=\int\frac{d^3p}{(2\pi)^3}p\left(\rho_p(t)+\bar{\rho}_p(t) \right),
\label{NO2}
\end{align}
where $E_e=\sqrt{p^2+m_e^2+\delta m^2_e(p,T)}$ and $f_{e^\pm}(p)$ is the distribution function of $e^\pm$.
$\delta m^2_e(p,T)$ is the QED finite temperature correction to $m_e$, which is given by eq.~(\ref{delta_me}) up to $\mathcal{O}(e^2)$.
Here we neglect the contributions of $\mu$ and $\tau$ since the densities of these charged particles are significantly suppressed.

In the following, we assume that electrons and positrons are always in thermal
equilibrium and follow the Fermi-Dirac distributions since electrons, positrons and photons interact with each
other through rapid electromagnetic interactions.
In addition we neglect lepton asymmetry since neutrino
oscillations leading to flavor equilibrium before the BBN imposes a stringent
constraint on this asymmetry \cite{Dolgov:2002ab, Wong:2002fa, Abazajian:2002qx, Mangano:2011ip,
Castorina:2012md, Oldengott:2017tzj,Escudero:2022okz}. The standard baryogenesis scenarios via the sphaleron process in leptogenesis models predict that the lepton asymmetry is of the order of the current baryon asymmetry, $n_b/n_\gamma \sim 10^{-10}$, which is much smaller than the above constraint. 
We also neglect any CP-violating phase in the PMNS matrix for simplicity. Note that from the recent global analysis of neutrino oscillation experiments \cite{Esteban:2020cvm,deSalas:2020pgw}, the CP-conserving PMNS matrix is excluded at approximately $3\sigma$ confidence level.
Strictly speaking, ignoring the CP-violating phase is inconsistent with the experimental results, but we adopt this assumption to save computational time. In fact, since effects of CP-violating phase on neutrino oscillations are sub-dominant, this ignorance will not affect the resultant neutrino spectra and $N_{\rm eff}$ significantly.
Under these assumptions, neutrinos and anti-neutrinos
satisfy the same density matrices and the same evolutions in the
Universe,\ $\rho_p(t) = \bar{\rho}_p(t)^{\mathrm{T}}$, and electrons and positrons follow the same Fermi-Dirac distributions with $T_{\gamma}$ and no chemical potential.

Note that without lepton asymmetry, $\bm{\mathrm{N}}_{\nu}-\bm{\mathrm{N}}_{\bar{\nu}}\neq 0$ due to $\rho_p(t) = \bar{\rho}_p(t)^{\mathrm{T}}\neq \bar{\rho}_p(t)$. However, in the following, we neglect it for  reducing computational time. 
We will discuss this uncertainty in section \ref{sec:3.3}.
In addition, as in refs.~\cite{Mangano:2005cc, deSalas:2016ztq, Gariazzo:2019gyi, Akita:2020szl}, we replace $\bm{\mathrm{E}}_{e^\pm}+\bm{\mathrm{P}}_{e^\pm}$ as $4/3\bm{\mathrm{E}}_{e^\pm}$ for simplicity. Strictly, this replacement is valid only in the ultra-relativistic limit \cite{Notzold:1987ik}. However, since in the non-relativistic region $\bm{\mathrm{E}}_{e^\pm}$ is suppressed by the Boltzmann factor, these difference would be quite small.
Ref. \cite{Bennett:2020zkv} reported this difference in $N_{\rm eff}$ is no more than $10^{-5}$.

The final term in the RHS of eq.~(\ref{BE}) represents both the momentum conserving and changing collisions of neutrinos
with neutrinos, electrons and their anti-particles. In this term, collisions are dominated by
two-body reactions $1+2\rightarrow3+4$, i.e., $C[\rho_p(t)]\propto G_F^2$, where $G_F$ is the Fermi coupling constant.
The detailed formula for $C[\rho_p(t)]$ is given by eq.~(\ref{CTS}) (see also eqs.~(\ref{Collision1}), (\ref{SCminus}), (\ref{SCplus}), (\ref{Ssc}), (\ref{Sann}) in this review and refs.~\cite{deSalas:2016ztq, Froustey:2020mcq}).
Nine integrals in the collision term (\ref{CI}) can be reduced analytically to two integrals as in appendix \ref{appb}.
We deal with both diagonal and
off-diagonal collision terms in eqs.~(\ref{Collision1}), (\ref{SCminus}) and (\ref{SCplus}) for the processes which involve electrons
and positrons,  $\nu e^\pm \leftrightarrow
\nu e^\pm$ and $\nu\bar{\nu} \leftrightarrow e^-e^+$.
On the other hand, we do not treat the off-diagonal terms in eqs.~(\ref{Ssc}) and (\ref{Sann})
for the self-interactions of neutrinos, $\nu\nu \leftrightarrow
\nu\nu$ and $\nu\bar{\nu} \leftrightarrow \nu\bar{\nu}$, since
the annihilations of electrons and positrons are important for the
heating process of neutrinos while the self-interactions of neutrinos less contribute
to this heating process. We treat this collision term from neutrino self-interaction in eq.~(\ref{CTSreduced}) of appendix~\ref{appa}.
In refs.~\cite{Froustey:2020mcq,Bennett:2020zkv}, the authors solve kinetic equations for neutrinos including the full collision term at tree level and reported almost the same results with very small difference in $N_{\rm eff}$, $\delta N_{\rm eff}\sim 2\times 10^{-4}$\cite{Bennett:2020zkv}. 
Here, we take into account finite temperature corrections to $m_e$ up to $\mathcal{O}(e^2)$ in the collision term as $E_e=\sqrt{p^2+m_e^2+\delta m^2_e(p,T)}$. However, we neglect other sub-leading contributions to the collision term, i.e., other QED corrections to weak interaction rates.
We also discuss these uncertainties in section \ref{sec:3.3}.

 \subsection{Computational method, initial conditions and values of neutrino masses and mixing}
 
We solve kinetic equations for neutrinos of eqs.~(\ref{BE}) and (\ref{EC}) with the following
comoving variables instead of the cosmic time $t$, the momentum $p$, and
the photon temperature $T_{\gamma}$,
\begin{align}
x = m_ea,\ \ \ \ \ \ \  y = pa,\ \ \ \ \ \ \ \  z = T_{\gamma}a, 
\end{align}
where we choose an arbitrary mass scale in $x$ to be the electron mass $m_e$ and
$a$ is the scale factor of the universe, normalized as $z \rightarrow 1\ (a \rightarrow 1/T_\gamma)$
in high temperature limit. 
The resultant kinetic equations for neutrinos in the comoving variables are described in appendix \ref{appa}.

Since the Boltzmann equations (\ref{BE}) are integro-differential
equations due to integrations in the collision terms, their equations were solved by a
discretization in a momentum grid $y_i$ in
refs.~\cite{Hannestad:1995rs, Dolgov:1997mb, Dolgov:1998sf,
Grohs:2015tfy, Mangano:2005cc, deSalas:2016ztq, Gariazzo:2019gyi}, by an
expansion of the distortions of neutrinos from the Fermi-Dirac distribution in
refs.~\cite{Esposito:2000hi, Mangano:2001iu, Birrell:2014uka}, or by a hybrid method combining the previous two methods in ref.~\cite{Froustey:2019owm}. In this
study, we adopt the discretization method we mentioned first and take 100 grid points
for $y_i$, equally spaced in the region $y_i \in [0.02,\
20]$ with the Simpson method.
We have used MATLAB ODE solver, in particular, ode15s with an absolute and relative tolerance of $10^{-6}$.
In these tolerances, we confirm that numerical errors for relic neutrino spectra and $N_{\rm eff}$ are typically $10^{-4}$ or less. 

We have numerically estimated the evolution of the density matrix for neutrinos and
the photon temperature in $x_{\rm in}\leq x \leq x_{f}$. We
have set $x_{\rm in}=m_e/10\ {\rm MeV}$ as an initial time. Since
neutrinos are kept in thermal equilibrium with the electromagnetic plasma at
$x_{\rm in}$, the initial values of density matrix $\rho_{y_i}^{\rm
in}(x)$ are regarded as
\begin{align}
\rho_{y_i}^{\rm in}(x) = {\rm diag}\left(\frac{1}{e^{y_i/z_{\rm in}}+1},\frac{1}{e^{y_i/z_{\rm in}}+1},\frac{1}{e^{y_i/z_{\rm in}}+1}  \right).
\end{align} 
The initial dimensionless photon temperature at $x_{\rm
in}$, $z_{\rm in}$, slightly deviates from 1 because a tiny amount of 
$e^\pm$-pairs have already been annihilated at $x_{\rm in}$.  Due to the entropy conservation of
electromagnetic plasma, neutrinos and anti-neutrinos, $z_{\rm in}$ is
estimated as in \cite{Dolgov:1998sf},
\begin{align}
z_{\rm in} = 1.00003.
\end{align}
We take $x_f = 30$ as a final time, when the neutrino density matrix and
$z$ can be regarded as frozen.

Finally we comment on values of neutrino masses and mixing we use in our numerical simulation.
We use the best-fit values in the global analysis in 2019 \cite{Esteban:2018azc}, but assume CP-symmetry, $\delta_{\rm CP}=0$.
We note that in 2020 their best-fit values are updated \cite{Esteban:2020cvm,deSalas:2020pgw} though their differences are very small. Their parameters include small uncertainties of about $10\%$ at $3\sigma$ confidence level. Effects of their uncertainties on $N_{\rm eff}$ is investigated in ref.~\cite{Bennett:2020zkv} and slightly change $N_{\rm eff}$ by $|\delta N_{\rm eff}|\sim 10^{-4}$.
In our numerical simulation, we confirmed that relic neutrino spectra and the value of $N_{\rm eff}$ with $10^{-3}$ precision are the same for both neutrino mass ordering. In the following, we show the results in the normal mass ordering, $\Delta m_{31}^2>0$, not in the inverted ordering, $\Delta m_{31}^2<0$, because the results do not change significantly.

\clearpage

\section{Effective number of neutrino species $N_{\rm eff}$}
\label{sec:3}
To describe the process of neutrino decoupling,
we first numerically solve a set of eqs.~(\ref{BE}) and
(\ref{EC}) and show relic neutrino spectra in the flavor basis.
Then we present a precise value of the effective number of neutrino species, $N_{\rm eff}=3.044$,
and discuss effects of neutrino oscillations and finite temperature corrections to $m_e,\ \rho$ and $P$ up to $\mathcal{O}(e^3)$ on $N_{\rm eff}$.
We also comment on uncertainties of ingredients we ignored in estimating $N_{\rm eff}$. 

\subsection{Relic neutrino spectra in the flavor basis}

\begin{figure}[htbp]
\begin{minipage}{0.5\hsize}
   \begin{center}
     \includegraphics[clip,width=8.8cm]{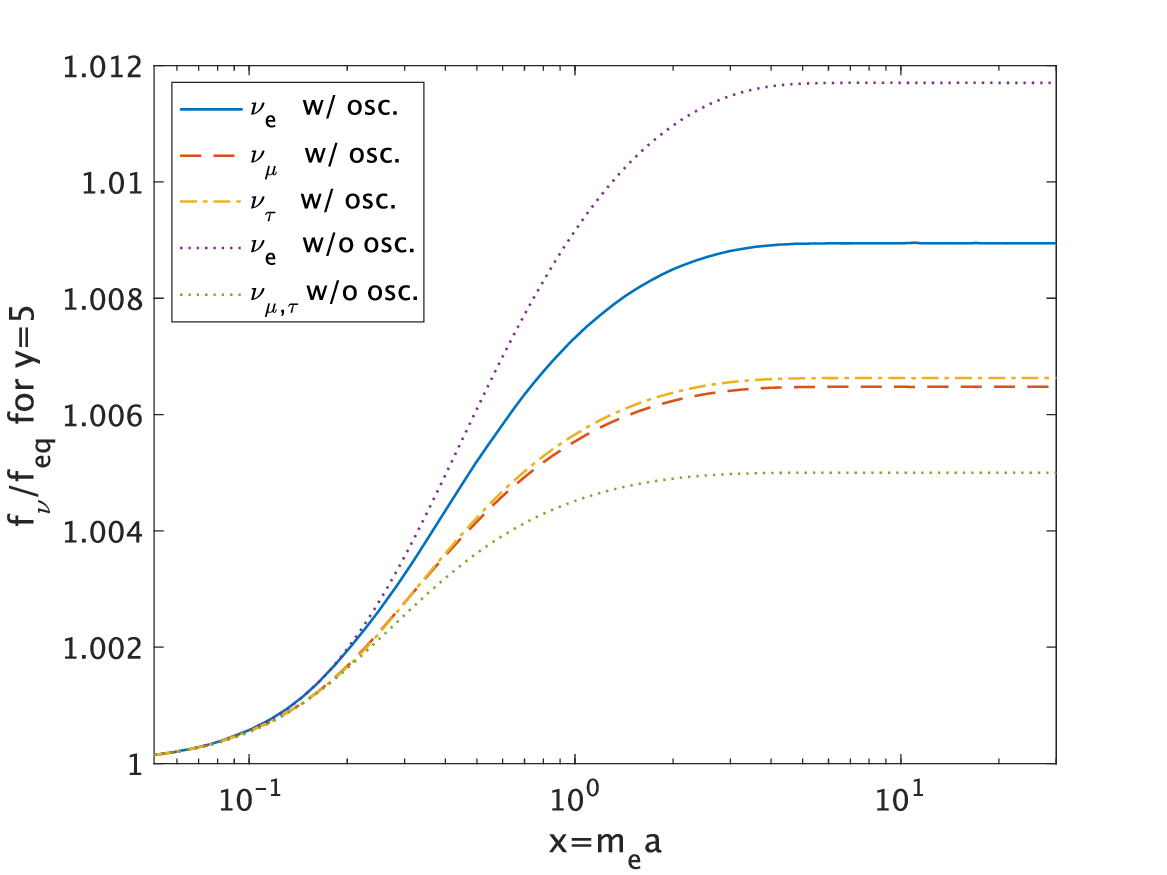}
    \end{center}
     \end{minipage}
      \begin{minipage}{0.5\hsize}
      \begin{center}
     \includegraphics[clip,width=8.8cm]{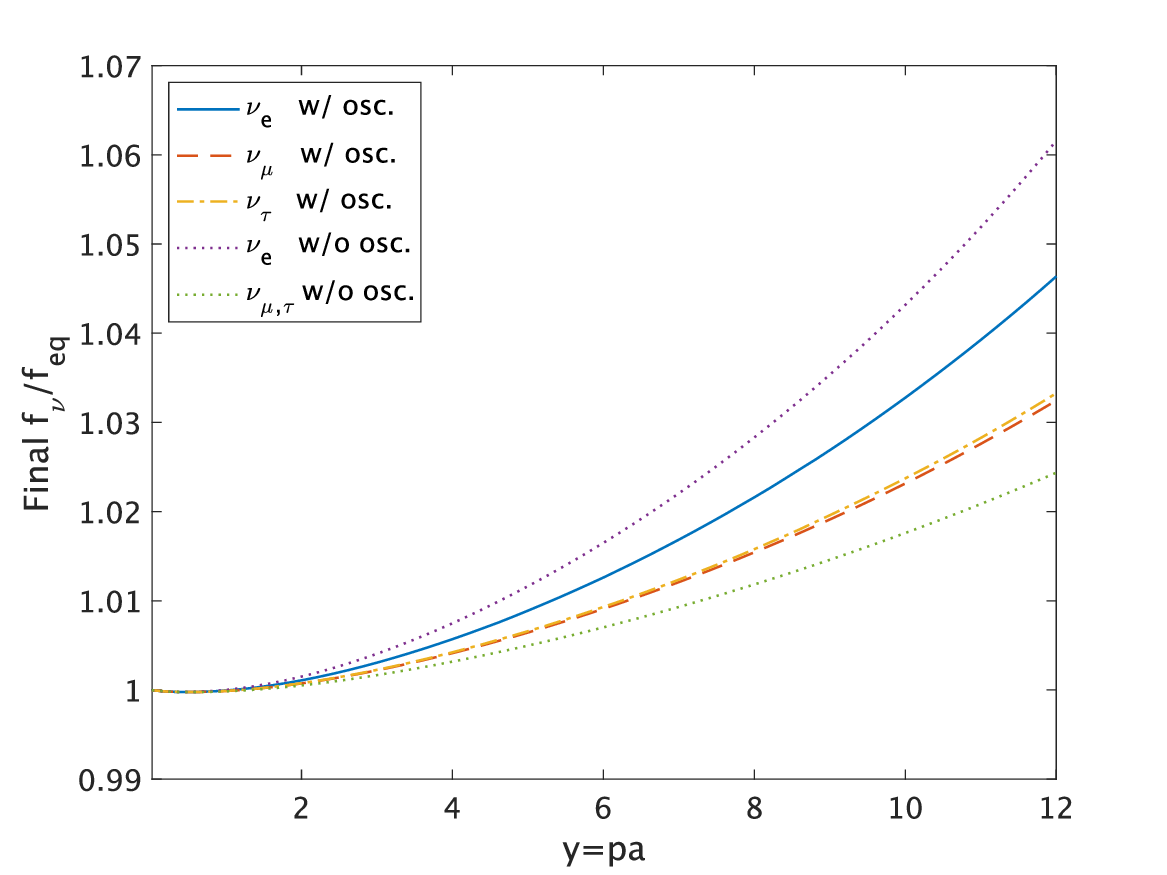}
    \end{center}
     \end{minipage}
    \vspace{-4mm}
 \caption{\underline{Left panel}: Time evolution of the distortions of flavor neutrinos for a
 fixed momentum ($y=5$) as a function of the normalized scale factor
 $x=m_ea$ with QED finite temperature corrections up to $\mathcal{O}(e^3)$.
 \underline{Right panel}: Final distortions of flavor neutrino spectra as a function
 of the comoving momentum $y$ with QED finite temperature corrections up to $\mathcal{O}(e^3)$. Upper (lower) dotted line is for $\nu_e\ (\nu_{\mu,\tau})$
 without neutrino oscillations, while inner solid and
 dashed lines represent those for flavor neutrinos with neutrino
 oscillations.} 
  \label{fig:flavorx-f}
 \end{figure}

In the left panel of figure~\ref{fig:flavorx-f}, we show the distortions of the flavor neutrino spectra
for a comoving momentum $(y=5)$, where we plot
the neutrino spectra $f_{\nu_\alpha}/f_{\rm eq}$ as a function of the normalized cosmic scale factor
$x$. $f_{\rm eq}(y)$ is the neutrino distribution function if neutrinos decoupled instantaneously and all $e^\pm$-pairs annihilated into photons,
\begin{align}
f_{\rm eq}(y)=\frac{1}{e^y+1}.
\end{align} 
At high temperature with $(x \lesssim 0.2)$, the temperature
differences between photons and neutrinos are negligible and neutrinos
are in thermal equilibrium with electrons and positrons. In the
intermediate regime with $(0.2 \lesssim x \lesssim 4)$, weak interactions
gradually become ineffective with shifting from small to large
momenta. In this period, the neutrino spectra are distorted since
the energies of electrons and positrons partially convert into those of
neutrinos coupled with electromagnetic plasma. Finally, at
low temperature with $(x \gtrsim 4)$, the collision term $C[\rho_p(t)]$
becomes ineffective and the distortions are frozen.

The
difference between the $\nu_e$ spectrum and the $\nu_{\mu,\tau}$ spectrum without flavor mixing
arises from the fact that only electron-type neutrinos interact with
electrons and positrons through the weak charged currents. On the other
hand, in the cases with neutrino mixing, neutrino oscillations
mix the distortions of the flavor neutrinos too.

In the right panel of figure~\ref{fig:flavorx-f}, we show the frozen values of the flavor
neutrino spectra $f_{\nu_\alpha}/f_{\rm eq}$ as a function of a comoving
momentum $y$ for both cases with and without neutrino mixing. This figure shows the fact
that neutrinos with higher energies interact with electrons and
positrons until a later epoch. In addition, we see neutrino
oscillations tend to equilibrate the flavor neutrino
distortions. Although the neutrino spectra $f_{\nu_\alpha}/f_{\rm eq}$
with low energies are very slightly less than unity, these
extractions of low energy neutrinos stem from an energy boost through
the scattering by electrons, positrons, (and neutrinos) with
sufficiently high energies, which are not yet annihilated and hence
still effective at neutrino decoupling process.

\subsection{Value of the effective number of neutrino species $N_{\rm eff}$}

The effective number of neutrinos
$N_{\rm eff}$ can be rewritten,
\begin{align}
N_{\rm eff} = \left( \frac{ (11/4)^{1/3}}{z} \right)^4 \left(3+ \frac{\delta \rho_{\nu_e}}{\rho_{\nu}^{\rm eq}}+ \frac{\delta \rho_{\nu_\mu}}{\rho_{\nu}^{\rm eq}}+ \frac{\delta \rho_{\nu_\tau}}{\rho_{\nu}^{\rm eq}}\right),
\label{Neff}
\end{align}
where  $\delta
\rho_{\nu_\alpha}=\rho_{\nu_\alpha}-\rho_{\nu}^{\rm eq}$ and $\rho_{\nu}^{\rm eq}=\int\frac{d^3p}{(2\pi)^3}pf_{\rm eq}$.
In tables~\ref{tb:flavor} and
\ref{tb:flavor2}, we present final values (at $x_f = 30$) of the
dimensionless photon temperature $z_{\rm fin}$, the difference of energy
densities and number densities of flavor neutrinos from those where neutrinos decoupled instantaneously denoted by $\rho_{\nu}^{\rm eq}=\int\frac{d^3p}{(2\pi)^3}pf_{\rm eq}$ and
$n_{\nu}^{\rm eq}=\int\frac{d^3p}{(2\pi)^3}f_{\rm eq}$, and the effective number of neutrinos $N_{\rm eff}$.

By comparing values of $N_{\rm eff}$ in the cases without QED corrections and with QED corrections to $m_e,\ \rho$ and $P$ up to $\mathcal{O}(e^2)$ and $\mathcal{O}(e^3)$ in table.~\ref{tb:flavor},
we find that the QED corrections at $\mathcal{O}(e^2)$ and $\mathcal{O}(e^3)$ shift $N_{\rm eff}$ by $+0.01$ and $-0.00095$, respectively, 
which is very close to the value estimated in the instantaneous decoupling limit \cite{Bennett:2019ewm}. 

In the cases with neutrino mixing,
table~\ref{tb:flavor2} shows that the energy densities of
$\mu,\tau$-type neutrinos increase more while those of
electron-type neutrinos increase less, compared to the cases without neutrino mixing. This modification leads to the enhancement of the total
energy density for neutrinos with final values of $N_{\rm eff}=3.04391\simeq 3.044$
with QED corrections  to $m_e,\ \rho$ and $P$ up to $\mathcal{O}(e^3)$. 
Since the blocking factor for electron neutrinos, $(1-f_{\nu_e}),$ is decreased
 by neutrino mixing, the annihilation of electrons and positrons into electron neutrinos increases. 
 Although the annihilation into the other neutrinos decreases, electron neutrinos contribute to the neutrino heating most efficiently, and
 neutrino oscillations enhance the annihilation of electrons and positrons into neutrinos.
From these processes, we
conclude that neutrino oscillations slightly promote neutrino
heating and the difference of $N_{\rm eff}$ is $0.00056$, which
agrees with the results of previous
works\cite{deSalas:2016ztq,Hannestad:2001iy,EscuderoAbenza:2020cmq}.

To conclude, our numerical calculation with neutrino oscillations and QED finite temperature corrections to $m_e,\ \rho$ and $P$ up to $\mathcal{O}(e^3)$ finds $N_{\rm eff}=3.044$. This value is in excellent agreement with later independent works \cite{Froustey:2020mcq, Bennett:2020zkv}.

\vspace{+1cm}

\begin{table}[htbp]
\begin{center}
\small
  \begin{tabular}{c|c|c} \hline \hline
    Case                                                              & $z_{\rm fin}$ & $N_{\rm eff}$ \\ \hline 
    Instantaneous decoupling                                        & 1.40102        & 3.00000  \\ \hline
    No mixing\hspace{-0.4mm} +\hspace{-0.4mm} No QED                                            & 1.39910        & 3.03404  \\
    \hspace{-0.8mm}No mixing\hspace{-0.4mm} +\hspace{-0.4mm} QED up to $\mathcal{O}(e^2)$\hspace{-0.8mm}        & 1.39789        & 3.04430  \\ 
     \hspace{-0.8mm}No mixing\hspace{-0.4mm} +\hspace{-0.4mm} QED up to $\mathcal{O}(e^3)$\hspace{-0.8mm}        & 1.39800        & 3.04335  \\ \hline
    mixing\hspace{-0.4mm} +\hspace{-0.4mm} QED up to $\mathcal{O}(e^2)$              & 1.39786        & 3.04486  \\
    mixing\hspace{-0.4mm} +\hspace{-0.4mm} QED up to $\mathcal{O}(e^3)$              & 1.39797        & 3.04391 \\ \hline \hline
  \end{tabular}
  \caption{Final values of comoving photon temperature and the effective number of neutrinos for flavor neutrinos in several cases.}
  \label{tb:flavor}
  \end{center}
\end{table}

\vspace{+0cm}

\begin{table}[htbp]
\begin{center}
\small
  \begin{tabular}{c|c|c|c|c|c|c} \hline \hline
    Case                                                              & $\delta \bar{\rho}_{\nu_e}(\%)$ & $\delta \bar{\rho}_{\nu_\mu}(\%)$ & $\delta \bar{\rho}_{\nu_\tau}(\%)$ & $\delta \bar{n}_{\nu_e}(\%)$ & $\delta \bar{n}_{\nu_\mu}(\%)$ & $\delta \bar{n}_{\nu_\tau}(\%)$ \\ \hline
    Instantaneous decoupling                                               &0   &0  &0  & 0  & 0  & 0\\ \hline
    No mixing\hspace{-0.4mm} +\hspace{-0.4mm} No QED                                                   &0.949   &0.397  &0.397       & 0.583 & 0.240 & 0.240  \\
    \hspace{-0.8mm}No mixing\hspace{-0.4mm} +\hspace{-0.4mm} QED up to $\mathcal{O}(e^2)$\hspace{-0.8mm}               &0.937   &0.391  &0.391       & 0.575 & 0.236 & 0.236 \\ 
    \hspace{-0.8mm}No mixing\hspace{-0.4mm} +\hspace{-0.4mm} QED up to $\mathcal{O}(e^3)$\hspace{-0.8mm}               &0.937   &0.391  &0.391       & 0.575 & 0.236 & 0.236   \\ \hline
    mixing\hspace{-0.4mm} +\hspace{-0.4mm} QED up to $\mathcal{O}(e^2)$                     &0.712   &0.511  &0.523       & 0.435 & 0.311 & 0.319   \\
    mixing\hspace{-0.4mm} +\hspace{-0.4mm} QED up to $\mathcal{O}(e^3)$                     &0.712   &0.511   &0.523      & 0.436 & 0.312 & 0.319\\ \hline \hline
  \end{tabular}
  \caption{Final values of the distortions of energy densities
  $\delta \bar{\rho}_{\nu_\alpha} \equiv  (\rho_{\nu_\alpha}-\rho_{\nu}^{\rm eq})/\rho_{\nu}^{\rm eq}$ and number densities $\delta
  \bar{n}_{\nu_\alpha} \equiv (n_{\nu_\alpha}-n_{\nu}^{\rm eq})/n_{\nu}^{\rm eq}$
  for flavor neutrinos in several cases.}  \label{tb:flavor2}
  \end{center}
\end{table}

\subsection{Discussions of uncertainties in $N_{\rm eff}$}
\label{sec:3.3}
We comment on possible errors of the results for relic neutrino spectra and $N_{\rm eff}$ due to approximations in eqs.~(\ref{BE}) and (\ref{EC}) and the choice of physical parameters. Our numerical calculations converge very well since we have directly computed $N_{\rm eff}$ in the mass basis as will be done in the next section and obtained $N_{\rm eff}=3.04388\simeq 3.044$.

First we neglect the off-diagonal parts for neutrino self-interactions in the collision term, $\nu\bar{\nu}\leftrightarrow \nu\bar{\nu}$ and $\nu\nu\leftrightarrow \nu\nu$. Later, in refs.~\cite{Froustey:2020mcq, Bennett:2020zkv}, the authors solve kinetic equations for neutrinos including their off-diagonal parts in the collision term and report the difference in $N_{\rm eff}$ is $\delta N_{\rm eff}\sim 2\times 10^{-4}$ \cite{Bennett:2020zkv}.
We also neglect the $\mathcal{O}(e^2)$ logarithmic terms and terms above $\mathcal{O}(e^4)$ in QED finite temperature corrections to $m_e$, $\rho$ and $P$. Their corrections to $\rho$ and $P$ are reported to contribute $\delta N_{\rm eff}<10^{-4}$ to $N_{\rm eff}$ in refs.~\cite{Bennett:2019ewm, Bennett:2020zkv}. Though their corrections to $m_e$ are not taken into account, the corrections to $m_e$ even at $\mathcal{O}(e^2)$ contribute $\delta N_{\rm eff}\lesssim 10^{-4}$ to $N_{\rm eff}$ \cite{Bennett:2020zkv} and we have also confirmed it. 

The neutrino masses and mixing parameters contain $10$-$20\%$ uncertainties at $3\sigma$ confidence level.
Since in our estimations, neutrino oscillations contribute $+0.0005$ to $N_{\rm eff}$, their uncertainties are expected to be quite small.\
In ref.~\cite{Bennett:2020zkv}, the authors report that their uncertainties are $\delta N_{\rm eff}\sim 10^{-4}$.
We also neglect the CP-violating phase $\delta_{\rm CP}$ in the PMNS matrix.
No one has yet computed precise neutrino evolution in the decoupling including three-flavor oscillations with CP violating phase. 
However, since effect of the CP-violating phase on neutrino oscillations is sub-dominant, we expect neutrino and anti-neutrino spectra might not change significantly. In addition, the total energy density, i.e., $N_{\rm eff}$ would change much less than $0.0005$ since the changes for the energy densities of neutrinos and anti-neutrinos would be canceled out. See also discussion in appendix F of ref.~\cite{Froustey:2020mcq} and ref.~\cite{Froustey:2021azz}.
Other physical parameters for electroweak interaction are measured very precisely and will not affect neutrino spectra and $N_{\rm eff}$.

However, QED corrections to weak interaction rates at order $\mathcal{O}(e^2G_F^2)$ and forward scattering of neutrinos via their self-interactions have not been precisely taken into account in the whole literature so far.


\subsubsection{QED corrections to weak interaction rates at order $e^2G_F^2$}
\label{sec:3.3.1}

QED interactions also modify the weak interaction rates in the collision term $C[\rho_p(t)]$ and the Hamiltonian for the forward scattering of neutrinos (\ref{HVM}) at order $e^2G_F^2$ in addition to the modification of the energy density and pressure for electromagnetic plasma, $\rho$ and $P$.
These corrections are partially taken into account by considering thermal QED corrections on $m_e$ so far. See also section 3.1.2 in ref.~\cite{Bennett:2020zkv}.

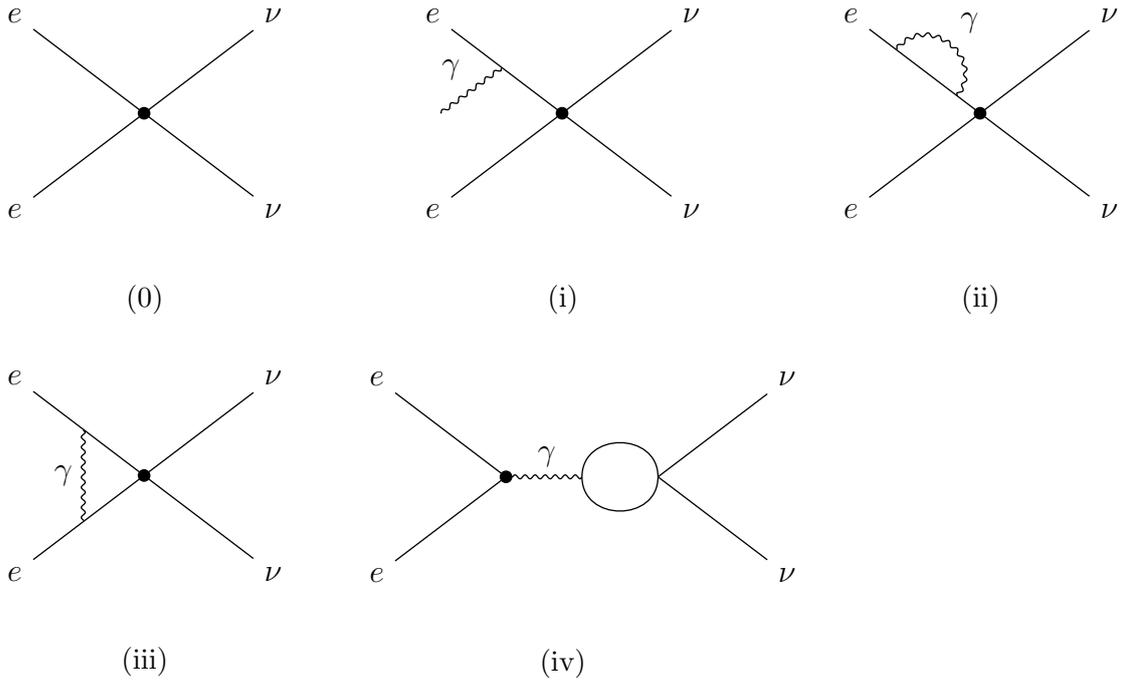
\begin{figure}[htpb]    
    \begin{minipage}{0.325\hsize}
\begin{center}      
\begin{tikzpicture}[baseline=+1.85cm]
\begin{feynhand}
   \vertex [particle] (iii1) at (8.3,4.3) {$e$};
    \vertex [particle] (iii2) at (8.3,1.7) {$e$};
    \vertex [particle] (fff1) at (11.7,4.3) {$\nu$};
    \vertex [particle] (fff2) at (11.7,1.7) {$\nu$};
    \vertex [dot](www1) at (10,3) {};
    \propag [plain] (iii1) to (www1);
    \propag [plain] (iii2) to (www1);
    \propag [plain] (www1) to (fff1);
    \propag [plain] (www1) to (fff2);
\end{feynhand}
\end{tikzpicture}
\end{center}
    \subcaption*{(0)}
    \end{minipage}
    \vspace{+0.5cm}
    \begin{minipage}{0.325\hsize}
    \begin{center} 
\begin{tikzpicture}[baseline=+1.85cm]
\begin{feynhand}
\vertex [particle] (iii1) at (8.3,4.3) {$e$};
    \vertex [particle] (iii2) at (8.3,1.7) {$e$};
    \vertex [particle] (fff1) at (11.7,4.3) {$\nu$};
    \vertex [particle] (fff2) at (11.7,1.7) {$\nu$};
     \vertex [particle] (p1) at (9.2,3.6);
      \vertex [particle] (p2) at (8.4,3);
    \vertex [dot](www1) at (10,3) {};
    \propag [plain] (iii1) to (www1);
    \propag [plain] (iii2) to (www1);
    \propag [plain] (www1) to (fff1);
    \propag [plain] (www1) to (fff2);
    \propag[boson] (p2) to [edge label=$\gamma$] (p1);
\end{feynhand}
\end{tikzpicture}
\end{center}
    \subcaption*{(i)}
    \end{minipage}
    \begin{minipage}{0.325\hsize}
    \begin{center} 
\begin{tikzpicture}[baseline=+1.85cm]
\begin{feynhand}
\vertex [particle] (iii1) at (8.3,4.3) {$e$};
    \vertex [particle] (iii2) at (8.3,1.7) {$e$};
    \vertex [particle] (fff1) at (11.7,4.3) {$\nu$};
    \vertex [particle] (fff2) at (11.7,1.7) {$\nu$};
     \vertex [particle] (p1) at (8.9,3.85);
      \vertex [particle] (p2) at (9.7,3.23);
    \vertex [dot](www1) at (10,3) {};
    \propag [plain] (iii1) to (www1);
    \propag [plain] (iii2) to (www1);
    \propag [plain] (www1) to (fff1);
    \propag [plain] (www1) to (fff2);
     \propag [boson, half left, looseness=1.64] (p1) to  [edge label=$\gamma$] (p2);
\end{feynhand}
\end{tikzpicture}
\end{center}
    \subcaption*{(ii)}
    \end{minipage} \\ \hspace{+1cm}
    \begin{minipage}{0.325\hsize}
    \begin{center} 
\begin{tikzpicture}[baseline=+1.85cm]
\begin{feynhand}
    \vertex [particle] (iii1) at (8.3,4.3) {$e$};
    \vertex [particle] (iii2) at (8.3,1.7) {$e$};
    \vertex [particle] (fff1) at (11.7,4.3) {$\nu$};
    \vertex [particle] (fff2) at (11.7,1.7) {$\nu$};
     \vertex [particle] (p1) at (9.2,3.6);
      \vertex [particle] (p2) at (9.2,2.4);
    \vertex [dot](www1) at (10,3) {};
    \propag [plain] (iii1) to (www1);
    \propag [plain] (iii2) to (www1);
    \propag [plain] (www1) to (fff1);
    \propag [plain] (www1) to (fff2);
    \propag[boson] (p2) to [edge label=$\gamma$] (p1);
\end{feynhand}
\end{tikzpicture}
\end{center}
    \subcaption*{(iii)}
    \end{minipage}
    \begin{minipage}{0.325\hsize}
    \begin{center} 
\begin{tikzpicture}[baseline=+1.85cm]
\begin{feynhand}
     \vertex [particle] (iii1) at (8.3,4.3) {$e$};
    \vertex [particle] (iii2) at (8.3,1.7) {$e$};
    \vertex [particle] (fff1) at (13.7,4.3) {$\nu$};
    \vertex [particle] (fff2) at (13.7,1.7) {$\nu$};
    \vertex [dot](www1) at (10,3) {};
    \vertex[particle] (www2) at (11,3);
    \vertex[particle] (www3) at (12,3);
    \propag [plain] (iii1) to (www1);
    \propag [plain] (iii2) to (www1);
    \propag [plain] (www3) to (fff1);
    \propag [plain] (www3) to (fff2);
    \propag [plain, half left, looseness=1.55]  (www2) to  (www3);
     \propag [plain, half right, looseness=1.55]  (www2) to  (www3);
    \propag [boson] (www1) to [edge label=$\gamma$] (www2);
\end{feynhand}
\end{tikzpicture}
\end{center}
    \subcaption*{(iv)}
    \end{minipage}
\caption{Feynman diagrams that contribute the weak interaction rates up to  $\mathcal{O}(e^2G_F^2)$ \cite{Esposito:2003wv, Bennett:2020zkv}. (0): 4-Fermi interactions. QED finite temperature corrections (i): additional photon emissions and absorptions, (ii): corrections to the dispersion relation for $e^\pm$, (iii): vertex corrections, (iv): corrections mediated by photon propagator. Matrix elements multiplied by (0) and one of (ii), (iii), (iv), and squared matrix elements for (i) contribute the weak interaction rates at $\mathcal{O}(e^2G_F^2)$.}
\label{fig:WRoneloop}
\end{figure}

QED corrections to the weak interaction rates (see also the diagrams in figure~\ref{fig:WRoneloop}) are categorized as
(i) additional photon emission and absorption, (ii) corrections to the dispersion relation for external $e^\pm$,
(iii) vertex corrections, and (iv) corrections mediated by photon propagator. 
The interference among the weak interaction at leading order $G_F$ and corrections (i)-(iv) produce modifications to the weak interaction rates at the next-to-leading order $\mathcal{O}(e^2G_F^2)$.

The correction (i) might be the most dominant contribution to $N_{\rm eff}$ since the photon emission processes, e.g. $e^+e^- \rightarrow \nu\bar{\nu}\gamma$, would not be suppressed by the distribution function of photons in the Boltzmann equations. The photon emission processes reduce $N_{\rm eff}$. However, there are many processes in the categories (ii), (iii) and (iv). In total, these contributions to $N_{\rm eff}$ might be as large as that from the correction (i).

For category (ii), corrections to the dispersion relation for $e^\pm$ produce a thermal electron mass as eq.~(\ref{delta_me}).
One can incorporate corrections (i) in the weak interaction rates by shifting $m_e^2\rightarrow m_e^2+\delta m_{e(2)}^2(p,T)$, but it is numerically difficult to take into account the momentum-dependent part of $\delta m_{e(2)}^2(p,T)$, which corresponds to the logarithmic $\mathcal{O}(e^2)$ corrections to $m_e$.
These logarithmic $\mathcal{O}(e^2)$ corrections to $m_e$  are less than $10\%$ of corrections at $\mathcal{O}(e^2)$ to $m_e$ around neutrino decoupling \cite{Lopez:1998vk},
and corrections even at leading $\mathcal{O}(e^2)$ to the weak interaction rates (i.e., $\delta m_{e(2)}(T)$) contributes $N_{\rm eff}<10^{-4}$ to $N_{\rm eff}$ \cite{Bennett:2020zkv} and we confirmed it. Thus, we would properly be able to incorporate corrections (i) to $N_{\rm eff}$ with $10^{-4}$ precision. But we should carefully derive these corrections to the weak interaction rates and consider effects of the logarithmic $\mathcal{O}(e^2)$ corrections and other sub-dominant neglected contributions in the collision term in the future.

For categories (i), (iii) and (iv), corrections to the weak interaction rates are typically momentum-dependent. It would be quite difficult to solve the Boltzmann equation, which is the integro-differential equation, including such momentum-dependent corrections.
In ref.~\cite{Esposito:2003wv}, the authors consider energy loss rate of a stellar plasma, including corrections on $e^-e^+\rightarrow \nu\bar{\nu}$ at order $\mathcal{O}(e^2G_F^2)$ and found such corrections modify the energy loss rate of a stellar plasma by a few percent. In ref.~\cite{EscuderoAbenza:2020cmq}, the author suggests $\delta N_{\rm eff}\simeq -0.0007$ due to correction (i) by roughly extrapolating the results in ref.~\cite{Esposito:2003wv} and using a precise and simple evaluation method of $N_{\rm eff}$ proposed in ref.~ \cite{EscuderoAbenza:2020cmq}.
The contributions of (i), (iii) and (iv) to $N_{\rm eff}$ should be evaluated in the future in a more precise way.

\subsubsection{Forward scattering of neutrinos via their self-interactions }

In the Hamiltonian (\ref{HVM}) in the Boltzmann equations (\ref{BE}), the forward scattering terms of neutrinos via their self-interactions correspond to
\begin{align}
\mathcal{H}_p\supset \sqrt{2}G_F(\bm{\mathrm{N}}_\nu-\bm{\mathrm{N}}_{\bar{\nu}}) -\frac{8\sqrt{2}G_Fp}{3m_Z^2}(\bm{\mathrm{E}}_\nu+\bm{\mathrm{E}}_{\bar{\nu}}).
\end{align}
Even in the case without lepton asymmetry, $\bm{\mathrm{N}}_{\nu}-\bm{\mathrm{N}}_{\bar{\nu}}\neq 0$ due to $\rho_p(t) = \bar{\rho}_p(t)^{\mathrm{T}}\neq \bar{\rho}_p$ in general, 
where  $\bm{\mathrm{N}}_{\nu}-\bm{\mathrm{N}}_{\bar{\nu}}$ is 
\begin{align}
\bm{\mathrm{N}}_{\nu}-\bm{\mathrm{N}}_{\bar{\nu}}=\int\frac{d^3p}{(2\pi)^3}\left(\rho_p(t)-\bar{\rho}_p(t) \right).
\end{align}
Though $\rho_p(t)-\bar{\rho}_p(t)$ might be small, forward scattering via neutrino self-interactions could be more dominant than neutrino oscillation in vacuum, with a typical dimensional analysis,
\begin{align}
&\sqrt{2}G_Fp^3\sim 10^{-11}\ {\rm MeV}\left(\frac{G_F}{10^{-5}\ {\rm GeV^{-2}}} \right)\left(\frac{p}{1\ {\rm MeV}} \right)^3 \nonumber \\
&\gg \frac{M^2}{2p}\sim 10^{-14}\ {\rm MeV}\left(\frac{M}{0.1\ {\rm eV}} \right)^2\left(\frac{1\ {\rm MeV}}{p} \right).
\end{align} 
In ref.~\cite{Hansen:2020vgm}, the authors suggest forward scattering of neutrinos via their self-interactions contributes $\delta N_{\rm eff}\simeq +(1- 5)\times 10^{-4}$ to $N_{\rm eff}$ by solving a simplified kinetic equations for neutrinos. 
In the future, relic neutrino spectra and $N_{\rm eff}$ should be estimated including the above forward scattering of neutrinos more precisely.

 Though recent estimations might contain uncertainties of $|\delta N_{\rm eff}| \lesssim (10^{-3}-10^{-4})$ in $N_{\rm eff}$,
$N_{\rm eff}=3.044$ would still be one of very good reference values in $N_{\rm eff}$.

\clearpage
\section{Relic cosmic neutrino spectra in the current homogeneous and isotropic universe}
\label{sec:4}

In the current universe, two neutrino species at least are non-relativistic.
Then relic neutrino spectra in the mass basis will be important observable to detect the C$\nu$B in a direct way as discussed in section \ref{sec:2.1}.
In this section we present the spectrum (as a function of comoving momenta) , number density and energy density of the C$\nu$B in the current homogeneous and isotropic universe,
including non-thermal distortions due to $e^\pm$-annihilation during neutrino decoupling.

\subsection{Relic neutrino spectra in the mass basis}

We present relic neutrino spectra in the mass basis by solving a
set of eqs.~(\ref{BE}) and (\ref{EC}) in the mass basis directly. 
We can also obtain the same result by transforming relic neutrino spectra in the flavor basis through eq.~(\ref{relation}).

In the mass basis, the neutral and charged currents including left-handed neutrino fields in eq.~(\ref{Currents}) are given by, using $\nu_\alpha=\sum_{i=1,2,3}U_{\alpha i}\nu_i$ as in eq.~(\ref{FOrelation}),
\begin{align}
    J_{\nu\nu}&=\frac{1}{4\cos\theta_W}\sum_{\alpha=e,\mu,\tau}\bar{\bm{\nu}}_\alpha\gamma^\mu(1-\gamma_5)\bm{\nu}_\alpha
= \frac{1}{4\cos\theta_W}\sum_{i = 1,2,3} \bar{\bm{\nu}}_i\gamma^\mu(1-\gamma_5)\bm{\nu}_i, \nonumber \\
J_{e\nu_e}^\mu&=\frac{1}{2\sqrt{2}}\bar{\bm{\nu}}_e\gamma^\mu(1-\gamma_5)\bm{e}
=\frac{1}{2\sqrt{2}}\sum_{i=1}^3U^*_{ei}\bar{\bm{\nu}}_i\gamma^\mu(1-\gamma_5) \bm{e}.
\label{Currentsmass}
\end{align}
Then, using the relations of eq.~(\ref{Currentsmass}) and (\ref{Fierztransformation}), we obtain the 4-point interaction Hamiltonian (\ref{enuint}) in the mass basis
\begin{align}
    H_{\rm int}^{e\nu}\bigl|_{\rm mass}&=\frac{G_F}{\sqrt{2}}\int dx^3 \left[\bar{\bm{\nu}}\gamma^\mu(1-\gamma_5)Z^L\bm{\nu}\bar{\bm{e}}\gamma_\mu(1-\gamma_5)\bm{e} +\bar{\bm{\nu}}\gamma^\mu(1-\gamma_5)Z^R\bm{\nu}\bar{\bm{e}}\gamma_\mu(1+\gamma_5)\bm{e}\right], \nonumber \\
    H_{\rm int}^\nu\bigl|_{\rm mass}&=\frac{G_F}{4\sqrt{2}}\int dx^3 \bar{\bm{\nu}}\gamma^\mu(1-\gamma_5)\bm{\nu}\bar{\bm{\nu}}\gamma_\mu(1-\gamma_5)\bm{\nu},
    \label{enuintmass}
\end{align}
with
\begin{align}
\bm{\nu}&=\begin{pmatrix}
\bm{\nu}_1  \\
\bm{\nu}_2  \\
\bm{\nu}_3 
\end{pmatrix}, \nonumber \\
    Z^L&=\begin{pmatrix}
\tilde{g}_L + U_{e1}^*U_{e1} & U_{e1}^*U_{e2} & U_{e1}^*U_{e3} \\
U_{e2}^*U_{e1} & \tilde{g}_L + U_{e 2}^*U_{e2} & U_{e 2}^*U_{e3} \\
U_{e3}^*U_{e1} & U_{e 3}^*U_{e2} & \tilde{g}_L + U_{e 3}^*U_{e3}
\end{pmatrix} 
,\ \ Z^R=Y^R=\sin^2\theta_W \times\bm{1}.
\label{Zvalue}
\end{align}
Then we obtain the Boltzmann equation for the neutrino density matrix in the mass basis after replacements of $Y^{L,R}\rightarrow Z^{L,R}$ and $\mathcal{H}_p \rightarrow U_{\rm PMNS}^\dag\mathcal{H}_pU_{\rm PMNS}$ analogous to $\bm{\mathrm{M}}_{\rm diag}^2 = U_{\rm PMNS}^\dag\bm{\mathrm{M}}^2U_{\rm PMNS}$ in eq.~(\ref{BE}) for the flavor basis.

In the left panel of figure~\ref{fig:massx-f}, we show the evolution of the
neutrino spectra, $f_{\nu_i}/f_{\rm eq}$, for a comoving momentum $(y=5)$
as a function of the normalized scale factor $x$.
In the right panel of
figure~\ref{fig:massx-f}, we show the asymptotic values of the
neutrino spectra
\footnote{The result in the right panel of figure~\ref{fig:massx-f} is quite different from figure~4 in ref.~\cite{deSalas:2016ztq}. Our results are confirmed by eq.~(\ref{relation}) and the numerical results in the flavor basis.}
$f_{\nu_i}/f_{\rm eq}$ as a function of $y$. 
 The differences of
distortions for each neutrino species arise from the charged current interactions between neutrinos and electrons weighted by the PMNS matrix with mass species $i$, $U_{ei}^{\ast}$, as in eq.~(\ref{enuintmass}).
Note that neutral currents between neutrinos in the mass basis are the same as that in the flavor basis except for the subscript, $J^{\mu}_{\nu\nu}=\sum_{\alpha=e,\mu,\tau}\bar{\bm{\nu}}_\alpha\gamma^{\mu}(1-\gamma_5)\bm{\nu}_\alpha=\sum_{\alpha=1,2,3}\bar{\bm{\nu}}_i\gamma^{\mu}(1-\gamma_5)\bm{\nu}_i$.
Then the scattering and annihilation among neutrinos and electrons and their anti-particles induce the spectral distortions in figure~\ref{fig:massx-f}.

Finally we comment on $N_{\rm eff}$.
After we directly solve a set of eqs.~(\ref{BE}) and (\ref{EC}) in the mass basis, including vacuum three-flavor neutrino oscillations, forward scatterings in $e^\pm$-background, and QED corrections to $m_e$, $\rho$ and $P$ up to $\mathcal{O}(e^3)$, we find $N_{\rm eff}=3.04388$, which is an excellent agreement with our calculation in the flavor basis.
The tiny difference from $N_{\rm eff}$ in the flavor basis may
come from ignoring the off-diagonal parts for
self-interaction processes in the Boltzmann
equations and/or numerical errors.

\subsection{Neutrino number density and energy density in the current homogeneous and isotropic universe}
\label{sec:4.2}

In table~\ref{tb:mass2}, we show the final
values of the dimensionless photon temperature $z_{\rm fin}$, the relativistic energy
densities $\rho_{\nu_i}/\rho_{\nu}^{\rm eq}$ and number densities $n_{\nu_i}/n_{\nu}^{\rm eq}$ of neutrinos in the mass basis after neutrino decoupling.
Note that the expression of energy density for a relativistic particle is not applicable to the first and second heaviest neutrinos today because they are non-relativistic in the current universe.

\begin{figure}[htbp]
\begin{minipage}{0.5\hsize}
   \begin{center}
     \includegraphics[clip,width=8.8cm]{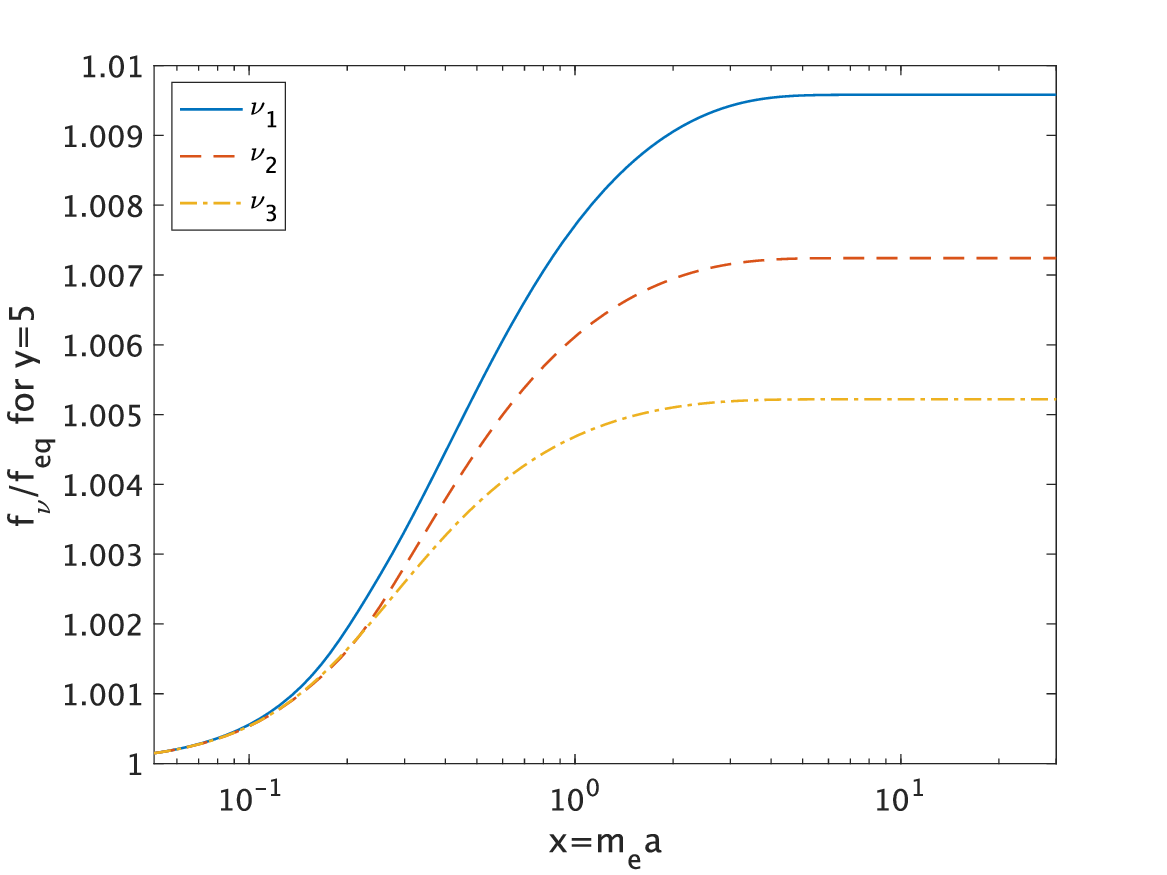}
    \end{center}
     \end{minipage}
      \begin{minipage}{0.5\hsize}
      \begin{center}
     \includegraphics[clip,width=8.8cm]{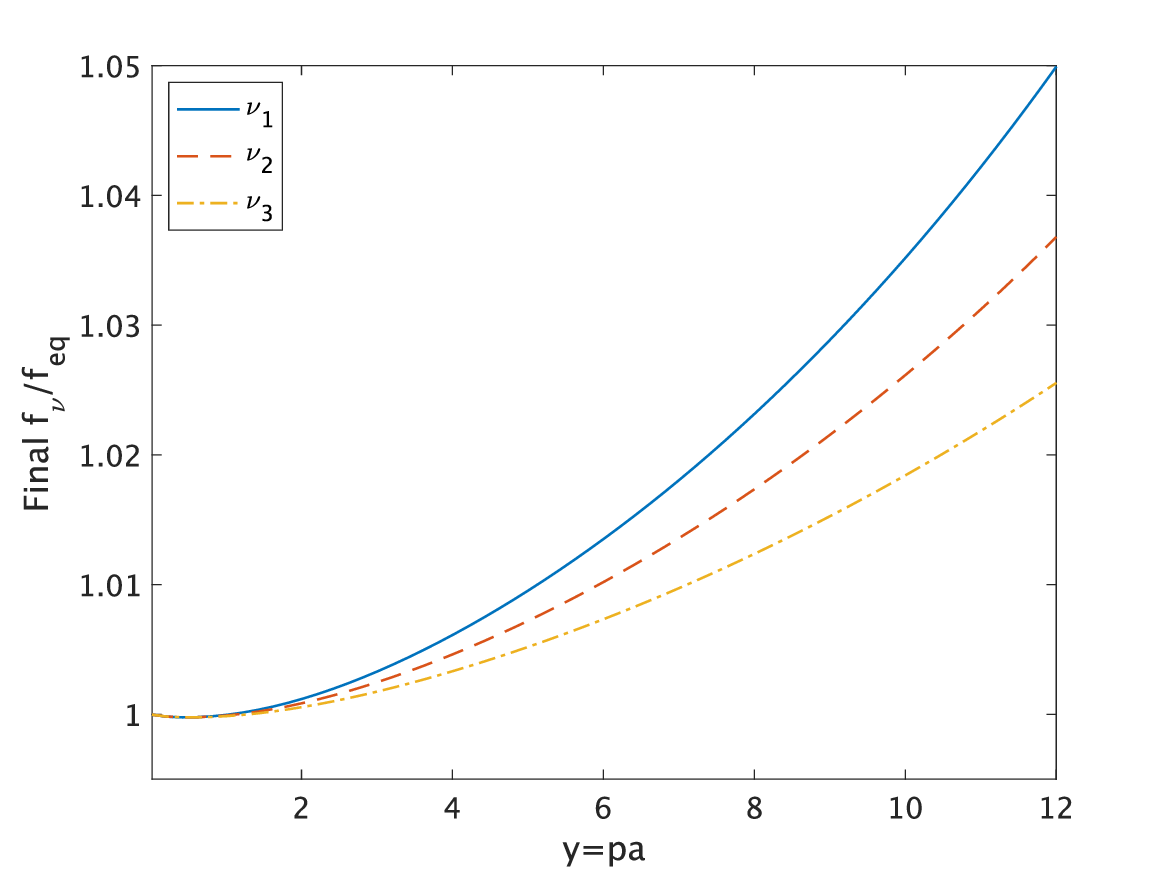}
    \end{center}
     \end{minipage}
    \vspace{-4mm}
 \caption{\underline{Left panel}: Time evolution of the distortions of neutrinos in the mass basis for a fixed momentum ($y=5$) with QED finite temperature corrections up to $\mathcal{O}(e^3)$.
\underline{Right panel}: Final distortions of neutrino spectra in the mass basis as a function of the comoving momentum $y$ with QED finite temperature corrections up to $\mathcal{O}(e^3)$. }
 \label{fig:massx-f}
 \end{figure}

After neutrino decoupling, the neutrino momentum distribution in the homogeneous and isotropic universe can be parametrized as
\begin{align}
f_{\nu_i}(\bm{p},t)=\frac{1}{e^{|\bm{p}|/\tilde{T}_{\nu}(t)}+1}\left(1+\delta f_{\nu_i}(\bm{p},t) \right).
\label{DFC}
\end{align}
$\tilde{T}_{\nu}(t)$ is the effective neutrino temperature, which is $\propto a(t)^{-1}$ and normalized as $\tilde{T}_{\nu}\rightarrow T_\gamma$ in high temperature limit. 
Under this definition of $\tilde{T}_{\nu}(t)$, neutrino spectral distortions, $\delta f_{\nu_i}(\bm{p},t)$, can be rewritten as $\delta f_{\nu_i}(y)$ given in the right panel of figure~\ref{fig:massx-f}.
At $t_0=4.35\times 10^{17}\ {\rm s}$ in the current universe, $\tilde{T}_{\nu}(t_0)$ satisfies
\begin{align}
\frac{T_\gamma(t_0)}{\tilde{T}_{\nu}(t_0)}=z_{\rm fin}=1.39797,
\end{align}
where $T_{\gamma}(t_0)\simeq 2.7255\ {\rm K}$ is the effective photon temperature in the current universe \cite{Fixsen:2009ug}. Then the effective neutrino temperature in the current universe is
\begin{align}
\tilde{T}_{\nu}(t_0)=1.9496\ {\rm K}.
\end{align}
Neutrino number density and energy density per one degree of freedom in the current universe are also parametrized as
\begin{align}
n_{\nu_i}(t_0)&=\int\frac{d^3p}{(2\pi)^3}f_{\nu_i}(\bm{p},t), \nonumber \\
&=\tilde{n}_0(1+\delta \bar{n}_{\nu_i}), \nonumber \\
\rho_{\nu_i}(t_0)&=\int\frac{d^3p}{(2\pi)^3}E_{\nu_i}f_{\nu_i}(\bm{p},t), \nonumber \\
&=\left\{
\begin{array}{l}
m_in_{\nu_i}\ \ \ \ \ \ \ \ \ \   {\rm for\ non\mathchar`- relativistic}\  \nu_i \\
\tilde{\rho}_0(1+\delta \bar{\rho}_{\nu_i})\ \ {\rm  for\ relativistic}\  \nu_i
\end{array}
\right.,
\end{align}
where $\tilde{n}_0$ and $\tilde{\rho}_0$ are given by
\begin{align}
\tilde{n}_0&=\int\frac{d^3p}{(2\pi)^3}\frac{1}{e^{|\bm{p}|/\tilde{T}_{\nu}(t_0)}+1}=\frac{3\zeta(3)}{4\pi^2}\tilde{T}_{\nu}(t_0)^3=56.376\ {\rm cm^{-3}},  \nonumber \\
\tilde{\rho}_0&=\int\frac{d^3p}{(2\pi)^3} \frac{|\bm{p}|}{e^{|\bm{p}|/\tilde{T}_{\nu}(t_0)}+1}=\frac{7\pi^2}{240}\tilde{T}_{\nu}(t_0)^4=29.848\ {\rm meV\ cm^{-3}}.
\end{align}
Then $\delta \bar{n}_{\nu_i}$ and $\delta \bar{\rho}_{\nu_i}$ are given in table~\ref{tb:mass2}.
The values of neutrino number density in the current universe are listed in table~\ref{tb:mass3}.

In the current universe, two species of cosmic relic neutrinos at least are non-relativistic because of $\tilde{T}_{\nu}(t_0)\ll \sqrt{\Delta m_{21}^2}\simeq 8.6\ {\rm meV},\ \sqrt{|\Delta m_{31}^2}|\simeq 50\ {\rm meV}$. On the other hand, the lightest neutrinos might be relativistic in the current universe because the lightest neutrino mass is not yet determined. In table~\ref{tb:mass4} we show energy density for the lightest neutrinos in the case of $m_{\rm lightest}\ll p_0\sim 3.15\tilde{T}_{\nu}(t_0)$. Here we consider both the normal mass ordering, $m_{\nu_3}>m_{\nu_2}>m_{\nu_1}$, and the inverted mass ordering, $m_{\nu_2}>m_{\nu_1}>m_{\nu_3}$.

To estimate the effects of $e^\pm$-annihilation into neutrinos during neutrino decoupling on neutrino number density and energy density, it is useful to compare the neutrino number density and relativistic energy density per one degree of freedom in the case when all $e^\pm$-pairs annihilate into photons, $n_0$ and $\rho_0$, respectively,
\begin{align}
n_0&=\frac{3\zeta(3)}{4\pi^2}T_{\nu}(t_0)^3=56.01\ {\rm cm^{-3}}, \\
\rho_0&=\frac{7\pi^2}{240}T_{\nu}(t_0)^4=29.65\ {\rm cm^{-3}},
\end{align}
where $T_{\gamma}(t_0)/T_{\nu}(t_0)=(11/4)^{1/3}$. We show the deviation of neutrino number density from the case when all $e^\pm$-pairs annihilate into photons, $\delta n_{\nu_i}^d\equiv n_{\nu_i}/n_0-1$, in table~\ref{tb:ND2}. The number densities for all neutrino species are enhanced by about $1\%$ due to $e^\pm$-annihilations to neutrinos during neutrino decoupling and the number density for $\nu_1$ is most efficiently enhanced.

\begin{table}[htbp]
\begin{center}
\small
  \begin{tabular}{c|c|c|c|c|c|c} \hline \hline
    $z_{\rm fin}$                                                              & $\delta \bar{\rho}_{\nu_1}(\%)$ & $\delta \bar{\rho}_{\nu_2}(\%)$ & $\delta \bar{\rho}_{\nu_3}(\%)$ & $\delta \bar{n}_{\nu_1}(\%)$ & $\delta \bar{n}_{\nu_2}(\%)$ & $\delta \bar{n}_{\nu_3}(\%)$ \\ \hline
   1.39797                    &0.764   &0.574   &0.409      & 0.468 & 0.350 & 0.248\\ \hline \hline
  \end{tabular}
 \caption{Final values of the distortions of ``relativistic'' energy densities $\delta \bar{\rho}_{\nu_i} \equiv \delta \rho_{\nu_i}/\rho_{\nu_0}$ and number densities $\delta \bar{n}_{\nu_i} \equiv  (n_{\nu_i}-n_{\nu_0})/n_{\nu_0}$ for  neutrinos in the mass basis after neutrino decoupling.}
  \label{tb:mass2}
  \end{center}
\end{table}

\begin{table}[htbp]
\begin{center}
\small
  \begin{tabular}{c|c|c} \hline \hline
  $n_{\nu_1}({\rm cm^{-3}})$ & $n_{\nu_2}({\rm cm^{-3}})$ & $n_{\nu_3}({\rm cm^{-3}})$ \\ \hline
      56.64 & 56.57 & 56.52\\ \hline \hline
  \end{tabular}
  \caption{Neutrino number density per one degree of freedom in the current homogeneous and isotropic universe including non-thermal distortions due to $e^\pm$-annihilation during neutrino decoupling.}
  \label{tb:mass3}
  \end{center}
\end{table}

\begin{table}[htbp]
\begin{center}
\small
  \begin{tabular}{c|c} \hline \hline
 {\rm Case} & $\rho_{\nu_{\rm lightest}}({\rm meV\ cm^{-3}})$ \\ \hline
       {\rm Normal\ Ordering}\ ($\nu_{\rm lightest}=\nu_1$,\ $m_{\nu_1}=0$) & 30.08\\ \hline 
        {\rm Inverted\ Ordering}\ ($\nu_{\rm lightest}=\nu_3$,\ $m_{\nu_3}=0$) & 29.97 \\ \hline \hline
  \end{tabular}
  \caption{Energy density per one degree of freedom for the lightest neutrinos with $m_{\nu_{\rm lightest}}=0$ in the current homogeneous and isotropic universe including non-thermal distortions due to $e^\pm$-annihilation during neutrino decoupling.}
  \label{tb:mass4}
  \end{center}
\end{table}

\begin{table}[h]
\begin{center}
	\begin{tabular}{c|c|c}
		\hline \hline
	  $\delta n_{\nu_1}^d$ (\%) & $\delta n_{\nu_2}^d$ (\%) & $\delta n_{\nu_3}^d$ (\%)   \\
		\hline
		 1.13 & 1.01 & 0.91  \\
		\hline \hline
	\end{tabular}
	\caption{Deviation of relic neutrino number density including non-thermal distortions during neutrino decoupling from the case when neutrinos decoupled instantaneously and all $e^\pm$-pairs annihilated into photons.}
  \label{tb:ND2}
\end{center}
\end{table}

\subsection{Helicity of relic neutrinos  --Majorana vs Dirac neutrinos--}
\label{sec:4.3}
The weak interaction is chiral, which is manifest in the Lagrangian. Due to its chirality, the left-chiral states for SM fermions interact with the weak bosons while the right-chiral states do not.
In the early universe, only left-chiral neutrinos and right-chiral anti-neutrinos, i.e., left-handed neutrinos and right-handed anti-neutrinos are produced via the weak interaction.
Note that chirality is different from helicity in general, which is defined as the projection of the spin vector onto the momentum vector.

During free streaming of relic neutrinos after their decoupling, the chirality for non-relativistic neutrinos is not conserved since the chiral symmetry in the free neutrino Lagrangian is broken due to their masses. On the other hand, the helicity for relic neutrinos is conserved in the homogeneous and isotropic universe.
Thus, we should estimate the spectrum for each helicity state of relic cosmic neutrinos in the current universe.

In the early universe, both chirality and helicity for relic neutrinos are conserved and then neutrino helicity and chirality have one-to-one correspondence since neutrinos are approximately massless in the early universe.
We define left (right) helical neutrinos with helicity $s_{\nu}=-1/2\ (+1/2)$ such that they correspond to left (right) handed neutrinos in the early universe. 
Then the spectra for the left-handed neutrinos (right-handed anti-neutrinos) produced in the early universe are translated into the left-helical neutrinos (right-helical anti-neutrinos) \cite{Long:2014zva},
\begin{align}
f_{\nu_i}(\bm{p}_{\nu},s_\nu=-1/2)&=
f_{\nu_i}(\bm{p}_\nu,t), \nonumber \\
f_{\nu_i}(\bm{p}_{\nu},s_\nu=+1/2)&\simeq 0, \nonumber \\
f_{\bar{\nu}_i}(\bm{p}_{\nu},s_\nu=-1/2)&\simeq
0, \nonumber \\
f_{\bar{\nu}_i}(\bm{p}_{\nu},s_\nu=+1/2)&= f_{\bar{\nu}_i}(\bm{p}_\nu,t)\simeq f_{\nu_i}(\bm{p}_\nu,t),
\end{align}
where $f_{\nu_i}(\bm{p}_{\nu}, t)$ is given by eq.~(\ref{DFC}) and $f_{\bar{\nu}_i}(\bm{p}_{\nu}, t)\simeq f_{\nu_i}(\bm{p}_\nu,t)$  if we neglect lepton asymmetry.
Here right-helical neutrinos, $\nu_i$ with $s_\nu=+1/2$, (left-helical anti-neutrinos, $\bar{\nu}_i$ with $s_\nu=+1/2$,) corresponds to right-handed neutrinos (left-handed anti-neutrinos), which are sterile states.
We assume sterile neutrinos are not produced in the early universe due to very weak interactions with the SM particles or have already decayed if sterile neutrinos are right-handed heavy Majorana particles as required for the see-saw mechanism.

For Majorana neutrinos, right-handed active anti-neutrinos are regarded as right-handed active neutrinos due to the lepton number violation. Then $f_{\nu_i}(\bm{p}_{\nu},s_\nu)$ for $\nu_i$ are given by
\begin{align}
f_{\nu_i}(\bm{p}_{\nu},s_\nu=-1/2)&=
f_{\nu_i}(\bm{p}_\nu,t), \nonumber \\
f_{\nu_i^s}(\bm{p}_{\nu},s_\nu=+1/2)&\simeq 0, \nonumber \\
f_{\nu_i^s}(\bm{p}_{\nu},s_\nu=-1/2)&\simeq
0, \nonumber \\
f_{\nu_i}(\bm{p}_{\nu},s_\nu=+1/2)&= f_{\bar{\nu}_i}(\bm{p}_\nu,t)\simeq f_{\nu_i}(\bm{p}_\nu,t),
\label{SNMa}
\end{align}
where $\nu_i^s$ denotes a sterile state of neutrino.
Note that even in the case of Majorana neutrinos lepton asymmetry can be interpreted as chiral asymmetry between left-handed and right-handed neutrinos. Then $f_{\bar{\nu}_i}(\bm{p}_\nu,t)$ and $f_{\nu_i}(\bm{p}_\nu,t)$ are different strictly speaking but almost the same approximately.

For Dirac neutrinos, since right-handed neutrinos and left-handed anti-neutrinos are sterile, $f_{\nu_i}(\bm{p}_{\nu},s_\nu)$ for $\nu_i$ are given by
\begin{align}
f_{\nu_i}(\bm{p}_{\nu},s_\nu=-1/2)&=
f_{\nu_i}(\bm{p}_\nu,t), \nonumber \\
f_{\nu_i^s}(\bm{p}_{\nu},s_\nu=+1/2)&\simeq 0, \nonumber \\
f_{\bar{\nu}_i^s}(\bm{p}_{\nu},s_\nu=-1/2)&\simeq
0, \nonumber \\
f_{\bar{\nu}_i}(\bm{p}_{\nu},s_\nu=+1/2)&= f_{\bar{\nu}_i}(\bm{p}_\nu,t)\simeq f_{\nu_i}(\bm{p}_\nu,t),
\label{SNDi}
\end{align}
where $\bar{\nu}_i^s$ denotes a sterile state of anti-neutrino.

From eqs.~(\ref{SNMa}) and (\ref{SNDi}), the  magnitude of relic neutrino spectra summed over helicity for Majorana and Dirac neutrinos differ by a factor of two, which is first pointed out in ref.~\cite{Long:2014zva},
\begin{align}
    \sum_{s_\nu=\pm 1/2}f_{\nu_i}(\bm{p}_{\nu},s_\nu)\simeq
    \left\{
    \begin{array}{ll}
   2f_{\nu_i}(\bm{p}_\nu,t) & {\rm for\ Majorana\ \nu_i} \\
   f_{\nu_i}(\bm{p}_\nu,t)  & {\rm for\ Dirac\ \nu_i}
    \end{array}
    \right..
\end{align}
Then number density and energy density summed over helicity for Majorana and Dirac neutrinos also differ by a factor of two,
\begin{align}
\sum_{s_\nu=\pm 1/2}n_{\nu_i}(s_\nu)\simeq
    \left\{
    \begin{array}{ll}
   2n_{\nu_i} & {\rm for\ Majorana\ \nu_i} \\
   n_{\nu_i}  & {\rm for\ Dirac\ \nu_i}
    \end{array}
    \right., \nonumber \\
    \sum_{s_\nu=\pm 1/2}\rho_{\nu_i}(s_\nu)\simeq
    \left\{
    \begin{array}{ll}
   2\rho_{\nu_i} & {\rm for\ Majorana\ \nu_i} \\
   \rho_{\nu_i}  & {\rm for\ Dirac\ \nu_i}
    \end{array}
    \right..
\end{align}

\clearpage
\section{Implications for the capture rates on cosmic neutrino capture on tritium} 
\label{sec:5}

Finally we discuss how neutrino spectral distortions from $e^\pm$-annihilations during neutrino decoupling affect direct detection of the C$\nu$B on tritium target, with emphasis on the PTOLEMY-type experiment \cite{Betti:2018bjv, PTOLEMY:2019hkd}, 
where cosmic neutrinos can be captured on tritium by the inverse beta decay process without threshold energy for neutrinos, $\nu_i+\mathrm{^3H}\rightarrow e^- + \mathrm{^3He}$.
Tritium is one of appropriate candidates for the target because of its availability, high capture rate for neutrinos, low Q-value and long half lifetime of $t_{1/2}=12.32$ years. Here we take 100 g of tritium as the target.
We take into account gravitational clustering for cosmic neutrinos in our Galaxy and nearby galaxies because we would observe the C$\nu$B directly inside our Galaxy.
We also comment on gravitational helicity flipping and annual modulation for the C$\nu$B.
Then we discuss the potential of direct measurements of such cosmological effects although it would be still extremely difficult to observe such effects directly.
In particular, we compute the capture rates of cosmic relic neutrinos on tritium, including such cosmological effects.

\subsection{Gravitational effects for the C$\nu$B}
\label{sec:5.1}


\subsubsection{Clustering for the C$\nu$B by our Galaxy and nearby galaxies}
\label{sec:5.1.1}

Near the Earth, non-relativistic relic neutrinos cluster locally in the gravitational potential of our Galaxy and nearby galaxies.
Then the local distribution function is distorted and the local number density is enhanced compared with the global distribution function and number density.
The local number density for relic neutrinos in the current universe is described as
\begin{align}
n_{\nu_i}^{\rm loc}=n_{\nu_i}(1+\delta n_{\nu_i}^c),
\end{align}
where $\delta n_{\nu_i}^c$ is an enhancement factor by the gravitational attraction by galaxies, which is estimated in refs.~\cite{Singh:2002de, Ringwald:2004np, deSalas:2017wtt, Zhang:2017ljh, Mertsch:2019qjv, Alvey:2021xmq}.
For reference, we display some of these values, estimated in a recent numerical study \cite{Mertsch:2019qjv}, in table~\ref{tb:CND}, where the authors consider the gravitational potential in the Milky Way, Virgo cluster, and Andromeda galaxy.
Note that so far, when evaluating values of $\delta n_{\nu_i}^c$, effects of $e^\pm$-annihilations into $\nu,\ \bar{\nu}$ during neutrino decoupling have not been taken into account simultaneously. 
For $m_{\nu_i}<0.15 {\rm eV}$, spectral distortions to the momentum distributions for relic cosmic neutrinos by the gravitational clustering have not also been explicitly estimated (see ref.~\cite{Ringwald:2004np} for spectral distortions by gravitational clustering for relic neutrinos with $m_{\nu_i}\geq 0.15 {\rm eV}$).  

In the following, we discuss only the case where $\delta n_{\nu_i}^c<1$ and the lightest neutrino mass is quite small because the Planck satellite suggests $\sum m_{\nu}<0.12\ {\rm eV}$. Then the local number density for relic neutrino can be parametrized as, using linear approximation,
\begin{align}
n_{\nu_i}^{\rm loc}\simeq n_{0}(1+\delta n_{\nu_i}^c+\delta n_{\nu_i}^d),
\end{align}
where $\delta n_{\nu_i}^d$ is the enhancement factor by $e^\pm$-annihilations into $\nu$ and $\bar{\nu}$ during neutrino decoupling given in table~\ref{tb:ND2}.

\begin{table}[h]
\begin{center}
	\begin{tabular}{cc}
		\hline \hline
		$m_{\nu_i}$ ({\rm meV}) & $\delta n^c_{\nu_i}\ (\%)$   \\
		\hline 
		10 & 0.53  \\
		50 & 12 \\
		100 & 50 \\
		200 & 300 \\
		\hline \hline
	\end{tabular}
	\caption{The enhancement factor, $\delta n^c_{\nu_i}$, due to neutrino clustering by our Galaxy and nearby galaxies for given values of neutrino masses \cite{Mertsch:2019qjv}.}
  \label{tb:CND}
\end{center}
\end{table}


\subsubsection{Helicity flipping and annual modulation for the C$\nu$B}
\label{sec:5.1.2}

We shortly comment on gravitational helicity flipping and annual modulations for relic neutrinos.
Gravitational clustering for massive neutrinos may induce mixing of relic neutrino helicity \cite{Long:2014zva, Roulet:2018fyh, Duda:2001hd} since the direction of neutrino momentum would change in the gravitational potential for our Galaxy whereas its spin does not. Although the quantitative calculations have not yet been achieved, the capture rates on tritium would not change since their capture rates depend on neutrino number density summed over helicities at leading order as we will see in the next section.
In addition, an annual modulation for relic neutrinos might occur in a direct detection experiment for the C$\nu$B since their velocity relative to the Earth could be anisotropic due to neutrino clustering and the gravitational focusing for the C$\nu$B by the Sun could also occur. The former effect is negligible since the capture rates on tritium target are independent of neutrino velocity as we will see in the next section. The latter effect is expected to change the capture rates by much less than 1$\%$ for $m_{\nu}<0.15\ {\rm meV}$ \cite{Safdi:2014rza}. In the following, we neglect helicity flipping and annual modulation for relic neutrinos.


\subsection{Precise capture rates on tritium including sub-dominant cosmological effects}
\label{sec:5.2}

In table~\ref{tb:mass4}, non-thermal distortions during neutrino decoupling enhance the number density of the C$\nu$B by about $1\%$.
To properly incorporate such effects into the capture rates of the C$\nu$B on tritium, we discuss the formula of their capture rate with $1\%$ precision.

Cosmic relic neutrinos can be captured on tritium by the following inverse beta decay process,
\begin{align}
\nu_i+\mathrm{^3H}\rightarrow \mathrm{^3He}+e^-.
\end{align}
The total capture rate for the C$\nu$B in this process, $\Gamma_{\rm C\nu B}$, can be written
\begin{align}
\Gamma_{\rm C\nu B}=\sum_{i=1}^{{\rm N}_\nu}\Gamma_i,
\label{Gi}
\end{align}
where ${\rm N}_{\nu}$ is the number of (mass) species of neutrinos. $\Gamma_i$ is the capture rate for a given mass-eigenstate of neutrino $\nu_i$, given by
\begin{align}
\Gamma_i=\mathrm{N}_{\rm T}\sum_{s_\nu=\pm 1/2}\int \frac{d^3p_{\nu}}{(2\pi)^3}\sigma_{\nu_i}(\bm{p}_\nu,s_\nu)v_{\nu_i}f^{\rm loc}_{\nu_i}(\bm{p}_\nu,s_\nu),
\label{CR}
\end{align} 
where $\mathrm{N}_{\rm T}=\mathrm{M}_{\rm T}/{\rm M}_{\rm ^3H}$ is the number of tritium, $\rm M_T$ is the total tritium mass in the experimental setup, and ${\rm M_{\rm ^3H}}\simeq 2809.432\ {\rm MeV}$ is the atomic mass of tritium.
$s_{\nu},\ v_{\nu_i}=|\bm{p_\nu}|/E_{\nu_i}$ and $\sigma_{\nu_i}$ are helicity, velocity and the total cross section in the inverse beta decay on tritium, respectively. 
$f_{\nu_i}^{\rm loc}(\bm{p}_{\nu},s_\nu)$ is the local momentum distribution for relic cosmic neutrinos around the Earth, which satisfies $n^{\rm loc}_{\nu_i}(s_\nu)=\int\frac{dp_\nu^3}{(2\pi)^3}f^{\rm loc}_{\nu_i}(\bm{p}_\nu,s_\nu)$.

In cosmic neutrino capture on tritium, the spins of the outgoing electron and nucleus would not be measured. In addition, the spin of the initial nucleus would not be identified either.
On the other hand, the helicity state for cosmic neutrinos in the Dirac case is polarized as in section~\ref{sec:4.3}.
Then we compute the spin-polarized cross section for $\nu_i$.
After averaging over the spin of $\mathrm{^3H}$ and summing over the spin of outgoing $e^-$ and $\mathrm{^3He}$ , the formulae of $\sigma_{\nu_i}(\bm{p}_\nu,s_{\nu})$ with $1\%$ precision reduces to (see appendix~\ref{appd} for detail calculations)
\begin{align}
\sigma_{\nu_i}(\bm{p}_\nu, s_{\nu})&\simeq\frac{G_F^2}{2\pi}|V_{ud}|^2|U_{ei}|^2\frac{m_{\rm ^3He}}{m_{\rm ^3H}v_{\nu_i}}\left(\langle f_F \rangle^2 +\frac{g_A^2}{g_V^2}\langle g_{GT} \rangle^2 \right) \nonumber \\
&\ \ \ \ \times F(2,E_e)E_e|\bm{p}_e|(1-2s_\nu v_{\nu_i}),
\label{PCS}
\end{align}
where $V_{ud}\simeq 0.9740$ is a component of the Cabibbo-Kobayashi-Maskawa (CKM) matrix, $m_{\rm ^3H}\simeq 2808.921\ {\rm MeV}$ and  $m_{\rm ^3He}\simeq 2808.391\ {\rm MeV}$ are the nuclear masses of ${\rm ^3H}$ and $\rm ^3He$, $g_A\simeq 1.2723$ and $g_V\simeq 1$ are the axial and vector coupling constant, and $\langle f_F \rangle\simeq 0.9998$ and $\langle g_{\rm GT} \rangle \simeq \sqrt{3}\times(0.9511\pm 0.0013)$ are the reduced matrix elements of the Fermi and Gamow-Teller (GT) operators, respectively. The Fermi function $F(Z, E_e)$ is an enhancement factor
by the Coulombic attraction of the outgoing electron and proton, which is approximately given by \cite{Primakoff:1959chj}
\begin{align}
F(Z,E_e)=\frac{2\pi\alpha ZE_e/|\bm{p}_e|}{1-e^{-2\pi\alpha Z E_e/|\bm{p}_e|}},
\label{FermiF}
\end{align}  
where $\alpha\simeq 137.036$ is the fine structure constant. $Z$ is the atomic number of the daughter nucleus and $Z=2$ for $\rm ^3He$.
The energy and momentum for an emitted electron $E_e$ and $\bm{p}_e$ depend on the neutrino masses and momenta strictly because of momentum conservation in the inverse $\beta$-decay process. However, since the contributions of the neutrino masses and momenta to $E_e$ and $\bm{p}_e$ are very small, $E_e$ and $|\bm{p}_e|$ are approximately given by (see appendix \ref{appc} for details)
\begin{align}
E_e&\simeq K^0_{\rm end}+m_e+E_{\nu_i}\simeq K_{\rm end}^0+m_e, \nonumber \\
|\bm{p}_e|&=\sqrt{E_e^2-m_e^2},
\label{Kinematics}
\end{align}
where $K_{\rm end}^0$ is the beta decay endpoint kinetic energy for massless neutrinos given by
\begin{align}
K_{\rm end}^0=\frac{(m_{\rm ^3H}-m_e)^2-m_{\rm ^3He}^2}{2m_{\rm ^3H}}\simeq 18.6\ {\rm keV}.
\end{align} 
$E_{\nu_i}$ is so small compared to $K_{\rm end}^0$ and $m_e$ that we can safely neglect $E_{\nu_i}$ in eq.~(\ref{Kinematics}).

Then we obtain $\Gamma_i$ with $1\%$ precision substituting eq.~(\ref{PCS}) into eq.~(\ref{CR}),
\begin{align}
    \Gamma_i&\simeq{\rm N_T}\frac{G_F^2}{2\pi}|V_{ud}|^2|U_{ei}|^2\frac{m_{\rm ^3He}}{m_{\rm ^3H}}\left(\langle f_F \rangle^2 +\frac{g_A^2}{g_V^2}\langle g_{ GT} \rangle^2 \right) \nonumber \\
&\ \ \ \ \times F(2,E_e)E_e|\bm{p}_e|\sum_{s_{\nu=\pm1/2}}\left(n_{\nu_i}(s_\nu)-2s_{\nu}\langle v_{\nu_i} \rangle \right),
\label{PCR}
\end{align}
where $\langle v_{\nu_i} \rangle$ is the (unnormalized) average magnitude of velocity for $\nu_i$ given by
\begin{align}
\langle v_{\nu_i} \rangle= \int \frac{d^3 p_{\nu}}{(2\pi)^3} f_{\nu_i}(\bm{p_{\nu}},s_\nu)v_{\nu_i}.
\end{align}
Typically, $\langle v_{\nu_i} \rangle$ contributes more than $1\%$ to $\Gamma_{\nu_i}$. If $m_{\nu_i}\gtrsim 100\ \rm meV$, due to $v_{\nu_i}\sim p_0/m_{\nu_i} \lesssim 0.01$,  we can drop $\langle v_{\nu_i} \rangle$ in the formula of eq.~(\ref{PCR}) with $1\%$ precision. Here $p_0 \sim 3.15T_{\nu}(t_0)\sim 0.53\ {\rm meV}$ is the average momentum of the C$\nu$B in the current universe.
We also comment on whether we can use further approximations with $1\%$ precision to write eq.~(\ref{PCR}) into a simpler form.
For massless neutrinos, due to $v_{\nu_i}=|\bm{p}_{\nu_i}|/E_{\nu_i}=1$, the (unnormalized) velocity is written as $\langle v_{\nu_i} \rangle=n_{\nu_i}$.
For non-relativistic neutrinos $(m_{\nu} \gtrsim 10\ {\rm meV})$, due to $v_{\nu_i}\ll1$, $\langle v_{\nu_i} \rangle$ is approximately written as $\langle v_{\nu_i} \rangle \simeq \int d^3p/(2\pi_{\nu}^3)f_{\nu}^0(\bm{p},t_0)|\bm{p}_{\nu}|/E_{\nu_i}$, where $f_{\nu}^0(\bm{p}_\nu,t_0)=[\exp(\bm{p}_{\nu}/T_{\nu}(t_0))+1]^{-1}$ and $T_{\nu}(t_0)/T_{\gamma}(t_0)=(4/11)^{1/3}$.
We note that gravitational helicity flipping for massive neutrinos by neutrino clustering would be negligible since the helicity-dependent part in $\Gamma_{i}$ is already suppressed by $v_{\nu_i}$.

\subsubsection{Majorana vs Dirac neutrinos}
\label{sec:5.2.1}

For non-relativistic neutrinos, i.e., $v_i \ll 1$, if we set $v_{\nu_i}=0$ in eq.~(\ref{PCR}), $\Gamma_i$ is porportional to $\sum_{s_\nu} n_{\nu_i}$ and left-helical and right-helical components for relic neutrinos interact with tritium with the same magnitude via the weak interaction. Then the capture rate on tritium for Majorana neutrinos $\Gamma_i^M$ is twice that for Dirac neutrinos \cite{Long:2014zva},
\begin{align}
    \Gamma_i^M\bigl|_{v_{\nu_i}\ll 1} \simeq 2\Gamma_i^D\bigl|_{v_{\nu_i}\ll 1}.
    \label{CRMD}
\end{align}
On the other hand, for relativistic neutrinos, i.e., $v_i\simeq 1$, only the left-helical neutrinos interact with tritium via the weak interaction since helicity coincides with chirality in the relativistic limit. Then in both Majorana and Dirac cases, the capture rates are the same \cite{Roulet:2018fyh},
\begin{align}
    \Gamma_i^M\bigl|_{v_{\nu_i}\simeq 1} \simeq \Gamma_i^D\bigl|_{v_{\nu_i}\simeq 1}.
    \label{CRMD2}
\end{align}
Note again that the approximations in eqs.~(\ref{CRMD}) and (\ref{CRMD2}) might not be valid for the capture rates with $1\%$ precision. To estimate the capture rates with $1\%$ precision, the term that depends on $v_{\nu_i}$ in eq.~(\ref{PCR}) should be included precisely.

\subsubsection{Values of the capture rates on tritium with $m_{\rm lightest}=0$}
\label{sec:5.2.2}

For references, we show values of the capture rates including cosmological effects discussed in sections \ref{sec:4.2} and \ref{sec:5.1} in the case of $m_{\rm lightest}=0$.
We choose other neutrino masses and their ordering to satisfy the observed values of neutrino squared-mass differences from neutrino oscillation experiments \cite{Esteban:2020cvm,deSalas:2020pgw},
\begin{align}
 {\rm Normal\ Ordering\ (NO)}:\ &\Delta m_{21}^2\simeq (8.6\ {\rm meV})^2\ \ \ \ \Delta m_{31}^2\simeq (50\ {\rm meV})^2  \nonumber \\
 {\rm Inverted\ Ordering\ (IO)}:\ &\Delta m_{21}^2\simeq (8.6\ {\rm meV})^2\ \ \ \ \ \Delta m_{32}^2\simeq -(50\ {\rm meV})^2
\end{align}
In both neutrino mass ordering we take the following values of the PMNS matrix,
\begin{align}
|U_{e1}|^2\simeq 0.681 ,\ \ \ \ |U_{e2}|^2\simeq 0.297,\ \ \ \ |U_{e3}|^2\simeq 0.0222.
\label{PMNS}
\end{align}
Note that neutrino squared-mass differences and neutrino mixing parameters currently include a few percent (about $10\%$) uncertainties even at $1\sigma$ ($3\sigma$) confidence level.

In table~\ref{tb:CR}, we show values of the capture rates on $100$ grams of tritium in both the cases of NO and IO for Majorana and Dirac neutrinos with $m_{\rm lightest}=0$.
$\delta \Gamma_i^d$ denotes the differences between the cases with and without effects of $e^\pm$-annihilation during neutrino decoupling and $\delta \Gamma_i^c$ denotes the differences with and without gravitational clustering for relic neutrinos in nearby galaxies.

For Majorana neutrinos, the capture rates for the first and second heaviest neutrinos are slightly less than twice those for Dirac neutrinos because of $v_{\nu_i}\simeq 0$. On the other hand, the capture rates for massless (or almost massless) neutrinos in the cases of Majorana and Dirac neutrinos are the same because of $v_{\nu_i}\simeq 1$.

\begin{table}[h]
\begin{center}
\small
	\begin{tabular}{cc|ccc|ccc|ccc}
		\hline \hline
		Ordering & Case & $\Gamma_1$ &  $\delta\Gamma_1^d$ & $\delta\Gamma_1^c$ & $\Gamma_2$ &$\delta\Gamma_2^d$ & $\delta\Gamma_2^c$ & $\Gamma_3$ & $\delta\Gamma_3^d$ & $\delta\Gamma_3^c$ \\
		\hline 
		\multirow{2}{*}{NO} & Majorana & 5.48 & 0.061 & 0 & 2.40 & 0.024 & 0.013 & 0.200 & $1.6{\tiny \times} 10^{-3}$ & 0.021  \\
		 &  Dirac & 5.48 & 0.061 & 0 & 1.27 & 0.012 & $6.3{\tiny \times} 10^{-3}$ & 0.101 & $8.0{\tiny \times} 10^{-4}$ & 0.011 \\
		\multirow{2}{*}{IO} & Majorana & 6.13 & 0.061 & 0.65 & 2.67 & 0.024 & 0.28 & 0.178 & $1.6{\tiny \times} 10^{-3}$ & 0 \\
		 & Dirac & 3.10 & 0.031 & 0.33 & 1.35 & 0.012 & 0.14 & 0.178 & $1.6{\tiny \times} 10^{-3}$ & 0 \\
		\hline \hline
	\end{tabular}
	\caption{ Capture rates of relic cosmic neutrinos on $100$ grams of tritium in unit of $\rm year^{-1}$ with $m_{\rm lightest}=0$. $\delta \Gamma_i^d$ is the differences between the cases with and without effects of $e^\pm$-annihilation during neutrino decoupling and $\delta \Gamma_i^c$ is the differences with and without gravitational clustering for relic neutrinos in nearby galaxies.}
  \label{tb:CR}
\end{center}
\end{table}
  
\subsubsection{Discussions on exposure and uncertainties in the capture rates}
\label{sec:5.2.3}
In this section we discuss the required amount of tritium to observe the sub-leading cosmological effects themselves, $\delta \Gamma_i^{c,d}$, and the estimated error of the capture rates for relic neutrinos on tritium in more detail. 

To observe $\delta \Gamma_i^{c,d}$, we need a large number of events to satisfy typically
\begin{align}
\frac{\sum_i\delta \Gamma_i^{c,d}\mathrm{T}}{\sqrt{\Gamma_{\rm C\nu B}\mathrm{T}+\Gamma_{\rm background}\mathrm{T}}} \gg 1,
\end{align}
where $\mathrm{T}$ is the exposure time and $\Gamma_{\rm background}$ is a background rate.
Even if the background is successfully removed, we need $10^2-10^4$ events of the C$\nu$B signal ($\Gamma_{\rm C\nu B}\mathrm{T}\sim10^2-10^4$) because of $\delta \Gamma_i^{c,d} \sim (0.1-0.01)\times\Gamma_i$ for $\sum_i m_{\nu_i}<0.12\ {\rm eV}$.
This requirement corresponds to the need for $10-10^3$ kg yr of exposure of tritium. Currently, it is extremely difficult to obtain such amount of the exposure. In the next section \ref{sec:5.4}, we comment on $\beta$-decay background, which is one of main background in cosmic neutrino capture on tritium. 

The estimated error of the neutrino capture rates mainly comes from the uncertainties of the neutrino mixing parameter, $|U_{ei}|^2$, and the undetermined value of the lightest neutrino mass, $m_{\rm lightest}$. 
The current errors of PMNS matrix are about a few percent (about $10\%$) at $1\sigma\ (3\sigma)$ confidence level \cite{Esteban:2020cvm,deSalas:2020pgw}. The current upper bound of $m_{\rm lightest}$ is $\lesssim 0.8\ {\rm eV}$ \cite{KATRIN:2021uub}.
Thus, unfortunately, it is still difficult to incorporate cosmological sub-dominant contributions into the value of $\Gamma_{\nu_i}$ precisely. However, $\delta \Gamma_i^{c,d}$ for $m_{\rm lightest}=0$ is correctly estimated since uncertainties of $|U_{ei}|$ are canceled out in $\delta \Gamma_i^{c,d}$.
Future neutrino oscillation experiments will reduce uncertainties of PMNS matrix (see ,e.g.,~\cite{Abe:2016tii, Hyper-Kamiokande:2018ofw, DUNE:2020ypp}).
In addition, measurement of large $\beta$-decay background in the PTOLEMY-type experiment might determine the value of $m_{\rm lightest}$ very precisely~\cite{PTOLEMY:2019hkd}. 

We also note that the theoretical calculation of $\langle g_{\rm GT} \rangle$ still includes the uncertainty of a few $\%$, although the estimation of $\langle g_{GT} \rangle$ through the observation of the tritium half-life and the value of the Fermi operator, $\langle f_{F} \rangle$, only involves uncertainty of $0.1\%$ \cite{Baroni:2016xll}.

For a large value of $m_{\rm lightest}$, gravitational clustering effects of relic neutrinos are typically more dominant than effects of $e^\pm$-annihilation during neutrino decoupling. 
Although the C$\nu$B itself with a large value of $m_{\rm lightest}$ would be easier to observe due to a large gravitational clustering, it is also a very difficult task to distinguish the effects of $e^\pm$-annihilation during neutrino decoupling from gravitational clustering effect of relic neutrinos.

Based on the evaluation in this section, it is still extremely difficult to observe $e^\pm$-annihilation during neutrino decoupling in the PTOLEMY-type experiment. But, the precise capture rates including cosmological sub-dominant contributions might be useful to distinguish the SM from physics beyond the SM properly in the future.


\subsection{$\beta$-decay background and the energy resolution of the detector to distinguish the C$\nu$B signal from it}
\label{sec:5.4}

Finally we comment on $\beta$-background and the required energy resolution of the detector to distinguish the C$\nu$B signal from this background, which is one of main difficulties to observe the C$\nu$B directly in the inverse $\beta$-decay process.

The main background comes from tritium $\beta$-decay process,
\begin{align}
{\rm ^3H}\rightarrow {\rm ^3He}+e^-+\bar{\nu}_i.
\end{align}
The $\beta$-decay spectrum and the capture rate for the $\beta$-decay process are given by \cite{Masood:2007rc} (see also appendix~\ref{appd})
\begin{align}
\frac{d\Gamma_\beta}{dE_e}&={\rm N_T}\frac{G_F^2}{2\pi^3}|V_{ud}|^2|U_{ei}|^2\frac{m_{\rm ^3He}}{m_{\rm ^3H}}\left(\langle f_F \rangle^2 +\frac{g_A^2}{g_V^2}\langle g_{ GT} \rangle^2 \right)\nonumber \\
&\ \ \ \   \times F(2,E_e)E_e|\bm{p}_e|  \sum_{i=1}^3|U_{ei}|^2H(E_e, m_{\nu_i}),
\end{align}
where 
\begin{align}
H(E_e,m_{\nu_i})&= \frac{1-m_e^2/(E_em_{\mathrm{^3H}})}{(1-2E_e/m_{\mathrm{^3H}}+m_e^2/m_{\mathrm{^3H}}^2)^2}
\sqrt{(E_{e}^{{\rm max},i}-E_e)\left(E_{e}^{{\rm max},i}-E_e+\frac{2m_{\nu_i}m_{\mathrm{^3He}}}{m_{\mathrm{^3H}}}\right)} \nonumber \\
&\ \ \ \ \times\left[E_{e}^{{\rm max},i}-E_e+\frac{m_{\nu_i}}{m_{\mathrm{^3H}}}(m_{\mathrm{^3He}}+m_{\nu_i})\right],
\end{align}
$E_e^{{\rm max},i}$ is the maximal energy of the emitted electron for ${\rm ^3H}\rightarrow {\rm ^3He}+e^-+\bar{\nu}_i$ , where the electron is emitted in opposite direction to both $\rm ^3He$ and $\bar{\nu_e}$ (see also appendix \ref{appc}),
\begin{align}
E_e^{{\rm max},i}\simeq K_{\rm end}^0+m_e-m_{\nu_{i}}.
\end{align}
Then the maximal energy for the emitted electron in the $\beta$-decay process called the energy at $\beta$-decay endpoint is
\begin{align}
E_e^{\rm end}\simeq K_{\rm end}^0+m_e-m_{\nu_{\rm lightest}},
\end{align}
where $m_{\rm lightest}$ is the lightest neutrino mass.
We can see that the $\beta$-decay spectrum $d\Gamma_{\beta}/dE_e$ vanishes for $E_e=E_e^{\rm end}$. Then the total tritium $\beta$-decay rate is obtained as
\begin{align}
\Gamma_{\beta}=\int^{E_e^{\rm end}}_{m_e}dE_e \frac{d\Gamma_{\beta}}{dE_e}\simeq 10^{24}\ \left( \frac{\rm M_T}{100\ {\rm g}}\right) {\rm yr^{-1}}.
\end{align}
Since the event number of $\beta$-decay background is extremely larger than that of the C$\nu$B signal, we must distinguish the two signals clearly.

To distinguish the C$\nu$B signal and $\beta$-decay background, we need a tiny energy resolution of the detector $\Delta$.
The energy resolution of a detector characterizes the smallest separation where two signals can be distinguished. The $\beta$-decay background closest to the C$\nu$B signal is the electron signal with the maximal energy $E_e^{\rm max}$. To distinguish the C$\nu$B signal for a mass species $\nu_i$ from $\beta$-decay background near the endpoint, the required energy resolution $\Delta_i$ is expected to be (see appendix \ref{appc} for details)
\begin{align}
\Delta_i \lesssim E_e^{{\rm C\nu B}, i}-E_e^{\rm end} \simeq m_{\rm lightest}+E_{\nu_i},
\end{align}
where $E_e^{{\rm C\nu B}, i}$ is the emitted electron energy from the C$\nu$B signal, $\nu_i+\mathrm{^3H}\rightarrow e^-+\mathrm{^3He}$ given by eq.~(\ref{Kinematics}). 

To take into account the energy resolution of the detector $\Delta$ in the spectrum and the number of events for the C$\nu$B signal and the $\beta$-decay background, we model the would-be observed spectrum of the emitted electron as a Gaussian-smeared version of the actual spectrum. This is achieved by convolving both the C$\nu$B signal and the $\beta$-decay background with a Gaussian of full width at half maximum (FWHM) equal to $\Delta=\sqrt{8\ln 2}\sigma$, where $\sigma$ is the Gaussian standard deviation,
\begin{align}
\frac{d\tilde{\Gamma}_i}{dE_e} &=\frac{1}{\sqrt{2\pi}\sigma}\int^{\infty}_{-\infty}dE_e'\  \Gamma_i(E_e')\ \delta[E_e'-(E_{\rm end}+E_{\nu_i}+m_{\rm lightest})]\exp\left[-\frac{(E_e'-E_e)^2}{2\sigma^2} \right], 
\label{tildeGi}
 \\
\frac{d\tilde{\Gamma}_{\beta}}{dE_e}&=\frac{1}{\sqrt{2\pi}\sigma}\int^{\infty}_{-\infty}dE_e' \ \frac{d \Gamma_{\beta}}{dE_e}(E_e')\ \exp\left[-\frac{(E_e'-E_e)^2}{2\sigma^2} \right],
\end{align}
Substituting eq.~(\ref{Gi}) into eq.~(\ref{tildeGi}), the smeared spectrum of the emitted electron from the C$\nu$B signal can be written as
\begin{align}
\frac{d\tilde{\Gamma}_i}{dE_e}&=\frac{\mathrm{N_T}}{\sqrt{2\pi}\sigma}\sum_{{s_{\nu}}=\pm\frac{1}{2}} \int \frac{d^3 p_\nu}{(2\pi)^3}\sigma_{\nu_i}(\bm{p}_\nu, s_{\nu})v_{\nu_i}f_{\nu_i}(\bm{p},s_{\nu}) \nonumber \\
&\ \ \ \ \ \ \ \  \times \exp \left\{-\frac{[E_e-(E_{\rm end}+m_{\rm lightest}+E_{\nu_i})]^2}{2\sigma^2} \right\},
\label{Fredholm}
\end{align}
where
\begin{align}
\sigma_{\nu_i}(\bm{p}_\nu,s_{\nu})&=\sigma_{\nu_i}(\bm{p}_\nu,s_{\nu},E_e') \nonumber \\
&=\sigma_{\nu_i}(\bm{p}_\nu,s_{\nu},E_{\rm end}+m_{\rm lightest}+E_{\nu_i}).
\end{align}
eq.~(\ref{Fredholm}) is a Fredholm integral equation of the first kind and $\frac{d\tilde{\Gamma}_i}{dE_e}$ is a would-be observed quantity. After solving eq.~(\ref{Fredholm}) inversely, the spectrum of the C$\nu$B, $f_{\nu_i}(\bm{p},s_\nu)$, can be in principle reconstructed though we might need a significantly large number of observations for the C$\nu$B events. We leave the detailed study for the reconstruction of the C$\nu$B spectrum $f_{\nu_i}(\bm{p},s_\nu)$ on tritium as future work.

\begin{figure}[htbp]
\begin{minipage}{0.5\hsize}
   \begin{center}
     \includegraphics[clip,width=8.8cm]{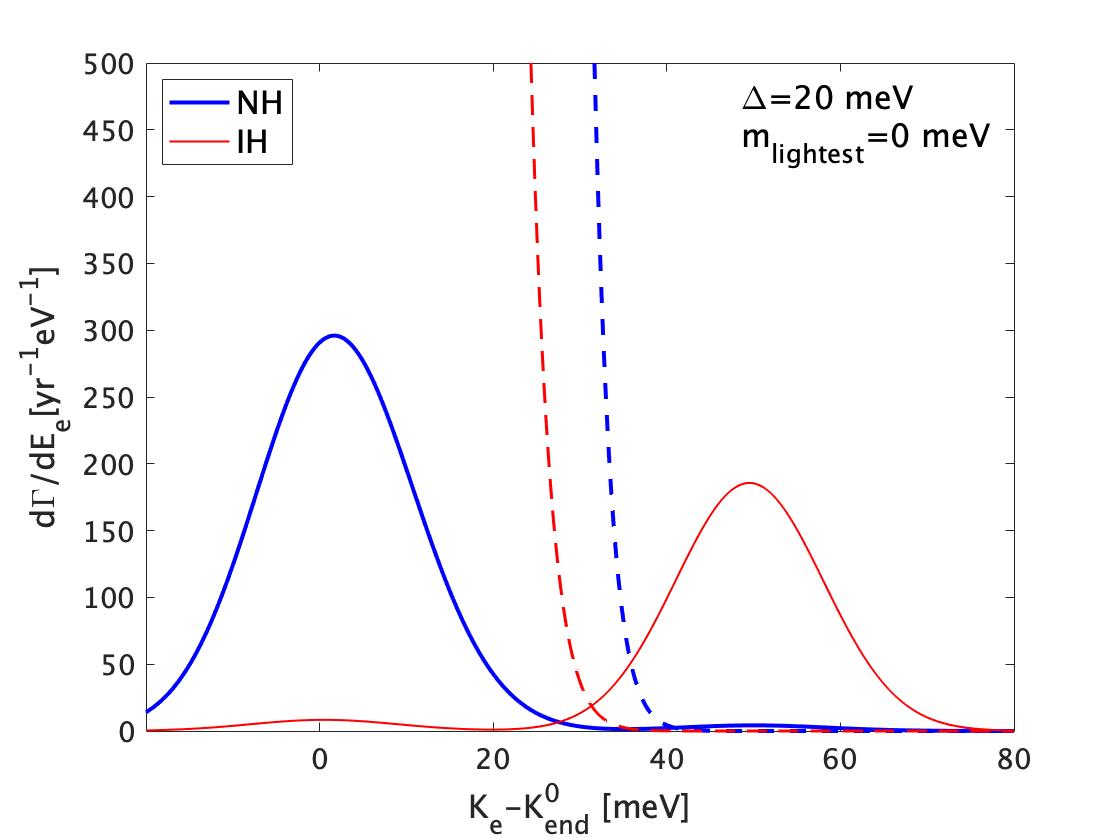}
    \end{center}
     \end{minipage}
      \begin{minipage}{0.5\hsize}
      \begin{center}
     \includegraphics[clip,width=8.8cm]{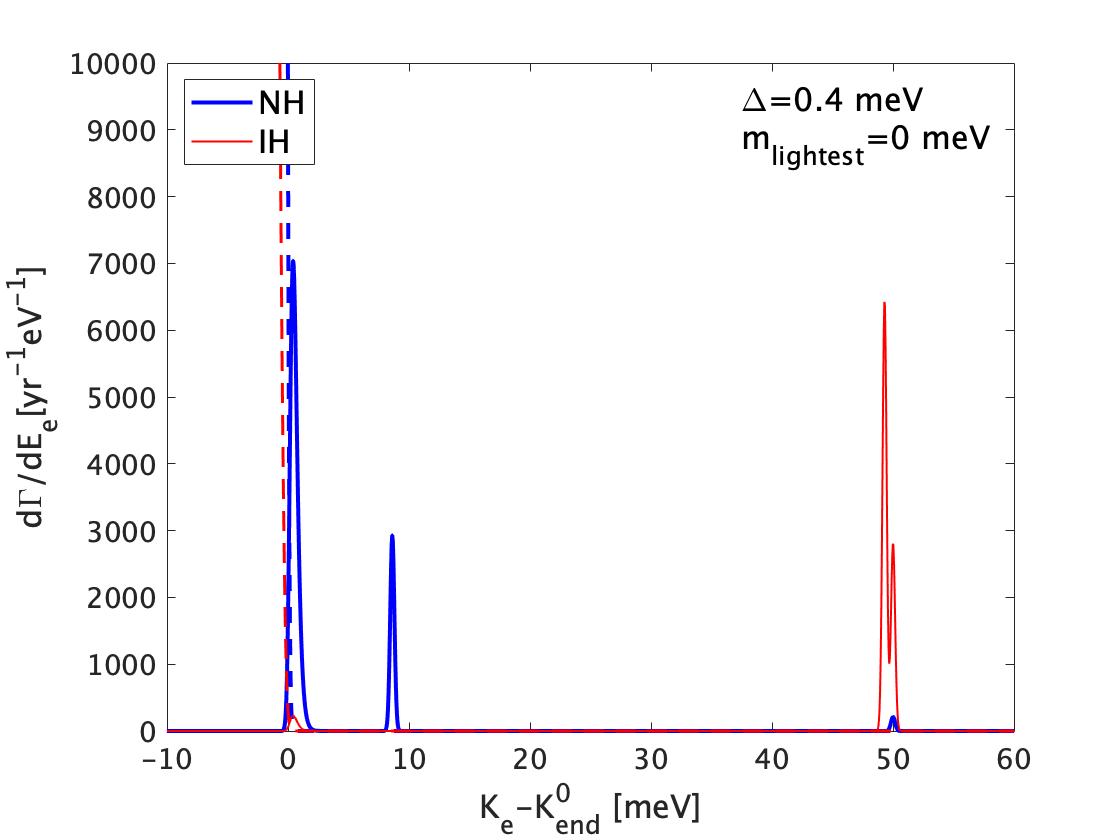}
    \end{center}
     \end{minipage}
    \vspace{-4mm}
 \caption{The expected spectra as a function of the electron kinetic energy, $K_e=E_e-m_e$, for the emitted electrons from the C$\nu$B signals (solid lines) and the $\beta$-decay background (dashed lines) in a tritium experiment, assuming $m_{\rm lightest}=0\ {\rm meV}$ and 100 g of tritium, with the energy resolution $\Delta=20\ {\rm meV}$ (left panel) and $\Delta=0.4\ {\rm meV}$ (right panel) in the case of Dirac neutrinos.
Bold blue lines represent the NH case and fine red lines represent the IH case.}
 \label{fig:SpectrumPTOLEMY}
 \end{figure}

In figure~\ref{fig:SpectrumPTOLEMY}, we show the expected spectra for the emitted electrons from the C$\nu$B signals (solid lines) and the $\beta$-decay background (dashed lines) with $m_{\rm lightest}=0\ {\rm meV}$ and 100 g of tritium, the energy resolution $\Delta=20\ {\rm meV}$ (left panel) and $\Delta=0.4\ {\rm meV}$ (right panel) considering the case of Dirac neutrinos and both the normal (fine red) and inverted (bold blue) mass hierarchies. In these figures, we neglect spectral distortions for the C$\nu$B from $e^\pm$-annihilation during their neutrino decoupling and the gravitational clustering for simplicity.
We can see that the C$\nu$B signal is distinguished from the $\beta$-decay background if $\Delta \gg E_{\nu_i}$. It is easier to distinguish the C$\nu$B signal from the $\beta$-decay background in the inverted mass ordering than the normal ordering. This is because we can obtain a larger number of events for the heaviest neutrinos in the inverted case due to the large value of $|U_{e1}|$.
In addition, $\beta$-decay spectrum near the endpoint is smaller in the inverted case because in the inverted case the $\beta$-decay spectrum near the endpoint is composed of $\nu_3$ with small $|U_{e3}|$ while in the normal ordering that is composed of $\nu_1$ with large $|U_{e1}|$.

\subsubsection{Comments on statistical analysis}

To estimate the required energy resolution of the detector $\Delta$ and exposure of tritium to discover the C$\nu$B in a qualitative way, we need statistical analysis.  
In ref.~\cite{PTOLEMY:2019hkd}, the authors estimated statistical significance for the detection of the C$\nu$B on tritium as a function of the lightest neutrino mass and the energy resolution in an exposure of 100 g yr of tritium using a $\chi^2$-analysis (see figure~5 in ref.~\cite{PTOLEMY:2019hkd}). Here a fiducial value of constant number events of background of $N_b=\Gamma_bT$, where $\Gamma_b=10^{-5} {\rm Hz}$ in the 15~eV region around the $\beta$-decay endpoint energy, is introduced in addition to the $\beta$-decay background.
If we would obtain a larger exposure of tritium, the result of figure~5 in ref.~\cite{PTOLEMY:2019hkd} will be improved. The reduction of the constant background $N_b$ might improve the result. 
A more quantitative discussion will be possible when the more concrete setup of the PTOLEMY-type experiment is decided, and the neutrino mass ordering and the lightest neutrino mass are constrained more severely from complementary future neutrino experiments.

We leave as future work the statistical analysis to estimate the required energy resolution $\Delta$ and exposures to observe the C$\nu$B spectral distortions due to $e^\pm$-annihilation in neutrino decoupling and gravitational clustering by nearby galaxies. However, the required energy resolution would not change drastically compared to observing the C$\nu$B itself since their spectral distortions are sub-leading contributions.
As discussed in the section \ref{sec:5.2.3}, to observe $1-10\%$ modifications in $\Gamma_i$ due to their spectral distortions, one will need $10^2-10^4$ events of the C$\nu$B. The required exposures correspond to $10-10^3$ kg yr of the exposure of tritium.
It is extremely difficult to achieve this exposure at present.
Note that here we consider neutrino masses small enough to satisfy $\sum m_{\nu}<0.12\ {\rm eV}$. 
If neutrino masses are enough large, the required exposure will be smaller due to large neutrino clustering.
However, it would be difficult to distinguish the C$\nu$B spectral distortions due to $e^\pm$-annihilation in neutrino decoupling from such large neutrino clustering experimentally.
We also leave as future work how to distinguish the two contributions to the C$\nu$B spectral distortions by numerical simulations and actual experiments.
\section{Conclusions}
\label{sec:con}

In the near future, CMB-S4 will determine $N_{\rm eff}$ with a very good precision of $\sim 0.03$ at $68\%$ C.L., and consequently confirm neutrino decoupling process in the SM and/or impose severe constraints on many scenarios in physics beyond the SM. In addition, in the future, a direct observation of the C$\nu$B might bring us more information about the early universe and neutrino physics. In both observations, the C$\nu$B spectrum is one of crucial ingredients to estimate $N_{\rm eff}$ and a direct detection rate.

In this article, we review the formula of kinetic equations for neutrinos in the early universe, which are the quantum Boltzmann equations for neutrinos and the continuity equation and the possible spectral distortions due to $e^\pm$-annihilation in neutrino decoupling. We also discuss the impact of the distortion of the C$\nu$B spectrum in neutrino decoupling on direct observation of the C$\nu$B on tritium, with emphasis on the PTOLEMY-type experiment.

We find $N_{\rm eff}= 3.044$ \cite{Akita:2020szl, Froustey:2020mcq, Bennett:2020zkv} by solving the kinetic equations for neutrino density matrix in the early universe, including vacuum three-flavor oscillations, oscillations in $e^\pm$-background, finite temperature corrections to $m_e$, $\rho$ and $P$ up to the next-to-leading order $\mathcal{O}(e^3)$ (see also ref.~\cite{Bennett:2019ewm} for the first suggestion on the importance of this contribution), and the collision term where we consider full diagonal parts and off-diagonal parts derived from charged current interactions but neglect off-diagonal parts derived from neutral current interactions. Later, the authors in refs.~\cite{Froustey:2020mcq, Bennett:2020zkv} also find $N_{\rm eff}=3.0440$ and $3.0440\pm 0.0002$, respectively, including off-diagonal parts in the collision term derived from neutrino neutral current interactions. Effects of their off-diagonal parts, and the choice of neutrino mass and mixing parameters on $N_{\rm eff}$ are quite small, $\delta N_{\rm eff}\sim \pm (1-2)\times 10^{-4}$ \cite{Bennett:2020zkv}.
In refs.~\cite{Akita:2020szl, Froustey:2020mcq, Bennett:2020zkv}, the Dirac CP-violating phase in neutrino mixing parameters is neglected. This contribution to $N_{\rm eff}$ is expected to be also quite small since increases and decreases for the energy densities of neutrinos and anti-neutrinos due to the Dirac CP-violating phase would be canceled out (see also ref.~\cite{Froustey:2021azz}).
However, QED corrections to weak interaction rates at order $\mathcal{O}(e^2G_F^2)$ and forward scattering of neutrinos via their self-interactions have not been precisely taken into account.
Recent studies \cite{EscuderoAbenza:2020cmq, Hansen:2020vgm} suggest that these neglects might still induce uncertainties of $ \pm(10^{-3}-10^{-4})$ in $N_{\rm eff}$.
Although we should consider their contributions to $N_{\rm eff}$ in the future, $N_{\rm eff}=3.044$ is still a very good reference value.

We have revealed the spectrum, number and energy density of the C$\nu$B in the current homogeneous and isotropic universe, including the spectral distortions in neutrino decoupling, as in the right panel of figure~\ref{fig:massx-f} and tables~\ref{tb:mass2} and \ref{tb:mass3}.
Then we have discussed the capture rates of the C$\nu$B on tritium with $1\%$ precision to observe effects of $1\%$ enhancement of the number density of the C$\nu$B by the spectral distortions due to $e^\pm$-annihilation during neutrino decoupling.
Unfortunately, it is extremely difficult to observe such sub-dominant effects since we will need more than 10 kg of tritium.
The precise capture rates of the C$\nu$B on tritium will be also useful to distinguish the SM from physics beyond the SM properly.

If observations and theoretical estimations of the C$\nu$B spectrum are improved significantly, we will obtain much richer information about neutrino physics and the early universe. 
Through direct observations of the C$\nu$B, one can impose significant constraints on neutrino decays and lifetimes in the region of the age of the universe, $t_0=4.35 \times 10^{17}\ {\rm s}$ \cite{Long:2014zva, Akita:2021hqn}. 
The C$\nu$B spectrum would also have fluctuations imprinted by inflationary perturbations.
Towards a precise estimation of anisotropy of the C$\nu$B as the CMB, one would need to solve kinetic equations for neutrinos in an anisotropic background, develop a detection method of the anisotropy, and reduce uncertainties of physical constants such as neutrino mass and mixing parameters, and Newton constant.


\section*{Acknowledgments}

We are grateful to Saul Hurwitz for the collaboration in the work~\cite{Akita:2020jbo} and Gaetano Lambiase for useful comments.
KA is supported by IBS under the project code, IBS-R018-D1. M.~Y. acknowledges financial support from JSPS Grant-in-Aid for Scientific Research No. JP18K18764, JP21H01080, JP21H00069. \\

\appendix

\section{Kinetic equations for neutrinos in comoving variables}
\label{appa}
In this appendix, we write the Boltzmann equations for the neutrino density matrix (\ref{BE}) and the continuity equation (\ref{EC}) in terms of the comoving variables, $x=m_ea,\ y=pa,\ z=T_{\gamma}a $. In terms of these variables, we can write the Boltzmann equations (\ref{BE}) as in ref.~\cite{deSalas:2016ztq},
\begin{align}
\frac{d\rho_y(x)}{dx}=m_{\rm Pl}\sqrt{\frac{3}{8\pi\bar{\rho}}}\left\{-i\frac{x^2}{m_e^3}\left[\bar{\mathcal{H}}_y(x), \rho_y(x)\right]+\frac{m_e^3}{x^4}\bar{C}[\rho_y(x)] \right\}.
\end{align}
where $\bar{\rho}$, $\bar{\mathcal{H}}_y(x)$, and $\bar{C}[\rho_y(x)]$ are quantities written in the comoving variables, $x,\ y,\ z$. 
Here we have used the following relations for the Hubble parameter, 
\begin{align}
H&=\frac{1}{m_{\rm Pl}}\sqrt{\frac{8\pi\rho}{3}}, \nonumber \\
\rho&=\bar{\rho}\left(\frac{m_e}{x}\right)^4.
\end{align}
The effective Hamiltonian for neutrino oscillations in vacuum and the forward scattering of neutrinos with the $e^\pm,\nu,\bar{\nu}$-background (multiplied by $m_e/x$), $\bar{\mathcal{H}}_y(x)$, is given by
\begin{align}
    \bar{\mathcal{H}}_y(x)&=\frac{\bm{\mathrm{M}}^2}{2y}+\sqrt{2}G_F\left(\frac{m_e}{x}\right)^4(\bar{\bm{\mathrm{N}}}_{e^-}-\bar{\bm{\mathrm{N}}}_{e^+})+\sqrt{2}G_F\left(\frac{m_e}{x}\right)^4(\bar{\bm{\mathrm{N}}}_\nu-\bar{\bm{\mathrm{N}}}_{\bar{\nu}}) \nonumber \\
&\ \ \ \ -\frac{2\sqrt{2}G_Fy}{m_W^2}\left(\frac{m_e}{x}\right)^6(\bar{\bm{\mathrm{E}}}_{e^-}+\bar{\bm{\mathrm{P}}}_{e^-}+\bar{\bm{\mathrm{E}}}_{e^+}+\bar{\bm{\mathrm{P}}}_{e^+})-\frac{8\sqrt{2}G_Fy}{3m_Z^2}\left(\frac{m_e}{x}\right)^6(\bm{\bar{\mathrm{E}}}_\nu+\bar{\bm{\mathrm{E}}}_{\bar{\nu}}),
\end{align}
where $\bar{\bm{\mathrm{N}}}_{e^\pm},\ \bar{\bm{\mathrm{N}}}_{\nu, \bar{\nu}},\ \bar{\bm{\mathrm{E}}}_{e^\pm}, \bar{\bm{\mathrm{P}}}_{e^\pm},\ \bar{\bm{\mathrm{E}}}_{\nu,\bar{\nu}}$ are written in the flavor basis around the temperature of MeV scale as
\begin{align}
&\bar{\bm{\mathrm{N}}}_{e^-}-\bar{\bm{\mathrm{N}}}_{e^+}={\rm diag}(\bar{n}_{e^-}-\bar{n}_{e^+},\ 0,\ 0),\  \ n_{e^\pm}=2\int\frac{d^3y}{(2\pi)^3}f_{e^\pm}(y), \nonumber \\
&\bar{\bm{\mathrm{N}}}_{\nu}-\bar{\bm{\mathrm{N}}}_{\bar{\nu}}=\int\frac{d^3y}{(2\pi)^3}\left(\rho_y(x)-\bar{\rho}_y(x) \right), \nonumber \\
&\bar{\bm{\mathrm{E}}}_{e^\pm}+\bar{\bm{\mathrm{P}}}_{e^\pm}={\rm diag}(\bar{\rho}_{e^\pm}+\bar{P}_e{^\pm},\ 0,\ 0),
\ \ \bar{\rho}_{e^\pm}+\bar{P}_{e^\pm}=\int\frac{d^3y}{(2\pi)^3}\left(\bar{E}_e+ \frac{y^2}{3\bar{E}_e} \right)f_{e^\pm}(y), \nonumber \\
&\bar{\bm{\mathrm{E}}}_{\nu}+\bar{\bm{\mathrm{E}}}_{\bar{\nu}}=\int\frac{d^3y}{(2\pi)^3}y\left(\rho_y(x)+\bar{\rho}_y(x) \right),
\end{align}
where, neglecting the chemical potential for $e^\pm$,
\begin{align}
    f_{e^\pm}(y)=\frac{1}{e^{\bar{E}_e/z}+1},\ \ \bar{E}_{e}=\sqrt{y^2+x^2+\delta \bar{m}^2_e}.
\end{align}
$\delta \bar{m}^2_e$ is the finite temperature correction to the electron mass up to $\mathcal{O}(e^2)$ in the comoving variables, ignoring the logarithmic term in eq.~(\ref{delta_me}) and the chemical potential for $e^\pm$,
\begin{align}
\delta \bar{m}_e^2 = \frac{e^2 z}{6}+\frac{e^2}{\pi^2}\int dy \frac{y^2}{\sqrt{y^2+x^2}}\frac{1}{\exp(\sqrt{y^2+x^2}/z)+1}.
\end{align}

The collision term in the comoving variables can be also decomposed as in eq.~(\ref{CTS})
\begin{align}
    \bar{C}[\rho_y(x)]=\bar{C}^{\nu\bar{\nu}\leftrightarrow e^-e^+}+\bar{C}^{\nu e^-\leftrightarrow \nu e^-}+\bar{C}^{\nu e^+\leftrightarrow \nu e^+}+\bar{C}^{\nu \nu\leftrightarrow \nu \nu}+\bar{C}^{\nu \bar{\nu}\leftrightarrow \nu\bar{\nu}}.
\end{align}
The collision terms from the annihilation and scattering processes including both $\nu$ and $e^\pm$ are, neglecting the chemical potential for $e^\pm$ and reducing nine-dimensional collision integrals in eq.~(\ref{CI}) to two integrals as in appendix~\ref{appb},
\begin{align}
&\bar{C}^{\nu\bar{\nu}\leftrightarrow e^-e^+}[\rho_{y_1}(x)] \nonumber \\
&=\frac{G_F^2}{2\pi^3y_1}\int dy_2dy_3\ y_2y_3\bar{E}_4 \nonumber \\
&\ \ \ \   \times \biggl[\Pi_{\rm ann}^1F^{LL}_{\rm ann}\left(\nu^{(1)}, \bar{\nu}^{(2)}, e^{-(3)}, e^{+(4)}\right) +\Pi_{\rm ann}^2F^{RR}_{\rm ann}\left(\nu^{(1)}, \bar{\nu}^{(2)}, e^{-(3)}, e^{+(4)}\right) \nonumber \\
&\ \ \ \ \ \ \ \   +\Pi_{\rm ann}^3\Bigl(F^{RL}_{\rm ann}\left(\nu^{(1)},\bar{\nu}^{(2)}, e^{-(3)}, e^{+(4)} \right)+ F^{LR}_{\rm ann}\left(\nu^{(1)}, \bar{\nu}^{(2)}, e^{-(3)}, e^{+(4)}\right)  \Bigl) \biggl], \label{CAC} \\
&\bar{C}^{\nu e^-\leftrightarrow \nu e^-}[\rho_{y_1}(x)]+\bar{C}^{\nu e^+\leftrightarrow \nu e^+}[\rho_{y_1}(x)] \nonumber \\
&=\frac{G_F^2}{2\pi^3y_1}\int dy_2dy_3\ y_2y_3\bar{E}_4 \nonumber \\
&\ \ \ \   \times \biggl[\Pi_{\rm sc}^1\Bigl(F^{LL}_{\rm sc}\left(\nu^{(1)},e^{(2)}, \nu^{(3)},e^{(4)}\right) +F^{RR}_{\rm sc}\left(\nu^{(1)},e^{(2)}, \nu^{(3)},e^{(4)}\right) \Bigl) \nonumber \\
&\ \ \ \ \ \ \ \ -\Pi_{\rm sc}^2 \Bigl(F^{LR}_{\rm sc}\left(\nu^{(1)},e^{(2)}, \nu^{(3)},e^{(4)}\right) +F^{RL}_{\rm sc}\left(\nu^{(1)},e^{(2)}, \nu^{(3)},e^{(4)}\right) \Bigl) \biggl],
\label{CSCC}
\end{align}
where $\bar{E}_i=\sqrt{y_i^2+x^2+\delta\bar{m}_e^2}$
and $F^{ab}_{\rm sc}\left(\nu^{(1)},e^{-(2)}, \nu^{(3)},e^{-(4)}\right)=F^{ab}_{\rm sc}\left(\nu^{(1)},e^{+(2)}, \nu^{(3)},e^{+(4)}\right)=F^{ab}_{\rm sc}\left(\nu^{(1)},e^{(2)}, \nu^{(3)},e^{(4)}\right)$ due to no lepton asymmetry. $F^{ab}_{\rm ann}$ and 
$F^{ab}_{\rm sc}$ are given by eqs.~(\ref{FAAA}) and (\ref{FSC}).
Similarly, the collision terms from the self-interaction processes in the comoving variables are
\begin{align}
&\bar{C}^{\nu\nu\leftrightarrow \nu\nu}[\rho_{y_1}(x)]+\bar{C}^{\nu\bar{\nu}\leftrightarrow \nu\bar{\nu}}[\rho_{y_1}(x)] \nonumber \\
&=\frac{G_F^2}{2\pi^3y_1}\int dy_2dy_3\ y_2y_3y_4 \nonumber \\
&\ \ \ \
\times \biggl[\Pi_{\rm self}^1F_{\rm sc}(\nu^{(1)},\nu^{(2)},\nu^{(3)},\nu^{(4)}) \nonumber \\
&\ \ \ \ \ \ \ \  +\Pi_{\rm self}^2\left(F_{\rm sc}(\nu^{(1)},\nu^{(2)},\nu^{(3)},\nu^{(4)}) +F_{\rm ann}(\nu^{(1)},\bar{\nu}^{(2)},\nu^{(3)},\bar{\nu}^{(4)})\right) \biggl].
\label{CSC}
\end{align}
$F_{\rm sc}\left(\nu^{(1)},\nu^{(2)}, \nu^{(3)},\nu^{(4)}\right)$, $F_{\rm sc}\left(\nu^{(1)},\bar{\nu}^{(2)}, \nu^{(3)},\bar{\nu}^{+(4)}\right)$ and $F_{\rm ann}\left(\nu^{(1)},\bar{\nu}^{(2)}, \nu^{(3)},\bar{\nu}^{(4)}\right)$ are given by eqs.~(\ref{Fself1})-(\ref{Fself3}).
The functions $\Pi_{\rm self,ann,sc}^{1,2,3}$ in eqs.~(\ref{CSC}), (\ref{CAC})
and (\ref{CSCC}) take the following forms,
\begin{align}
\Pi_{\rm ann}^1&= 2D_1-\frac{2D_2(y_2,y_3)}{y_2\bar{E}_3}-\frac{2D_2(y_1,y_4)}{y_1\bar{E}_4}+\frac{2D_3}{y_1y_2\bar{E}_3\bar{E}_4}, \nonumber \\
\Pi_{\rm ann}^2&= 2D_1-\frac{2D_2(y_2,y_4)}{y_2\bar{E}_4}-\frac{2D_2(y_1,y_3)}{y_1\bar{E}_3}+\frac{D_3}{y_1y_2\bar{E}_3\bar{E}_4}, \nonumber \\
\Pi_{\rm ann}^3&= (x^2+\delta \bar{m}_e^2)\left(D_1+\frac{D_2(y_1,y_2)}{y_1y_2}\right)\frac{1}{\bar{E}_3\bar{E}_4}, \nonumber \\
\Pi_{\rm sc}^1&=4D_1-\frac{2D_2(y_2,y_3)}{\bar{E}_2y_3}-\frac{2D_2(y_1,y_4)}{y_1\bar{E}_4}+\frac{2D_2(y_3,y_4)}{y_3\bar{E}_4}+\frac{2D_2(y_1,y_2)}{y_1\bar{E}_2}+\frac{4D_3}{y_1\bar{E}_2y_3\bar{E}_4}, \nonumber \\
\Pi_{\rm sc}^2&= 2(x^2+\delta \bar{m}_e^2)\left(D_1-\frac{D_2(y_1,y_3)}{y_1y_3} \right)\frac{1}{\bar{E}_2\bar{E}_4}, \nonumber \\
\Pi_{\rm self}^1&=D_1+\frac{D_2(y_1,y_2)}{y_1y_2}+\frac{D_2(y_3,y_4)}{y_3y_4}+\frac{D_3}{y_1y_2y_3y_4}, \nonumber \\
\Pi_{\rm self}^2&=D_1-\frac{D_2(y_2,y_3)}{y_2y_3}-\frac{D_2(y_1,y_4)}{y_1y_4}+\frac{D_3}{y_1y_2y_3y_4}.
\end{align}
The functions of $D_{1,2,3}$ are written as,
\begin{align}
D_1&=\frac{4}{\pi}\int^{\infty}_0\frac{d\lambda}{\lambda^2} \sin(\lambda y_1) \sin(\lambda y_2) \sin(\lambda y_3) \sin(\lambda y_4), \nonumber \\
D_2(y_3,y_4)&=\frac{4y_3y_4}{\pi}\int^{\infty}_0\frac{d\lambda}{\lambda^2} \sin(\lambda y_1) \sin(\lambda y_2)\left[\cos(\lambda y_3)-\frac{\sin(\lambda y_3)}{\lambda y_3} \right]\left[\cos(\lambda y_4)-\frac{\sin(\lambda y_4)}{\lambda y_4} \right], \nonumber \\
D_3&=\frac{4y_1y_2y_3y_4}{\pi}\int^{\infty}_0\frac{d\lambda}{\lambda^2}\left[\cos(\lambda y_1)-\frac{\sin(\lambda y_1)}{\lambda y_1} \right]\left[\cos(\lambda y_2)-\frac{\sin(\lambda y_2)}{\lambda y_2} \right] \nonumber \\
&\ \ \ \ \ \ \ \ \ \ \ \ \ \ \ \ \ \ \ \  \times \left[\cos(\lambda y_3)-\frac{\sin(\lambda y_3)}{\lambda y_3} \right]\left[\cos(\lambda y_4)-\frac{\sin(\lambda y_4)}{\lambda y_4} \right],
\label{Dfunc}
\end{align}
which can be integrated out analytically as in appendix~\ref{appb}.

If we neglect the off-diagonal components of $\rho_y(x)$ in the collision terms from neutrino self-interactions, which could have a negligible effect on $N_{\rm eff}$ with $10^{-3}$ precision, their collision terms are reduced to
\begin{align}
    &\bar{C}^{\nu\nu\leftrightarrow \nu\nu}[\rho_{y_1}(x)]+\bar{C}^{\nu\bar{\nu}\leftrightarrow \nu\bar{\nu}}[\rho_{y_1}(x)]\bigl|_{\rm diag} \nonumber \\
    &=\frac{G_F^2}{2\pi^3y_1}\int dy_2dy_3\ y_2y_3y_4
     \biggl[\left(2\Pi_{\rm self}^1+4\Pi_{\rm self}^2\right)(\nu_{\alpha}^{(1)},\nu_{\alpha}^{(2)},\nu_{\alpha}^{(3)},\nu_{\alpha}^{(4)}) \nonumber \\
&\ \ \ \ \ \ \ \ \ \ \ \  +\left(\Pi_{\rm self}^1+\Pi_{\rm self}^2\right)F(\nu_{\alpha}^{(1)},\nu_{\beta}^{(2)},\nu_{\alpha}^{(3)},\nu_{\beta}^{(4)}) +\Pi_{\rm self}^2F(\nu_{\alpha}^{(1)},\nu_{\alpha}^{(2)},\nu_{\beta}^{(3)},\nu_{\beta}^{(4)}) \nonumber \\
&\ \ \ \ \ \ \ \ \ \ \ \  +\left(\Pi_{\rm self}^1+\Pi_{\rm self}^2\right)F(\nu_{\alpha}^{(1)},\nu_{\gamma}^{(2)},\nu_{\alpha}^{(3)},\nu_{\gamma}^{(4)}) +\Pi_{\rm self}^2F(\nu_{\alpha}^{(1)},\nu_{\alpha}^{(2)},\nu_{\gamma}^{(3)},\nu_{\gamma}^{(4)}) \biggl].
\label{CTSreduced}
\end{align}
where 
\begin{align}
   F(\nu_\alpha ^{(1)}, \nu_\beta^{(2)}, \nu_\gamma^{(3)}, \nu_\delta^{(4)} ) &= f_{\nu_\gamma}(y_3)f_{\nu_\delta}(y_4)\left(1-f_{\nu_\alpha}(y_1)\right)\left(1-f_{\nu_\beta}(y_2)\right) \nonumber \\
&-f_{\nu_\alpha}(y_1)f_{\nu_\beta}(y_2)\left(1-f_{\nu_\gamma}(y_3)\right)\left(1-f_{\nu_\delta}(y_4)\right).
\end{align}

Finally, the continuity equation (\ref{EC}) is translated into the evolution equation for $z$, including finite temperature corrections from QED up to $\mathcal{O}(e^3)$ but neglecting the logarithmic $\mathcal{O}(e^2)$ corrections \cite{Mangano:2001iu, Bennett:2019ewm}, 
\begin{align}
\frac{dz}{dx}=\frac{\frac{x}{z}J(x/z)-\frac{1}{2\pi^2z^3}\int^{\infty}_0dy\ y^3\left(\frac{df_{\nu_e}}{dx}+\frac{df_{\nu_\mu}}{dx}+\frac{df_{\nu_\tau}}{dx} \right)+G^{(2)}_1(x/z)+G^{(3)}_1(x/z)}{\frac{x^2}{z^2}J(x/z)+Y(x/z)+\frac{2\pi^2}{15}+G^{(2)}_2(x/z)+G^{(3)}_2(x/z)},
\end{align} 
where
\begin{align}
G_1^{(2)}(\omega)&= 2\pi\alpha \left[\frac{1}{\omega}\biggl(\frac{K(\omega)}{3} +2K(\omega)^2-\frac{J(\omega)}{6}-K(\omega)J(\omega)\right) \nonumber \\
&\ \ \ \ \ \ \ \ +\left(\frac{K'(\omega)}{6}-K(\omega)K'(\omega)+\frac{J'(\omega)}{6}+J'(\omega)K(\omega)+J(\omega)K'(\omega) \right) \biggl], \nonumber \\
G_2^{(2)}(\omega)&=-8\pi\alpha \left(\frac{K(\omega)}{6}+\frac{J(\omega)}{6}-\frac{1}{2}K(\omega)^2+K(\omega)J(\omega) \right) \nonumber \\
&\ \ \ \ \ \ \ \ +2\pi\alpha\omega \left(\frac{K'(\omega)}{6}-K(\omega)K'(\omega)+\frac{J'(\omega)}{6} +J'(\omega)K(\omega)+J(\omega)K'(\omega) \right),  \nonumber \\
G_1^{(3)}(\omega)&=\frac{e^3}{4\pi}\left(K(\omega)+\frac{\omega^2}{2}k(\omega) \right)^{1/2} \biggl[ \frac{1}{\omega} \left(2J(\omega)-4K(\omega) \right)-2J'(\omega)-\omega^2j'(\omega) \nonumber \\
&\ \ \ \ \ \ \ \ \ \ \ \ \ \ \ \ \ \ \ \ \ \   -\omega\left(2k(\omega)+j(\omega)\right)-\frac{\left(2J(\omega)+\omega^2j(\omega)\right)\left(\omega\left(k(\omega)-j(\omega)\right)+K'(\omega)\right)}{2\left(2K+\omega^2k(\omega)\right)} \biggl], \nonumber \\
G_2^{(3)}(\omega)&=\frac{e^3}{4\pi}\left(K(\omega)+\frac{\omega^2}{2}k(\omega) \right)^{1/2}\left[\frac{(2J(\omega)+\omega^2j(\omega))^2}{2(2K(\omega)+\omega^2k(\omega))}-\frac{2}{\omega}Y'(\omega)-\omega\left(3J'(\omega)+\omega^2j'(\omega) \right) \right], \nonumber \\
\end{align} 
with
\begin{align}
K(\omega)&=\frac{1}{\pi^2}\int^{\infty}_0du~\frac{u^2}{\sqrt{u^2+\omega^2}}\frac{1}{\exp\left(\sqrt{u^2+\omega^2}\right)+1}, \nonumber \\
J(\omega)&=\frac{1}{\pi^2}\int^{\infty}_0du~u^2\frac{\exp\left(\sqrt{u^2+\omega^2}\right)}{\left(\exp\left(\sqrt{u^2+\omega^2}\right)+1\right)^2}, \nonumber \\
Y(\omega)&=\frac{1}{\pi^2}\int^{\infty}_0du~u^4\frac{\exp\left(\sqrt{u^2+\omega^2}\right)}{\left(\exp\left(\sqrt{u^2+\omega^2}\right)+1\right)^2}, \nonumber \\
k(\omega)&=\frac{1}{\pi^2}\int^{\infty}_0du \frac{1}{\sqrt{u^2+\omega^2}}\frac{1}{\exp\left(\sqrt{u^2+\omega^2}\right)+1}, \nonumber \\
j(\omega)&=\frac{1}{\pi^2}\int^{\infty}_0du \frac{\exp\left(\sqrt{u^2+\omega^2}\right)}{\left(\exp\left(\sqrt{u^2+\omega^2}\right)+1\right)^2}.
\end{align}
The prime represents the derivative with respect to $\omega$.
$G^{(2)}(\omega)$ and $G^{(3)}(\omega)$ denote QED finite temperature corrections at $\mathcal{O}(e^2)$ and $\mathcal{O}(e^3)$, respectively.


\section{Reduction of the collision integrals}
\label{appb}
In this appendix, we analytically perform seven out of nine integrations in the collision terms for four-Fermi interaction processes at order of $\mathcal{O}(G_F^2)$ in the homogeneous and isotropic universe, following refs.~\cite{Dolgov:1997mb,Blaschke:2016xxt}.  We consider the general form of the collision term in this case,
\begin{align}
C_{\rm coll}=\frac{1}{2E_1}\int(2\pi)^4\delta^{4}(\sum_ip_i)\left(\mathcal{|M|}^2\right)_{\rm part}F\left(\rho_p\right)\prod_{i=2}^4\frac{d^3\bm{p}_i}{(2\pi)^32E_i},
\label{CC}
\end{align}
where $E_i$ is the energy of $i$-th particle. The matrix $F\left(\rho_p \right)$ is a function of neutrino density matrix and $\left(|\mathcal{M}|^2\right)_{\rm part}$ is a part of the possible squared matrix elements summed over spin degrees of freedom of all particles except for the first particle $|\mathcal{M}|^2$. We change the delta function for 3-momentum into the exponential representation:
\begin{align}
\delta^{(3)}(\sum_i \bm{p}_i)=\int e^{\bm{\lambda}\cdot(\bm{p}_1+\bm{p}_2-\bm{p}_3-\bm{p}_4)}\frac{d^3\bm{\lambda}}{(2\pi)^3},
\label{delta}
\end{align}
and decompose momentum integrations into the radial and angle components,
\begin{align}
d^3\bm{p}_i=p_i^2dp_i\sin\theta_id\theta_id\phi_i \equiv p_i^2dp_id\Omega_i.
\label{d^3p}
\end{align}
Using eqs.~(\ref{delta}) and (\ref{d^3p}), we rewrite the general collision term (\ref{CC}) to
\begin{align}
C_{\rm coll}=\frac{1}{64\pi^3E_1p_1}\int \delta(E_1+E_2-E_3-E_4)F(\rho_p(t))D(p_1,p_2,p_3,p_4)\frac{p_2dp_2}{E_2}\frac{p_3dp_3}{E_3}\frac{p_4dp_4}{E_4},
\end{align}
where
\begin{align}
D(p_1,p_2,p_3,p_4)&=\frac{p_1p_2p_3p_4}{64\pi^5}\int^{\infty}_0\lambda^2d\lambda \int e^{i\bm{\lambda}\cdot \bm{p}_1}d\Omega_\lambda\int e^{i\bm{\lambda}\cdot \bm{p}_2}d\Omega_{p_2} \nonumber \\
&\ \ \ \ \ \ \ \ \ \ \ \times \int e^{-i\bm{\lambda}\cdot \bm{p}_3}d\Omega_{p_3}\int e^{-i\bm{\lambda}\cdot \bm{p}_4}d\Omega_{p_4}|\mathcal{M}|^2.
\label{Dlambda}
\end{align}
For four-Fermi interaction processes at order of $\mathcal{O}(G_F^2)$, all of $|\mathcal{M}|^2$ have two kinds of forms,
\begin{align}
K_1({q_1}_\mu q_2^\mu)({q_3}_{\nu}q_4^\nu)&=K_1(E_1E_2-\bm{q}_1\cdot\bm{q}_2)(E_3E_4-\bm{q}_3\cdot\bm{q}_4), \label{K1} \\
K_2m^2({q_3}_\mu q_4^\mu)&=K_2m^2(E_3E_4-\bm{q}_3\cdot \bm{q}_4),
\label{K2}
\end{align}
where $q_i$ corresponds to one of $p_j$ and the angle between $\bm{q}_i$ and $\bm{q}_j$ is written in terms of the integration variables of angle,
\begin{align}
\cos\psi_{ij}=\sin\theta_i\sin\theta_j\cos(\phi_i-\phi_j)+\cos\theta_i\cos\theta_j.
\end{align}
In both cases of eqs.~(\ref{K1}) and (\ref{K2}), we can perform all integrals for angle components in eq.~(\ref{Dlambda}) so that $D(q_1,q_2,q_3,q_4)$ in the case of eq.~(\ref{K1}) reduces to
\begin{align}
D=K_1[E_1E_2E_3E_4D_1 + E_1E_2D_2(q_3,q_4)+E_3E_4D_2(q_1,q_2) + D_3],
\end{align}
while in the case of eq.~(\ref{K2}),  $D(q_1,q_2,q_3,q_4)$ is given by 
\begin{align}
D=K_2E_1E_2[E_3E_4D_1+D_2(q_3,q_4)],
\end{align}
where $D_{1,2,3}$ are defined in eq.~(\ref{Dfunc}).

In the following we only consider $D_1,\ D_2(q_3,q_4),\ D_3$.
For simplicity we assume that $q_1>q_2$ and $q_3>q_4$ without loss of generality
though we can perform the integrals in $D_{1,2,3}$ without this assumption and obtain the exact expressions given in ref.~{\cite{Blaschke:2016xxt}}.
Then we obtain the simplified expressions of $D_{1,2,3}$ in four cases: \\
\\
$(1)\ q_1+q_2>q_3+q_4$, $q_1+q_4>q_2+q_3$ and $q_1 \leq q_2+q_3+q_4$
\begin{align}
D_1&=\frac{1}{2}(q_2+q_3+q_4-q_1), \nonumber \\
D_2(q_3,q_4)&=\frac{1}{12}\left((q_1-q_2)^3+2(q_3^3+q_4^3)-3(q_1-q_2)(q_3^2+q_4^2) \right), \nonumber \\
D_3&=\frac{1}{60}\bigl(q_1^5-5q_1^3q_2^2+5q_1^2q_2^3-q_2^5 \nonumber \\
&\ \ \ \ -5q_1^3q_3^2+5q_2^3q_3^2+5q_1^2q_3^3+5q_2^2q_3^3-q_3^5 \nonumber \\
&\ \ \ \ -5q_1^3q_4^2+5q_2^3q_4^2+5q_3^3q_4^2+5q_1^2q_4^3+5q_2^2q_4^3+5q_3^2q_4^3-q_4^5\bigl).
\label{Case1}
\end{align}
Note that the case $q_1>q_2+q_3+q_4$ is unphysical so that $D_1=D_2=D_3=0$ in this case.
\\
\\
$(2)\ q_1+q_2>q_3+q_4$ and $q_1+q_4<q_2+q_3$
\begin{align}
D_1&=q_4, \nonumber \\
D_2(q_3,q_4)&=\frac{1}{3}q_4^3, \nonumber \\
D_3&=\frac{1}{30}q_4^3\left(5q_1^2+5q_2^2+5q_3^2-q_4^2\right).
\end{align}
\\
$(3)\ q_1+q_2<q_3+q_4$, $q_1+q_4<q_2+q_3$ and $q_3 \leq q_1+q_2+q_4$
\begin{align}
D_1&=\frac{1}{2}(q_1+q_2+q_4-q_3), \nonumber \\
D_2(q_3,q_4)&=\frac{1}{12}\left(-(q_1+q_2)^3-2q_3^3+2q_4^3+3(q_1+q_2)(q_3^3+q_4^3)\right).
\end{align}
$D_3$ is equal to that in eq.~(\ref{Case1}) with the replacement of variables $q_1 \leftrightarrow q_3$ and $q_2 \leftrightarrow q_4$ and the case of $q_3>q_1+q_2+q_4$ is unphysical so that $D_1=D_2=D_3=0$ in this case.
\\
\\
$(4)\ q_1+q_2<q_3+q_4$ and $q_1+q_4>q_2+q_3$
\begin{align}
D_1&=q_2, \nonumber \\
D_2(q_3,q_4)&=\frac{1}{6}q_2\left(3q_3^2+3q_4^2-3q_1^2-q_2^2\right), \nonumber \\
D_3&=\frac{1}{30}q_2^3\left(5q_1^2+5q_3^2+5q_4^2-q_2^2 \right).
\end{align}
After we have integrated the $\delta$-function, we obtain the simplified expression of the collision term, leaving two integrals,
\begin{align}
C_{\rm coll}=\frac{1}{64\pi^3E_1p_1}\int \int F\left(\rho_p(t)\right)D(p_1,p_2,p_3,p_4)\frac{p_2dp_2}{E_2}\frac{p_3dp_3}{E_3},
\end{align}
where $E_4=E_1+E_2-E_3$ and $p_4=\sqrt{E_4^2-m_4^2}$. 


\section{Kinematics for $\nu_i+\mathrm{^3H} \rightarrow e^-+\mathrm{^3He}$ and $\mathrm{^3H}\rightarrow e^-+\mathrm{^3He}+\bar{\nu}_i$}
\label{appc}

In this appendix, we estimate the kinematics of inverse tritium $\beta$-decay for the C$\nu$B, $\nu_i+\mathrm{^3H} \rightarrow e^-+\mathrm{^3He}$, and tritium $\beta$-decay $\mathrm{^3H}\rightarrow e^-+\mathrm{^3He}+\bar{\nu}_i$. We also discuss the kinematic relations between the two processes.
In particular, we investigate the maximal energy of the electron emitted from $\beta$-decay, called the $\beta$-decay endpoint energy, and the energy of the electron emitted from the inverse $\beta$-decay process for the C$\nu$B.
Here we consider the nuclear process and use the nuclear masses of $\mathrm{^3H}$ and $\mathrm{^3He}$, $m_{\mathrm{^3H}}$ and $m_{\mathrm{^3He}}$.

We first consider the kinematics of tritium beta decay, $\mathrm{^3H}\rightarrow \mathrm{^3He}+e^-+\bar{\nu}_i$, in the rest frame of $\mathrm{^3H}$.
From 4-momentum conservation, the energy of the electron is
\begin{align}
E_e=\frac{m_{\mathrm{^3H}}^2+m_e^2-m_{\nu_i}^2-m_{\mathrm{^3He}}^2-2E_{\nu_i}E_{\mathrm{^3He}}+2|\bm{p}_{\nu}||\bm{p}_{\mathrm{^3He}}|\cos \theta_{\nu \mathrm{^3He}}}{2m_{\mathrm{^3H}}}.
\end{align}
The maximal energy, $E_{\rm end}$, is achieved when the emitted anti-neutrino is the lightest and $\cos \theta_{\nu \mathrm{^3He}}=1\ (\theta_{\nu \mathrm{^3He}}=0)$. 
When the neutrino and the helium-3 nucleus are emitted in parallel, the electron is produced in opposite direction. 
In addition, the maximization condition of the electron energy corresponds to the minimization condition of $(E_{\nu}+E_{\mathrm{^3He}})$, which yields 
\begin{align}
\frac{E_{\nu_i}}{E_{\mathrm{^3He}}}=\frac{|\bm{p}_{\nu}|}{|\bm{p}_{\mathrm{^3He}}|}={\frac{m_{\nu_i}}{m_{\mathrm{^3He}}}}.
\end{align}
From these conditions, the maximal energy of the electron for $\mathrm{^3H}\rightarrow e^-+\mathrm{^3He}+\bar{\nu}_i$ is given by
\begin{align}
E_e^{{\rm max}, i}=\frac{m_{\mathrm{^3H}}^2+m_e^2-(m_{\nu_i}+m_{\mathrm{^3He}})^2}{2m_{\mathrm{^3H}}}.
\label{Emaxi}
\end{align}  
The endpoint energy of the electron for the tritium $\beta$-decay is also given by
\begin{align}
    E_e^{{\rm end}}=\frac{m_{\mathrm{^3H}}^2+m_e^2-(m_{\rm lightest}+m_{\mathrm{^3He}})^2}{2m_{\mathrm{^3H}}}.
    \label{Eend}
\end{align}
If the lightest neutrino is massless, the endpoint energy is identified as
\begin{align}
E_e^{\rm end,0}=\frac{m_{\mathrm{^3H}}^2+m_e^2-m_{\mathrm{^3He}}^2}{2m_{\mathrm{^3H}}}.
\end{align}
Due to $m_{\mathrm{^3H}}\simeq m_{\mathrm{^3He}}$, the difference between the endpoint energy for the massive and massless lightest neutrinos is
\begin{align}
 E_e^{\rm end}-E_e^{\rm end,0} \simeq-m_{\rm lightest}.
\end{align} 

Next we investigate the kinematics of inverse tritium beta decay for relic cosmic neutrinos, $\nu_i + \mathrm{^3H}\rightarrow \mathrm{^3He}+e^-$. In the rest-frame of $\mathrm{^3H}$, we similarly obtain the energy of the electron as
\begin{align}
E_e^{{\rm{C\nu B},i}}&=\frac{(E_{\nu_i}+m_{\mathrm{^3H}})^2+m_e^2-|\bm{p}_{\nu}|^2+2|\bm{p}_{\nu}||\bm{p}_e|\cos \theta_{e\nu}-m_{\mathrm{^3He}}^2}{2(E_{\nu_i}+m_{\mathrm{^3H}})} \nonumber \\
&\simeq \frac{(E_{\nu_i}+m_{\mathrm{^3H}})^2+m_e^2-m_{\mathrm{^3He}}^2}{2(E_{\nu_i}+m_{\mathrm{^3H}})}.
\end{align}
where we neglect the terms proportional to $|\bm{p}_{\nu}|^2$ and $|\bm{p}_{\nu}||\bm{p}_e|$ and leave the term proportional to $E_{\nu_i}m_{\mathrm{^3H}}$ because of $m_{\mathrm{^3H}} \gg |\bm{p}_e| \gg |\bm{p}_\nu|$.
For $m_{\mathrm{^3H}} \gg m_e$, the difference between $E_e^{{\rm C\nu B},i}$ and $E_{\rm end}$ is
\begin{align}
E_e^{{\rm C\nu B},i}-E_e^{\rm end}\simeq E_{\nu_i}+m_{\rm lightest}.
\end{align}
Since $E_e^{{\rm C\nu B},i}-E_{\rm end}$ is (approximately) not function of any nuclear masses, it is insensitive to the uncertainties in the nuclear masses which are calculated from the measured values of atomic masses.

\section{Cross section for $\nu_i+\mathrm{^3H} \rightarrow e^-+\mathrm{^3He}$ and decay rate for $\mathrm{^3H}\rightarrow e^-+\mathrm{^3He}+\bar{\nu}_i$}
\label{appd}

In this section we derive the cross section with $1\%$ precision for $\nu_i+\mathrm{^3H} \rightarrow e^-+\mathrm{^3He}$, $\sigma_{\nu_i}$, following ref.~\cite{Long:2014zva} and the decay rate for $\mathrm{^3H}\rightarrow e^-+\mathrm{^3He}+\bar{\nu}_i$, $\Gamma_\beta$. We also discuss the spectrum for the tritium $\beta$-decay, $d\Gamma_\beta/dE_e$.


\subsection{Cross section for $\nu_i+\mathrm{^3H} \rightarrow e^-+\mathrm{^3He}$}
\label{appd1}
In this section, we follow ref.~\cite{Long:2014zva}.
The differential cross section for $\nu_i+\mathrm{^3H} \rightarrow e^-+\mathrm{^3He}$ takes the following Lorentz invariant form:
\begin{align}
    \frac{d\sigma_{\nu_i}}{dt}=\frac{1}{16\pi}\frac{|\mathcal{M}_i|^2}{[s-(m_{\nu_i}+m_{\mathrm{^3H}})^2][s-(m_{\nu_i}-m_{\mathrm{^3H}})^2]},
\end{align}
where $s=(p_{\nu_i}+p_{\mathrm{^3H}})^2$ and $t=(p_{\nu_i}-p_e)^2$ are the Mandelstam variables, and $|\mathcal{M}_i|^2$ is the squared matrix element for the inverse $\beta$-decay.
In the rest frame of $\mathrm{^3H}$, $s$ and $t$ are expressed as
\begin{align}
    s&=(m_{\mathrm{^3H}}+E_{\nu_i})^2-|\bm{p}_{\nu}|^2=m_{\mathrm{^3H}}^2 + 2m_{\mathrm{^3H}}E_{\nu_i} +m_{\nu_i}^2, \nonumber \\
    t&=(E_e-E_{\nu_i})^2-|\bm{p}_e-\bm{p}_\nu|^2\simeq (m_e-m_{\nu_i})^2+2|\bm{p}_e||\bm{p}_{\nu}|\cos\theta.
\end{align}
Using also $dt/d\cos\theta=2|\bm{p}_e||\bm{p}_{\nu}|$, we obtain
\begin{align}
    \frac{d\sigma_{\nu_i}}{d\cos\theta}=\frac{1}{32\pi}\frac{1}{m_{\mathrm{^3H}}^2}\frac{|\bm{p}_e|}{|\bm{p}_\nu|}|\mathcal{M}_i|^2.
\end{align}
The matrix element for $\nu_i+\mathrm{^3H} \rightarrow e^-+\mathrm{^3He}$ is effectively given by
\begin{align}
    i\mathcal{M}_i = -i\frac{G_F}{\sqrt{2}}V_{ud}U^*_{ei}\left[\bar{u}_e\gamma^\mu(1-\gamma^5)u_{\nu_i}\right]\left[\bar{u}_{\mathrm{^3He}}\gamma_{\mu}\left(F -G \gamma^5 \right)u_{\mathrm{^3H}}\right],
\end{align}
where
\begin{align}
    F&=\langle f_F \rangle ,\ \ \ \ G=\frac{g_A}{\sqrt{3}g_V}\langle g_{GT} \rangle.
\end{align}
$u_{\alpha}$ denotes the Dirac spinor for species $\alpha$, $g_A\simeq1.2723$ and $g_V\simeq1$ are the axial and vector coupling constants respectively, and $\langle f_F \rangle\simeq 0.9998$ and $\langle g_{GT} \rangle \simeq \sqrt{3} \times (0.9511 \pm 0.0013)$ denote the reduced matrix elements of the Fermi and Gamow-Teller (GT) operators respectively \cite{Baroni:2016xll}.

After averaging over the spins of $\mathrm{^3H}$ and summing over the spins of the outgoing $e^-$ and $\mathrm{^3He}$, the squared matrix element is given by
\begin{align}
    \frac{1}{2}\sum_{s_e, s_{\mathrm{^3H}}, s_{\mathrm{^3He}}=\pm\frac{1}{2}}|\mathcal{M}_i|^2=
    \frac{G_F^2}{4}|V_{ud}|^2|U_{ei}|^2\mathcal{T}_1^{\alpha\beta}\mathcal{T}_{2\alpha\beta},
\end{align}
where
\begin{align}
    \mathcal{T}_1^{\alpha\beta}&= \sum_{s_e=\pm 1/2}{\rm tr}\left[\gamma^{\alpha}(1-\gamma^5)u_{\nu_i}\bar{u}_{\nu_i}\gamma^{\beta}(1-\gamma^5)u_e\bar{u}_e \right], \nonumber \\
    \mathcal{T}_2^{\gamma \delta}&=
     \sum_{s_\mathrm{^3H},s_{\mathrm{^3He}}=\pm 1/2}{\rm tr}\biggl[\gamma^{\gamma}\left(F -G \gamma^5 \right)u_{\mathrm{^3H}}\bar{u}_{\mathrm{^3H}} \gamma^{\delta}\left(F -G \gamma^5 \right)u_{\mathrm{^3He}}\bar{u}_{\mathrm{^3He}} \biggl].
     \label{mathT}
\end{align}
Using the completeness relations, we obtain the relation of Dirac spinors for $\mathrm{^3H}$, $\mathrm{^3He}$, and $e^-$,
\begin{align}
\sum_{s_j=\pm 1/2}u_j\bar{u}_j=(\slashed{p}_j+m_j),
\end{align}
and for neutrinos with their helicity $s_\nu$,
\begin{align}
u_{\nu_i}\overline{u}_{\nu_i}=\frac{1}{2}\bigl(\slashed{p}_{\nu_i}+m_{\nu_i}\bigl)\bigl(1+2s_{\nu}\gamma^5\slashed{S}_{\nu_i}\bigl),
\end{align}
where $S_{\nu_i}$ is the spin vector for neutrinos given by
\begin{align}
(S_{\nu_i})^{\alpha}=\left(\frac{|\bm{p}_{\nu}|}{m_{\nu_i}}, \frac{E_{\nu}}{m_{\nu_i}}\frac{\bm{p}_{\nu}}{|\bm{p}_\nu|}\right).
\end{align}
In the massless limit, the previous relation of the Dirac spinor for neutrinos becomes
\begin{align}
u_{\nu_i}\overline{u}_{\nu_i}=\frac{1}{2}\slashed{p}_{\nu_i}\left(1-2s\gamma^5 \right),
\end{align}
where we used $mS^{\mu}=p^{\mu}$ and $p_{\mu}S^{\mu}=0$.
Using the above relations, we rewrite eq.~(\ref{mathT}) as
\begin{align}
\mathcal{T}_1^{\alpha\beta}&=\frac{1}{2}\mathrm{tr}\left[\gamma^\alpha\bigl(1-\gamma^5\bigl)\bigl(\slashed{p}_{\nu_i}+m_{\nu_i}\bigl)\bigl(1+2s_\nu\gamma^5\slashed{S}_{\nu_i}\bigl)\gamma^\beta\bigl(1-\gamma^5\bigl)\bigl(\slashed{p}_e+m_e\bigl)\right], 
\label{N2}  \\
\mathcal{T}_2^{\gamma\delta}&=\mathrm{tr}\biggl[\gamma^\gamma\left(F -G \gamma^5 \right)\bigl(\slashed{p}_n+m_n\bigl) \gamma^\delta\left(F -G \gamma^5 \right)\bigl(\slashed{p}_p+m_p\bigl)\biggl].
\label{N3}
\end{align}
Then we obtain $\mathcal{T}_1^{\alpha\beta}\mathcal{T}_{2\alpha\beta}$ as
\begin{align}
\mathcal{T}_1^{\alpha\beta}\mathcal{T}_{2\alpha\beta}&= 32\bigl\{\left(G+F\right)^2\left[\left(p_e\cdot p_{\mathrm{^3He}}\right)\left(p_{\nu_i} \cdot p_{\mathrm{^3H}}\right)\right] 
+\left(G-F\right)^2\left[\left(p_e\cdot p_{\mathrm{^3H}}\right)\left(p_{\nu_i}\cdot p_{\mathrm{^3He}}\right)\right] \nonumber \\
&\ \ \ \ +\left(G^2-F^2\right)m_{\mathrm{^3H}}m_{\mathrm{^3He}}\left(p_e\cdot p_{\nu_i}\right)\bigl\} \nonumber \\
&\ \ \ \ -64s_\nu m_{\nu_i} \bigl\{\left(G+F\right)^2\left[\left(p_e\cdot p_{\mathrm{^3He}}\right)\left(S_{\nu_i} \cdot p_{\mathrm{^3H}}\right)\right]+\left(G-F\right)^2\left[\left(p_e\cdot p_{\mathrm{^3H}}\right)\left(S_{\nu_i}\cdot p_{\mathrm{^3He}}\right)\right] \nonumber \\
&\ \ \ \ +\left(G^2-F^2\right)m_{\mathrm{^3H}}m_{\mathrm{^3He}}\left(p_e\cdot S_{\nu_i}\right)\bigl\}.
\end{align}
In the rest frame of $\mathrm{^3H}$, neglecting the momentum of 3-helium $|\bm{p}_{\mathrm{^3He}}|/m_{\mathrm{^3He}}\sim (m_{\mathrm{^3H}}-m_{\mathrm{^3He}})/m_{\mathrm{^3He}}$ $\sim \mathcal{O}(10^{-4})$, $\mathcal{T}_1^{\alpha\beta}\mathcal{T}_{2\alpha\beta}$ is given by
\begin{align}
    &\mathcal{T}_1^{\alpha\beta}\mathcal{T}_{2\alpha\beta} \nonumber \\
& =32m_{\mathrm{^3H}}E_{\mathrm{^3He}}E_eE_{\nu_i}\left\{\left(F^2+3G^2 \right)\left(1-2s_{\nu}v_{\nu_i}\right) +\left(F^2-G^2\right)\left(v_{\nu_i}-2s_{\nu}\right)v_e\cos\theta  \right\}.
\label{NN}
\end{align}
We note that $\theta$ is the angle between $\bm{p}_e$ and $\bm{p}_\nu$.
Finally we obtain the differential cross section for $\nu_i+\mathrm{^3H} \rightarrow e^-+\mathrm{^3He}$, including the enhancement factor due to the Coulombic attraction between $e^-$ and $\mathrm{^3He}$, $F(2,E_e)$, and using also $F=\langle f_A \rangle$ and $G=\frac{g_A}{\sqrt{3}g_V}\langle g_{GT} \rangle$,
\begin{align}
    \frac{d\sigma_{\nu_i}}{d\cos\theta}&=\frac{G_F^2}{4\pi}|V_{ud}|^2|U_{ei}|^2F(2,E_e)\frac{m_{\mathrm{^3He}}}{m_{\mathrm{^3H}}v_{\nu_i}}E_e|\bm{p}_e| \nonumber \\
    & \times \left[\left(\langle f_A \rangle^2  +\frac{g_A^2}{g_V^2}\langle g_{GT} \rangle^2\right)(1-2s_\nu v_{\nu_i}) +\left(\langle f_A \rangle^2 -\frac{g_A^2}{3g_V^2}\langle g_{GT} \rangle^2\right)(v_{\nu_i}-2s_\nu)v_e\cos\theta \right].
\end{align}


\subsection{Decay rate for $\mathrm{^3H}\rightarrow e^-+\mathrm{^3He}+\bar{\nu}_i$}

The decay rate of the $\beta$-decay follows the standard formula at the rest frame of tritium,
\begin{align}
\Gamma_{\beta}&=\frac{1}{2^9\pi^5 m_{\mathrm{^3H}}}\int  \frac{d^3p_e d^3p_{\nu_i}d^3p_{\mathrm{^3He}}}{E_{e}E_{\nu_i}E_{\mathrm{^3He}}}|\mathcal{M}|^2\delta^{4}(p_{\mathrm{^3H}}-p_{e}-p_{\nu_i}-p_{\mathrm{^3He}}),\nonumber \\
&=\frac{1}{2^6\pi^4m_{\mathrm{^3H}}}\int dE_{e}dE_{\nu_i}|\mathcal{M}_\beta|^2,
\label{betadecayformula}
\end{align}
where $|\mathcal{M}_\beta|^2$ is the effective squared matrix element for $\beta$-decays summed over spins for the final states and averaged over spins for the initial state,
\begin{align}
|\mathcal{M}_\beta|^2= \frac{1}{2}\sum_{i=1}^3\sum_{s_{\mathrm{^3H}}, s_{\mathrm{^3He}}, s_{\nu_i}=\pm1/2} |\mathcal{M'}_i|^2,
\end{align}
where
\begin{align}
i\mathcal{M'}_i=-i\frac{G_F}{\sqrt{2}}V_{ud}U^{\ast}_{ei}\biggl[\bar{u}_e\gamma^{\mu}(1-\gamma^5)v_{\nu_i} \biggl]\left[\bar{u}_{\mathrm{^3H}}\gamma_{\mu}\left(\langle f_F \rangle -\frac{g_A}{\sqrt{3}g_V}\langle g_{GT} \rangle \gamma^5 \right)u_{\mathrm{^3He}} \right].
\end{align}
Then we integrate over $E_{\nu_i}$ for each $E_e$ in eq.~(\ref{betadecayformula}). The upper (lower) limit of the integral denotes $E_{\nu_i}^{\rm max}\ (E_{\nu_i}^{\rm min})$.
After some calculations, $E_{\nu_i}^{\rm max}-E_{\nu_i}^{\rm min}$ and  $E_{\nu_i}^{\rm max}+E_{\nu_i}^{\rm min}$ are given by
\begin{align}
E_{\nu_i}^{\rm max}-E_{\nu_i}^{\rm min}&=\frac{2m_{\mathrm{^3H}}|\bm{p}_{e}|}{M^2}(E_e^{{\rm max},i}-E_e)^{1/2}\left[E_e^{{\rm max}, i}-E_e+\frac{2m_{\nu_i}m_{\mathrm{^3He}}}{m_{\mathrm{^3H}}} \right]^{1/2}, \nonumber \\
E_{\nu_i}^{\rm max}+E_{\nu_i}^{\rm min}&=\frac{2m_{\mathrm{^3H}}}{M^2}(m_{\mathrm{^3H}}-E_e)\left[E_e^{{\rm max}, i}-E_e+\frac{m_{\nu_i}}{m_{\mathrm{^3H}}}(m_{\mathrm{^3He}}+m_{\nu_i}) \right],
\label{Emaxmin}
\end{align}
where $E_e^{{\rm max}, i}$ is the maximal energy of the emitted electron for $\mathrm{^3H}\rightarrow e^-+\mathrm{^3He}+\bar{\nu}_i$ given by eq.~(\ref{Emaxi}) in appendix \ref{appc}.
\begin{align}
M^2=m_{\mathrm{^3H}}^2-2m_{\mathrm{^3H}}E_e+m_e^2.
\end{align}
Then $d\Gamma_{\beta}/dE_e$ is given by
\begin{align}
\frac{d\Gamma_{\beta}}{dE_e}=\frac{1}{2^6\pi^3m_{\mathrm{^3H}}}\int^{E_{\nu_i}^{\rm max}}_{E_{\nu_i}^{\rm min}}dE_{\nu_i}|\mathcal{M}_\beta|^2.
\label{betaspectrum1}
\end{align}
After similar calculations in appendix~\ref{appd1}$, |\mathcal{M}_\beta|^2$ for $\beta$-decays at rest of tritium is written as
\begin{align}
|\mathcal{M}_\beta|^2&\simeq 16G_F^2|V_{ud}|^2\sum_{i=1}^3|U_{ei}|^2m_{\mathrm{^3H}}m_{\mathrm{^3He}}E_eE_{\nu_i} \nonumber \\
&\ \ \ \ \times \left[ \left(\langle f_F^2 \rangle +\frac{g_A^2}{g_V^2}\langle g_{GT}^2 \rangle \right) + \left(\langle f_F^2 \rangle -\frac{g_A^2}{3g_V^2}\langle g_{GT}^2 \rangle \right)\frac{\bm{p}_{\nu_i}\cdot \bm{p}_e}{E_{\nu_i}E_e} \right],
\label{M^2beta}
\end{align}
where we neglect the momentum of $\mathrm{^3He}$ due to $\bm{p}_{\mathrm{^3He}}\ll m_{\mathrm{^3He}}$. In addition, we neglect the second term in eq.~(\ref{M^2beta}) since $|\bm{p}_e|\sim m_{\mathrm{^3H}}-m_{\mathrm{^3He}} \ll E_e$. Thus, $|\mathcal{M}_\beta|^2$ approximately becomes
\begin{align}
|\mathcal{M}_\beta|^2\simeq 16G_F^2|V_{ud}|^2\sum_{i=1}^3|U_{ei}|^2m_{\mathrm{^3H}}m_{\mathrm{^3He}}E_eE_{\nu_i}
 \left(\langle f_F^2 \rangle +\frac{g_A^2}{g_V^2}\langle g_{GT}^2 \rangle \right).
 \label{M^2beta2}
\end{align}
Plugging eq.~(\ref{M^2beta2}) into eq.~(\ref{betaspectrum1}), we obtain
\begin{align}
\frac{d\Gamma_{\beta}}{dE_e}&=\frac{G_F^2}{8\pi^3}|V_{ud}|^2m_{\mathrm{^3He}}E_e \left(\langle f_F^2 \rangle +\frac{g_A^2}{g_V^2}\langle g_{GT}^2 \rangle \right) \nonumber \\
&\ \ \ \ \times \sum_{i=1}^3|U_{ei}|^2(E_{\nu_i}^{\rm max}+E_{\nu_i}^{\rm min})(E_{\nu_i}^{\rm max}-E_{\nu_i}^{\rm min}).
\label{betaspectrum2}
\end{align}
Finally, substituting eq.~(\ref{Emaxmin}) into eq.~(\ref{betaspectrum2}), we obtain the electron spectrum from the $\beta$-decays as 
\begin{align}
\frac{d\Gamma_\beta}{dE_e} = \frac{\bar{\sigma}}{\pi^2}N_T\sum_{i=1}^3|U_{ei}|^2H(E_e,m_{\nu_i}),
\end{align}
where $\bar{\sigma}$ is the average cross section at the leading order for neutrino capture, including the enhancement due to the Coulombic attraction between $e^-$ and $\mathrm{^3He}$, $F(2, E_e)$,
\begin{align}
\bar{\sigma}=\frac{G_F^2}{2\pi}|V_{ud}|^2\frac{m_{\mathrm{^3He}}}{m_{\mathrm{^3H}}}\left(\langle f_F \rangle^2+\frac{g_A^2}{g_V^2}\langle g_{GT} \rangle^2 \right)F(2, E_e)E_e|\bm{p}_e|.
\end{align}
$F(Z, E_e)$ is given in eq.~(\ref{FermiF}) and $H(E_e,m_{\nu_i})$ takes the following form,
\begin{align}
H(E_e, m_{\nu_i}) &= \frac{1-E_e/m_{\mathrm{^3H}}}{(1-2E_e/m_{\mathrm{^3H}}+m_e^2/m_{\mathrm{^3H}}^2)^2}
\sqrt{(E_e^{{\rm max}, i}-E_e)\left(E_e^{{\rm max}, i}-E_e+\frac{2m_{\nu_i}m_{\mathrm{^3He}}}{m_{\mathrm{^3H}}}\right)} \nonumber \\
&\ \ \ \ \times \left[E_e^{{\rm max}, i}-E_e+\frac{m_{\nu_i}}{m_{\mathrm{^3H}}}(m_{\mathrm{^3He}}+m_{\nu_i})\right].
\end{align}
Then we obtain $\Gamma_\beta$,
\begin{align}
    \Gamma_\beta=\int_{m_e}^{E_e^{\rm end}}dE_e\frac{d\Gamma_\beta}{dE_e},
\end{align}
where $E_e^{\rm end}=\max\{{E_e^{{\rm max}, 1}}, E_e^{{\rm max}, 2}, E_e^{{\rm max}, 3}\}$ is the endpoint energy of the tritium $\beta$-decay given by eq.~(\ref{Eend}) in appendix \ref{appc}.




\bibliographystyle{JHEP}
\bibliography{reference}

\end{document}